\documentclass[]{article}
\usepackage[margin=1in]{geometry} 
\usepackage{graphicx}
\usepackage{caption}
\usepackage{subcaption}
\usepackage{algorithm}
\usepackage{pdfpages}
\usepackage[noend]{algpseudocode}
\usepackage{graphicx}              
\usepackage{amsmath}               
\usepackage{amsfonts}              
\usepackage{amsthm}                
\usepackage{dsfont}
\usepackage{url}
\usepackage{hyperref}
\usepackage{epstopdf}
\usepackage{setspace}
\usepackage{rotating}
\usepackage{calrsfs}
\usepackage{mathrsfs}
\usepackage{amsmath}
\usepackage{amssymb}
\usepackage{amsthm}
\usepackage{enumitem}
\usepackage{mathtools}
\usepackage{longtable}
\usepackage{upgreek}
\usepackage{mathrsfs}
\usepackage{bm}

\DeclareMathAlphabet{\pazocal}{OMS}{zplm}{m}{n}

\newcommand{\superscript}[1]{^{^{#1}}}
\newcommand{\subscript}[1]{_{#1}}
\makeatother

\begin{document}

\doublespacing 

\title{\textbf{Mechanics and Modeling of Cold Rolling of Polymeric Films at Large Strains - A Rate-Independent  Approach}}

\author{Nikhil Padhye} 

\date{}
\maketitle

\tableofcontents
\newpage

\abstract


Recently, a new phenomenon of bonding polymeric films in solid-state, via symmetric rolling, at ambient temperatures ($\approx$ $20$ $^o$C) well below the glass transition temperature 
(T$_g$ $\approx$ $78$ $^o$C) of the polymer, has been reported. In this new type of bonding, polymer films are subject to plane strain active bulk plastic compression between the rollers 
during deformation. Here, we analyze these plane strain cold rolling processes, at large strains but slow strain rates, by finite element modeling.  We find that at low temperatures, slow strain rates, and moderate thickness reductions during rolling (at which Bauschinger effect can be neglected for the particular 
class of polymeric films studied here), the task of material modeling is greatly simplified, and enables us 
to deploy a computationally efficient, yet accurate, finite deformation rate-independent elastic-plastic material behavior (with inclusion of  isotropic-hardening) for analyzing these rolling 
processes. The finite deformation elastic-plastic material behavior based on (i) the additive decomposition of stretching tensor ($\mathbf{D=D^e+D^p}$, i.e., a hypoelastic formulation) with 
incrementally objective time integration and, (ii) multiplicative decomposition of deformation gradient ($\mathbf{F=F^eF^p}$) into elastic and plastic parts, are programmed and carried out 
for cold rolling within Abaqus Explicit. Frictional interactions are modeled using a consistent rate-independent Coulombic law.
Predictions from both the formulations, i.e., hypoelastic and multiplicative decomposition, exhibit a close match
with the experimentally observed rolling loads. We find that no specialized hyperlastic/visco-plastic material model is required to describe the behavior of the particular 
blend of polymeric films, under the conditions described here, thereby significantly speeding up the computation for steady-state rolling simulations. It is revealed that under the deformation conditions when principal axes show negligible rotation, hypoelastic formulation is valid at large elastic stretches. 
Moreover, the use of classical rigid-plastic 
modeling (which is often applicable to metals) is found to greatly underestimate the rolling loads for polymers, due to large elastic stretches in the polymer films at large strains. 
Deformation aspects of solid-state polymeric sheets presented here are expected to facilitate the development of new processes involving (or related to) cold roll-bonding of polymers. \\


\noindent \textbf{Notation}\\
$a$   half contact width in rolling;\\
$da$  differential area in current configuration;\\
B  reference body;\\
$\mathbf{b}\subscript{0}$  body force (per unit volume) in current configuration;\\
$\mathbf{b}$  body force (per unit volume) in current configuration with inertia effect included;\\
$c$  dilatational wave speed;\\
$\mathbf{C}\superscript{e}$ elastic right Cauchy--Green strain; \\
${\mathds{C}}$ Fourth order elasticity tensor;\\
$\mathds{C}\superscript{ep}$  Elasto-plastic tangent modulus for hypoelasticity;\\
$2d$  total compression in the film-stack;	\\
$dv$  differential volume in current configuration;\\
$dv^*$ differential volume of current configuration with respect to a rotated frame;\\
$dv\subscript{R}$  differential volume in the reference configuration;\\
$\mathbf{D}$ stretching-rate tensor, symmetric part of velocity gradient $\mathbf{L}$;\\
$\mathbf{D}\superscript{e}$  elastic stretching-rate tensor, symmetric part of $\mathbf{L}\superscript{e}$;\\
$\mathbf{D}\superscript{p}$  plastic stretching-rate tensor, symmetric part of $\mathbf{L}\superscript{p}$;\\
$\bar{\mathbf{D}}\superscript{p}$  plastic stretching-rate tensor in current configuration for hypoelasticity;\\
$E$  Young's modulus;\\
$E_{roller}$  Young's modulus of roller;\\
$E_{film}$  Young's modulus of polymer film;\\
$\mathbf{E}\superscript{e}$  Hencky strain given by $ln(\mathbf{U}\superscript{e})$;\\
$E\subscript{e}^{i}$ eigen values of Hencky strain given by $ln({\lambda}\superscript{e}\subscript{i})$ with ${i=1,2,3}$ ;\\
$\triangle \mathbf{E}$  spatial incremental strain in hypoelastic formulation;\\
$\triangle \mathbf{E}\superscript{e}$  spatial incremental elastic strain in hypoelastic formulation;\\
$\triangle \mathbf{E}\superscript{p}$  spatial incremental plastic strain in hypoelastic formulation;\\
$E_{roller}$ Young's modulus of the rollers;                                   \\
$\Delta\overline{\varepsilon}\superscript{pl}$  incremental equivalent plastic strain;\\
$\overline{\varepsilon}\superscript{pl}$  equivalent plastic strain;\\
$e^p$ accumulated plastic strain;\\
$f$  yield function for multiplicative plasticity;\\
$f\subscript{h}$  yield function for hypoelasticity;\\
$\mathbf{\mathscr{F}}$  vector of external nodal point forces for the finite element model;\\
$\mathbf{F}$  deformation gradient;\\
$\mathbf{F}\superscript{e}$  elastic part of deformation gradient;\\
$\mathbf{F}\superscript{p}$  plastic part of deformation gradient;\\
$F_{friction}$ measured force during measurement of coefficient of friction;\\
$G$  elastic shear modulus;\\
$h_1$ 	 thickness of film-stack at the inlet; \\
$h_o$    thickness of film-stack at the outlet; \\
$\bar{h}$  average of the inlet and outlet thicknesses ($h_1$, $h_o$);\\
$h$  thickness of film-stack (at $x$ coordinate) in the roller bite;\\
 $H$ hardening modulus;\\
$\mathbf{I}$ second order identity tensor ;\\
${\mathds{I}}$  fourth order identity tensor; \\
$\mathbf{\mathscr{I}}$  vector of internal nodal point forces for finite element model;\\
$J$ determinant of the deformation gradient $\mathbf{F}$;\\
$J\superscript{e}$  determinant of deformation gradient $\mathbf{F}\superscript{e}$;\\
$J\superscript{p}$  determinant of deformation gradient $\mathbf{F}\superscript{p}$;\\
$k$  yield strength in shear; \\
$K$  elastic bulk modulus;\\
$l_e$ smallest element dimension in finite element mesh;\\
$L$    total load in rolling;       \\
$\mathbf{L}$  spatial velocity gradient; \\
$\mathbf{L}\superscript{e}$  elastic part of velocity gradient; \\
$\mathbf{L}\superscript{p}$  plastic part of velocity gradient; \\
$\tilde{\mathbf{L}}\superscript{e}$  virtual elastic velocity gradient; \\
$\tilde{\mathbf{L}}\superscript{p}$  virtual plastic velocity gradient; \\
$m'g$ weight of the steel block used in friction measurement experiment;\\
$m$  velocity-dependent friction factor in range between 0 and 1;\\
$M$  moment per-unit width during rolling;\\
$\mathbf{M}\superscript{e}$  Mandell stress, work conjugate to $\mathbf{L}\superscript{p}$;\\
$\mathbf{\mathscr{M}}$  diagonal lumped mass matrix for global finite elements;\\
$\mathbf{n}$ unit vector $\mathbf{n}$ in the deformed configuration;\\
$\mathbf{N}\superscript{p}$  direction of plastic flow (determining $\mathbf{D}\superscript{p}$);\\
$\bar{\mathbf{N}}\superscript{p}$ direction of plastic flow transformed into the current configuration
(determining $\bar{\mathbf{D}}\superscript{p}$); \\
$P$  material region in the reference configuration;\\
$P\subscript{t}$  material region in the current configuration;\\
$P_l$  line loading in rolling;    \\
$\mathbf{Q}$  frame-rotation;\\
 $\mathbf{r}\subscript{i}^{e}$ orthonormal eigen-vectors of $\mathbf{C}\superscript{e}$ and $\mathbf{U}\superscript{e}$ with $i=1,2,3$;\\
$R$      radius of rollers;                                    \\
$\mathbf{R}$  rotation tensor obtained by the decomposition $\mathbf{F}=\mathbf{R}\mathbf{U}$;\\
$\mathbf{R}\superscript{e}$  rotation tensor obtained by the decomposition $\mathbf{F}\superscript{e}=\mathbf{R}\superscript{e}\mathbf{U}\superscript{e}$;\\ 
$\mathbf{R}\superscript{p}$ rotation tensor obtained by the decomposition $\mathbf{F}\superscript{p}=\mathbf{R}\superscript{p}\mathbf{U}\superscript{p}$;\\
$\mathbf{S}\superscript{e}$  microstress that is work conjugate to elastic velocity gradient $\mathbf{L}\superscript{e}$;\\
$\mathbf{T}$  Cauchy stress;\\
$\mathbf{T}\superscript{p}$ microstress that is work conjugate to plastic velocity gradient $\mathbf{L}\superscript{p}$;\\
$T\subscript{g}$  Glass transition temperature;\\
$\mathbf{t}(\mathbf{n})$  Traction field (associated with $\mathbf{n}$);\\
$\mathbf{T}\superscript{J}$  Jaumman rate of Cauchy stress;\\
$\mathbf{U}$  symmetric and positive-definite stretch tensor tensor from right polar decomposition of $\mathbf{F}$ ;\\
$\mathbf{U}\superscript{e}$  symmetric and positive-definite right elastic stretch tensor from right polar 
decomposition of $\mathbf{F}\superscript{e}$;\\
$\mathbf{U}\superscript{p}$  symmetric and positive-definite right plastic stretch tensor from right polar decomposition of $\mathbf{F}\superscript{p}$;\\
$\dot{\mathbf{\mathscr{U}}}$  vector of nodal point velocities for the finite element model;\\
$v_o$	 exit speed of the film-stack; 	\\
$\mathbf{v}$  spatial velocity;\\
$\tilde{\mathbf{v}}$  virtual spatial velocity; \\
$v\subscript{x}$  velocity at any section (at a distance of $x$ units from the exit);\\
$\mathbf{V}$  symmetric and positive-definite stretch tensor tensor from left polar decomposition of $\mathbf{F}$;\\
$\mathbf{V}\superscript{e}$  symmetric and positive-definite elastic stretch tensor from left polar 
decomposition of $\mathbf{F}\superscript{e}$;\\
$\mathbf{V}\superscript{p}$ symmetric and positive-definite plastic stretch tensor from left polar 
decomposition of $\mathbf{F}\superscript{p}$;\\ 
$\mathscr{V}$  set of virtual velocity fields;\\
$\mathbf{W}$  continuum spin, the anti-symmetric part of $\mathbf{L}$;\\
$\mathbf{W}\superscript{e}$  elastic spin, the anti-symmetric part of $\mathbf{L}\superscript{e}$; \\
$\mathbf{W}\superscript{p}$  plastic spin, the anti-symmetric part of $\mathbf{L}\superscript{p}$; \\
$x$, $y$, $z$  coordinate variables;\\
$\mathbf{x}$  position of a material point in current configuration;\\
$\mathbf{X}$  position of a material of a body in reference configuration;\\
$Y$  scalar yield-strength and radius of the spherical yield strength; \\
$Y\superscript{'}$  hardening modulus;\\
$\tau_{shear}$ tangential traction due to friction;\\
$\sigma\subscript{v}$  equivalent (von Mises) stress;\\
$\sigma_{y,film}$  yield-strength of film;                          \\
$\mathbf{\mathscr{U}}$ Nodal point displacements in the finite element solution;\\
$\nu$  Poisson's ratio;\\
$\tau$  duration of contact in rollers during rolling;\\			
$\mu$ coefficient of friction; \\
$\dot{(\,\,)}$  time rate of a quantity;\\
$|(\,\,) |$  norm of the quantity; \\
$\ddot{(\,\,)}$  double derivative w.r.t. time;\\
$\mathbf{A}:\mathbf{B}$  inner product of two second order tensors $\mathbf{A}$ and $\mathbf{B}$;\\
Sym $(\,\,)$  symmetric part of the operand;\\
$\triangledown$  gradient with respect to material coordinates $\mathbf{X}$;\\
$(\,\,)\superscript{-1}$  inverse of the tensor operand;\\
$(\,\,)\superscript{\top}$  transpose of the tensor operand;\\
det $(\,\,)$  determinant of the tensor operand;\\
tr $(\,\,)$  trace of the tensor operand;\\
grad $(\,\,)$  gradient of the quantity with respect to spatial coordinates $\mathbf{x}$;\\
$(\,\,)\superscript{*}$  transformed quantity in rotated-frame;\\
div $(\,\,)$  divergence of the quantity with respect to spatial coordinates $\mathbf{x}$;\\
$(\,\,)\subscript{0}$  deviatoric part of the tensor quantity;\\
$\nu$  Poisson's ratio;\\
$\bm{\chi}$  deformation, mapping a material point to a spatial point;\\
$\rho$  density at a material point in current configuration ;\\
$\mathscr{W} (P\subscript{t})$ external power delivered to a body part $P_t$ in current configuration;   \\
$\mathscr{I} (P\subscript{t})$ internal power expended inside a body part $P_t$ in current configuration;   \\
$\mathscr{W}\subscript{int}$  Virtual internal power; \\
$\mathscr{W}\subscript{ext}$ Virtual external power; \\
$\varphi$ free energy per-unit intermediate volume; \\
$\tilde{\varphi}$ free energy function expressed in terms of invariants of $\mathbf{C}
\superscript{e}$; \\
$\hat{\psi}$ free energy function expressed in terms of invariants of $\mathbf{U}
\superscript{e}$; \\
$\delta$ dissipation per-unit intermediate volume; \\
$\bm{\Omega}$  frame-spin;\\
$\lambda$  consistency parameter for the yield condition;\\
$\dot{\lambda}$  equivalent plastic strain rate;\\
$\lambda\subscript{i}^{e}$ positive eigen values of $\mathbf{U}\superscript{e}$ with ${i=1,2,3}$  ;\\
$\omega\subscript{i}^{e}$ positive eigen values of $\mathbf{C}\superscript{e}$ with ${i=1,2,3}$  ;\\
$\omega$ angular speed of the rollers.


\section{Introduction}

Several methods, usually referred to as \textit{calendering}, are in practice for processing
flat sheet-like materials such as paper, polymers, metals, blankets, cardboard laminates, etc. The central idea in 
calendering (or alike) methods is to subject the incoming stock of material under compression through rollers with a desired processing step. 
For centuries the metal industries have also been producing flat products such as metal sheets, plates, strips, bars, etc. through rolling processes.  
These processes fall in the category of hot or cold \textit{forming} (indicating  a net change in the shape of input stock material). Rolling processes have also been used in cold bonding or welding of metals, 
and cladding processes (joining of dissimilar metals)
\cite{movahedi2008influence, coules2012effect,govindaraj2013tensile,zhang1996numerical, maleki2013analysis}. 
In context to polymers, cold rolling processes have been employed in studying the time-dependent evolution of 
micro-structural properties after plastic deformation \cite{cangialosi2005amorphous}, improving the tensile strength and 
toughness of certain thermoplastics (by unidirectional and biaxial rolling) 
\cite{broutman1971cold,broutman1974cold,mcglamery1963cold}, enhancing the yield strength 
of cold-forged products \cite{matsuoka1998effects}, and analyzing the tensile properties of 
rubbers, and (nano)composites  \cite{grancio1972cold,phua2011mechanical}.
Recently, a new  process for manufacturing of pharmaceutical tablets from thin polymeric films 
has been reported \cite{padhye-thesis-2015,trout2015layer,padhye2021deformation,padhye2021molecular,padhye2022dilatational,padhye2014roll}, 
where polymer films are roll-bonded at temperatures well-below their glass transition temperatures.
These polymeric films are engineered to exhibit unique mechanical and physical properties, and active plastic deformation triggers requisite molecular mobility to cause macromolecular interdiffusion across polymeric interfaces held in intimate contact, thereby leading to bonding through formation of new entanglements. Such a methodology 
alleviates the need of adhesives, surface modifications, heat treatments, etc., and is therefore well-suited
for pharmaceutical manufacturing.  Key requirements for this rolling (bonding) process are:
uniform and homogeneous compression of the incoming film-stack
at linear feed rates of $5-30$ mm/min, induction of through-thickness plastic deformation up to 20\%, 
and minimal shearing between the polymeric interfaces during deformation. Details of a rolling machinery to carry out 
this roll-bonding process are given in \cite{padhye-thesis-2015,padhye2014roll}. To ensure the
product quality and develop a process control for roll-bonding of polymeric films, an accurate and efficient scheme 
is required for predicting the optimal rolling loads, deformation fields, and resulting stock geometries and properties. 

Over the course of the last century there have been extensive theoretical, experimental, and  computational efforts to study deformations in rolling processes, and development of models in predicting the forming loads and torques, and distribution of pressures along the arc of contact in the roller bite
\cite{roberts1978cold,roberts1983hot,ginzburg2000flat,wusatowski2013fundamentals,lenard2013primer,mukerjee1973critical,Hartly1992,muniz2007non}.
Broadly speaking, the methods of analysis for rolling, both
analytical and numerical, can be categorized into: (i) slab methods, 
(ii) slip line analysis, (iii) upper bound analysis, and (iv) computational techniques. 
Approaches (i)-(iii) make several simplifying approximations on the deformation patterns in the roller bite, and constitutive material behavior, and were largely relied upon when the finite element methods were non-existent. 
In slab methods the object of deformation is divided into multiple slabs, 
and for each slab simplifying assumptions are made for the stress distributions. 
Then, for the resulting system the approximate equilibrium equations are solved
with imposition of strain compatibility between the neighboring slabs
and boundary tractions, and approximate loads and stress distributions are derived. 
Often homogeneous compression, constant yield strength, negligible elastic strains, and
kinetic friction between the roller and strip are assumed.
See \cite{siebel1924berichte,siebel1925werkstoffausschuss,von1925theory,nadai1939forces,tselikov1939influence,
ekelund1933analysis,smith1952pressure,ford1957theory,orowan1943calculation,bland1948calculation,ford1948researches,
al1973experimental,henningsen2006measurements,tieu2004friction}
for different attempts employing slab methods. Slip line analysis, which also makes assumptions of perfectly-plastic and rate-independent 
material behavior, yielded only limited success \cite{alexander1955slip,alexander1964simplified,crane1968slip,denton1972roll} in rolling, and
incorporation of realistic frictional or traction boundary conditions between the stock and the roller
remained a challenge. Finally, the upper-bound analysis approaches \cite{avitzur1964upper}, in which 
material is again treated as rigid-plastic, and kinematically admissible velocity fields are 
are derived from experimental evidence or experience, have not yielded any greater success. 
Summarily, the classical techniques are unable to handle geometric and material nonlinearities, 
rate- and temperature-dependent effects, arbitrary strains in presence of plastic deformation, 
and complex frictional conditions, and it is to this end modern numerical techniques for solving 
boundary value problems have emerged as a powerful tool, and have 
largely superseded these classical approaches. Even though classical techniques cannot capture 
complexities of an arbitrary rolling process, they are, however, useful sometime in making an order of 
magnitude estimates for the rolling loads. As we demonstrate in this work, we shall utilize the predictions 
from rigid-plastic rolling  to guide our finite-element analysis.

Finite Element Methods (FEM) have emerged as a powerful computational tool for a variety of purposes. Nowadays  most complex, elasto-plastic deformation analyses in solids, where exact analytical solutions are not available, are carried out using FEM. Notable contributions on the development of Total Lagrangian \cite{hibbitt1970finite}, and Updated Lagrangian \cite{mcmeeking1975finite} formulations for the displacement based finite elements, addressing
issues related to large deformations and rotations, and material non-linearities, have greatly facilitated the 
study of metal-forming problems, see \cite{liu1985elastic,rowe2005finite,lee1998large,maniatty1991eulerian,dixit1996finite,dixit1997finite}.
The use of objective stress rate measure has been widely used in nonlinear problems, and plasticity is incorporated through additive decomposition of spatial strain-rate, or multiplicative decomposition of the deformation gradient into the elastic and plastic parts. Locking phenomenon exhibited by the displacement based finite elements due to volume preserving plastic flow constraint in fully plastic regimes has been avoided through special arrangements of the elements \cite{nagtegaal1974numerically}, or by use of the selective reduced integration method \cite{zienkiewicz1971reduced}. Other notable approaches of treating 
material as rigid-plastic, or visco-plastic, and deploying `Eulerian' flow-type formulations, with velocity field as an unknown \cite{cornfield1975theoretical, zienkiewicz1974flow,dawson1978viscoplastic,
zienkiewicz1978flow,mori1982simulation,kobayashi1982rigid,hwu1988finite,oh1975approximate}, are not applicable 
to large strain elasto-plastic analysis of polymer rolling, and are therefore ruled out from further consideration here. A detailed overview of the numerical solutions in context to rolling and forming methods can be found in \cite{oh1989metal,Rowe1991,dixit2008modeling}.

While traditional nonlinear elasto-plastic finite element analyses, based on incremental kinematics, employ implicit iterative schemes for solving equilibrium or momentum equations, explicit dynamics-based solution strategies have gained significant popularity in a variety of metal forming applications   
 \cite{prior1994applications,jung1998study,marusich1995modelling}. Unlike implicit schemes, explicit methods have smaller memory footprint as the storage of tangent stiffness matrix is not required.
 Explicit integration schemes, however, are only conditionally stable, and the critical time step for stability purposes 
 is approximated  as $l_e/c$, where $l_e$ is the smallest element size and $c$ is the dilatational wave 
 speed,
also known as the CFL limit. 
Explicit dynamics have been quite successful for problems where the inertial effects are negligible, because mass scaling can be adopted to increase the stable time increments,
and they also show robust performance in handling contact conditions  (impenetrability and slip/stick),
since contact tangent stiffness matrix, as required for implicit methods, is not needed. Another drawback of implicit methods is that the size of the stiffness matrix can change depending upon the contact state, and the stiffness matrix may loose symmetry due to friction. In presence of complex loading and contact conditions, the implicit methods can also face convergence difficulties (even though, when they are successful, they can provide quadratic convergence, and contact conditions can be satisfied exactly using the method of Lagrange multipliers). For explicit methods, predictor-corrector, or penalty based methods are commonly employed to tackle contact conditions. 
To best of our knowledge, currently there are no research attempts on efficient and accurate computational modeling of cold rolling of polymers. Thus, in this paper, we resort to the commercial finite element software Abaqus for developing computational procedures to accurately and efficiently model such rolling processes. 
In general, glassy polymers exhibit a time-dependent viscoplastic mechanical behavior, and therefore 
quasi-static rolling simulations, adopting a finite deformation time-dependent behavior with sufficiently fine mesh, would require large physical computational times to arrive at a steady-state (long-time) solution. On the other hand, glassy polymers  are known to exhibit rate-independent behavior at slow strain-rates and low temperatures 
\cite{mulliken2006mechanics,ames2009thermo}, and if operating conditions are such that polymer behavior can be modeled without any rate effects, one can make faster computations. 
For this purpose, using a classical rigid-plastic 
analysis, we first estimate the upper bound strain rates in the roller-bite, at the specified feed rates, and find that the average strain-rates during rolling are relatively small, so that a rate-independent material model for the polymer films can in fact be adopted. Accordingly, elasto-plastic modeling at large plastic strains with moderate elastic stretches is carried out using formulations based on 
hypoelasticity, and multiplicative decomposition of deformation gradient into elastic and plastic parts. Consistent with the rate-independent material behavior, Coulombic friction is used between the rollers and polymer films through the kinematic-constraint algorithm. Quasi-static rolling simulations, with the explicit time stepping scheme, and mass-scaling are then used to make predictions for rolling loads. 

The rest of the paper is structured as follows: In Section ~\ref{sec:mechanics-and-modeling}, we describe the 
cold rolling of polymeric films, and utilize a classical rolling model to make preliminary predictions for 
various deformation fields. Then, a rate-independent polymer deformation model with finite elastic effects, and isotropic hardening is proposed based on hypoelasticity and multiplicative decomposition of deformation gradient in Section ~\ref{sec:material-modeling}. 
Section ~\ref{sec:performance-results} presents the numerical simulations, and compares the 
experimentally observed rolling loads with the classical rolling theory, and steady-state finite element rolling simulations. Key features of the deformation mechanics of polymer films in the roller bite are revealed here. 
Finally, conclusions and directions for future 
work are summarized in Section ~\ref{sec:conclusion}.


\section{Mechanics and Modeling of Cold Roll-Bonding of Solid-State Polymeric Films}
\label{sec:mechanics-and-modeling}

\subsection{Materials}
\begin{figure}[htbp]
\centering
\includegraphics[scale=.35]{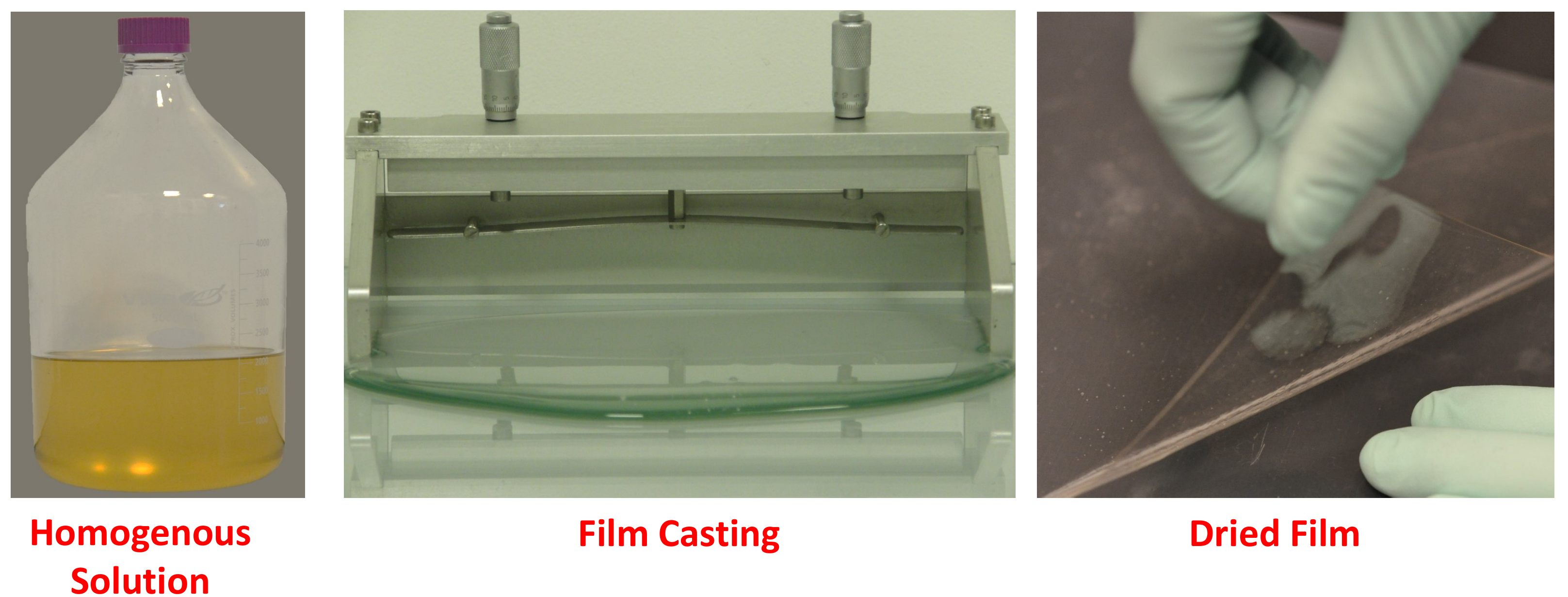}
\caption{Preparation of polymeric films through solution mixing, film casting, and peeling.}
\label{fig:Solution-Casting-Film_cropped}
\end{figure}

\begin{figure}[htbp]
\centering
\includegraphics[scale=.4]{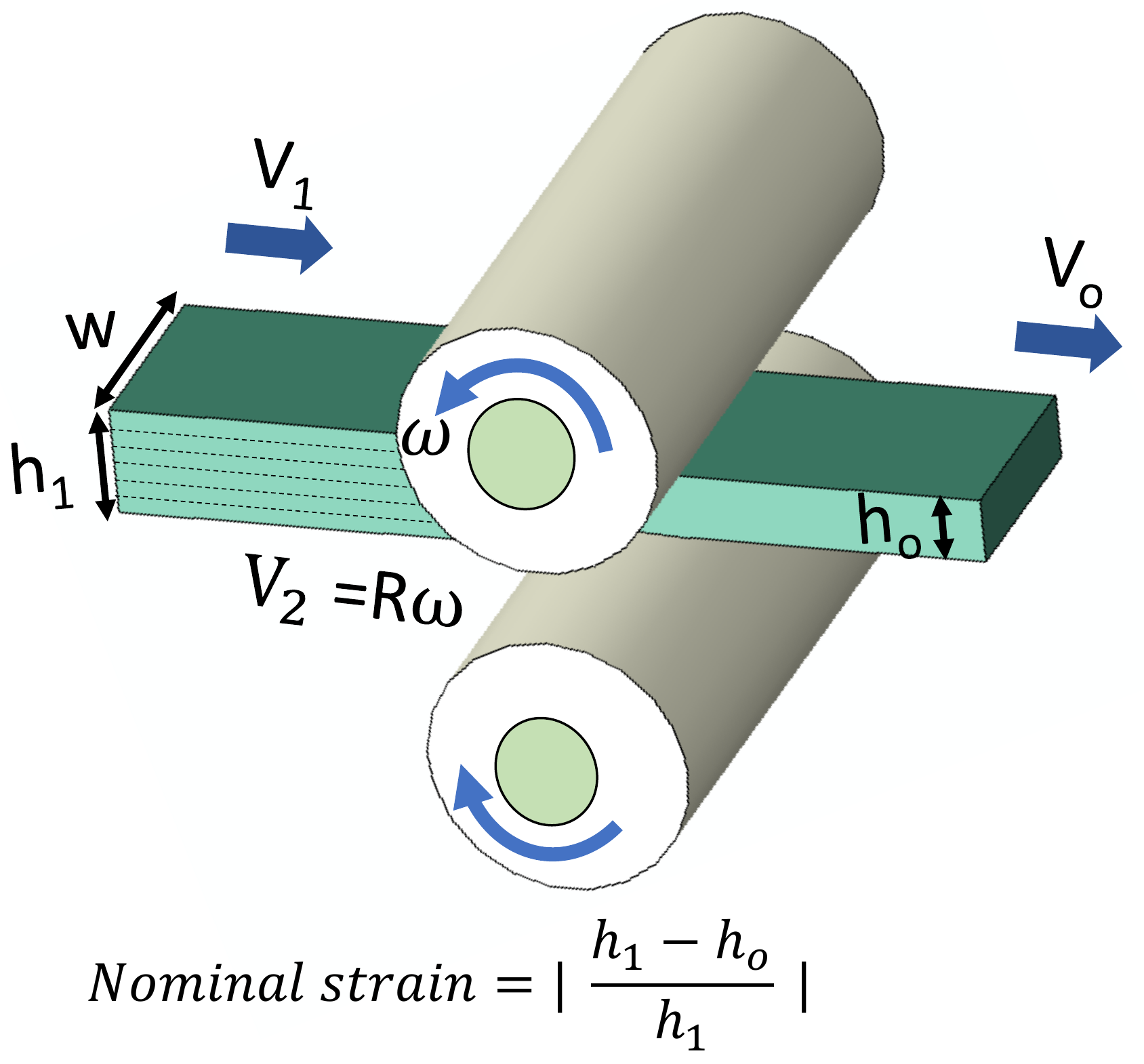}
\caption{Schematic of a rolling scheme, where multiple layers are fed through rollers
and bonded via induced plastic deformation. 
}
\label{fig:Rigid-rolling-plastic.pdf}
\end{figure}

Figure ~\ref{fig:Solution-Casting-Film_cropped} shows synthesis of polymeric films through solvent casting used in this study. Polymer films contained a base polymer METHOCEL E15 and a compatible plasticizer PEG-400. 
METHOCEL is the trade name for Hydroxypropyl-Methyl-cellulose (HPMC), and PEG-400
is polyethylene glycol with a molecular weight of 400.  E15, PEG, ethanol, and water were mixed
in a weight ratio of $1.32:4.36:4.36:1$, respectively, and a homogeneous solution was obtained through
mixing with an electric stirrer for over 24 hours. After completion of the blending process,
the solution was carefully stored in glass bottles at rest for 12 hours to get rid of any
air bubbles. Solvent casting was carried out using a casting knife applicator from Elcometer
on a Heat-resistant Borosilicate Glass. All steps were carried out in a chemical laboratory
where ambient conditions of 20$^o$C$\pm$ 2$^o$C and R.H. 20\%$\pm$5\% were noted. After drying,
glassy polymeric films of E15 were obtained that contained 42.3\% of PEG by weight. 
The glass transition temperature of the polymeric films was found to be 78 $^o$C.  Figure ~\ref{fig:Rigid-rolling-plastic.pdf} shows bonding of multiple layers
of films when subjected to plastic deformation in simple compression. 
See \cite{padhye-thesis-2015,padhye-main-bonding} 
for further details on synthesis, preparation, and characterization of polymeric
films, and solid-state bonding through plastic deformation. 

\begin{figure}[htbp]
\centering
\includegraphics[scale=.8]{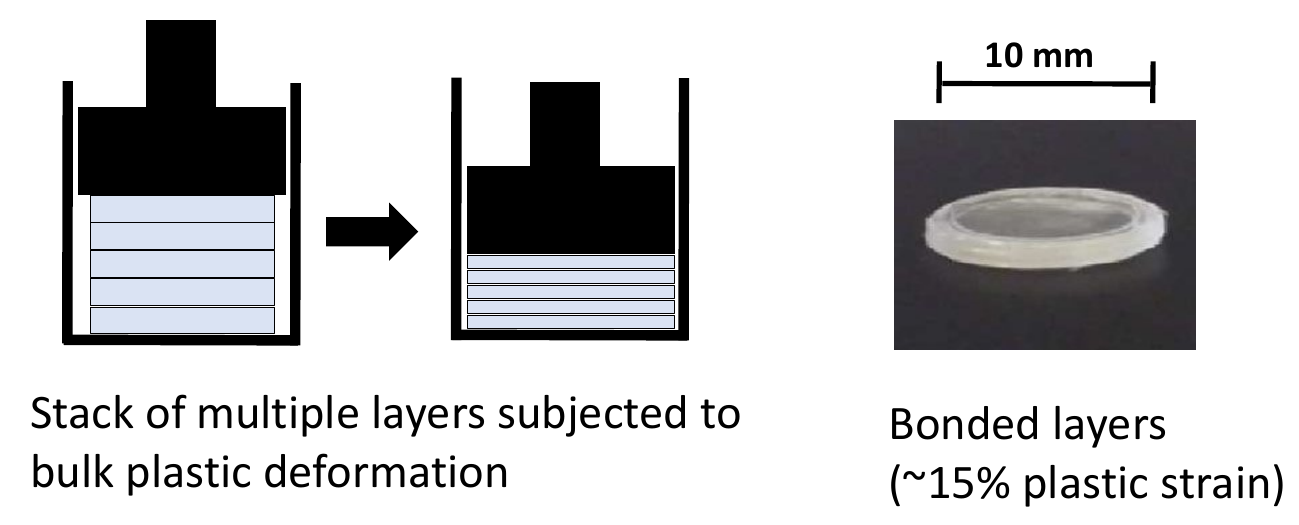}
\caption{Bonding of polymeric films in compression due to bulk plastic deformation.}
\label{fig:sample-bonding}
\end{figure}

\begin{figure}[htp]
     \centering
   \includegraphics[scale=.5]{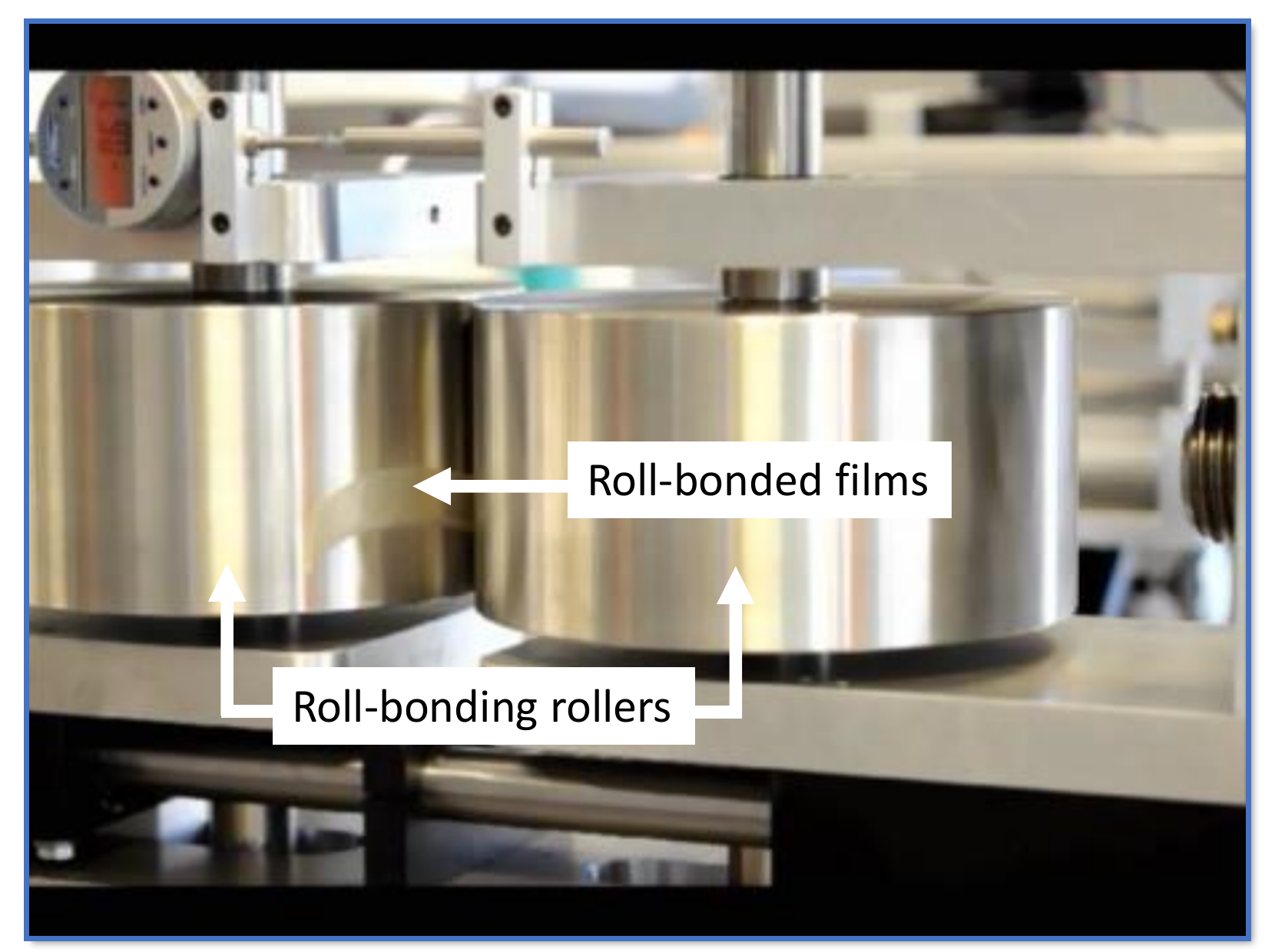}
   \caption{Rolling of polymeric films \cite{padhye-thesis-2015,padhye2014roll}.}
     \label{fig:roll-bonding-of-films}
    \end{figure}

A schematic of roll-bonding of multiple polymeric layers is shown in Figure ~\ref{fig:Rigid-rolling-plastic.pdf},
and the actual process is shown in Figure ~\ref{fig:Rigid-rolling-plastic.pdf}. A key requirement in this roll-bonding is to
apply sufficient levels of loads to induce through-thickness plastic deformation 
on the incoming stack of polymer films. The desired levels of plastic strains must be 
attained over the interval of time $\tau$ spent under the compression rollers.
The time $\tau$ is set by size of the rollers, specified feed rates, and material mechanical response of the films. 
During active plastic deformation in the roller bite, the polymer molecules interpenetrate across film interfaces and cause bonding. If the stack thickness is \textit{small} compared to the roller radius (R), i.e., $h<<R$, an element in the roller bite will experience almost homogeneous straining through its thickness, and negligible shear stresses will develop between the film interfaces. On the other hand, if there is a non-uniform strain across thickness then the interfaces (other than the symmetry plane of the film-stack in thickness direction) will have a tendency to exhibit relative tangential motion, which will hinder molecular interpenetration of polymer chains, and thus diminish bonding. In an extreme limit if the film-stack is quite thick compared to the radius of the rollers, then no through-thickness (plastic) deformation will incur, and only local indentation on polymer films will occur during roller compression.  
This type of bonding is a multi-scale process in the sense that polymer mobility and interpenetration occurs at a molecular scale, while plastic deformation is triggered at the macroscopic continuum scale. 
In order to achieve homogeneous and through-thickness plastic deformation,
rollers of radius $R$ with $R/h >> 1 $ were selected in the designed machinery \cite{padhye-thesis-2015,padhye2014roll}. Later through finite element simulations we will verify that under 
conditions of large roller radii, the through thickness shear stresses are in fact negligible and that the deformation is homogeneous.
Next, we resort to a classical analytical model based on 
rigid-plastic analysis to estimate various fields associated with rolling under prescribed processing conditions, and accordingly make simplifying, yet accurate, assumptions to carry out finite element modeling. 

%

\subsection{Estimates based on rigid perfectly-plastic material model}

\begin{figure}[tbp]
\centering
\includegraphics[scale=.5]{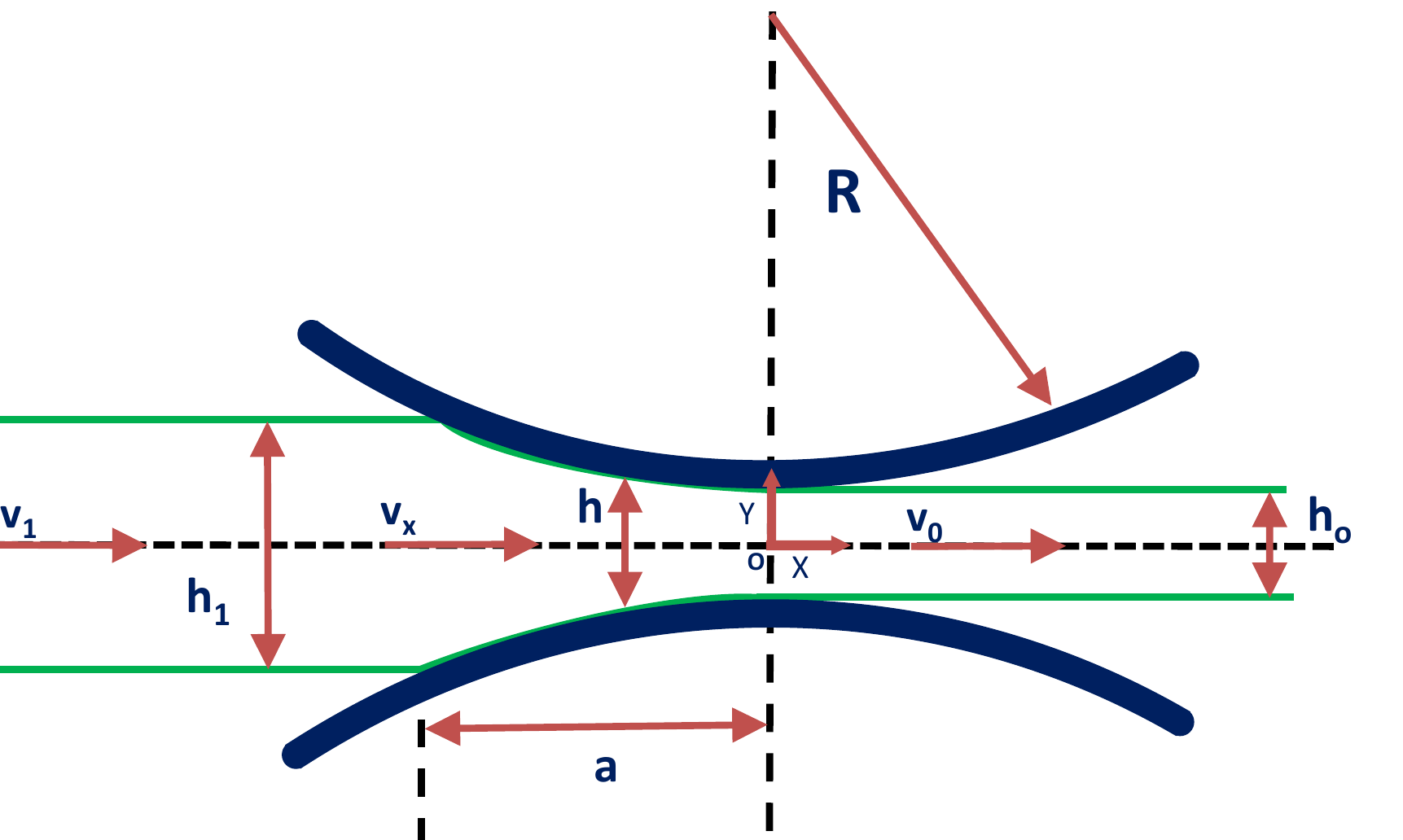}
\caption{Rigid-perfectly plastic rolling scheme.}
\label{fig:Rigid-rolling-plastic-free-body}
\end{figure}
\begin{figure}[tbp]
\centering
\includegraphics[scale=.45]{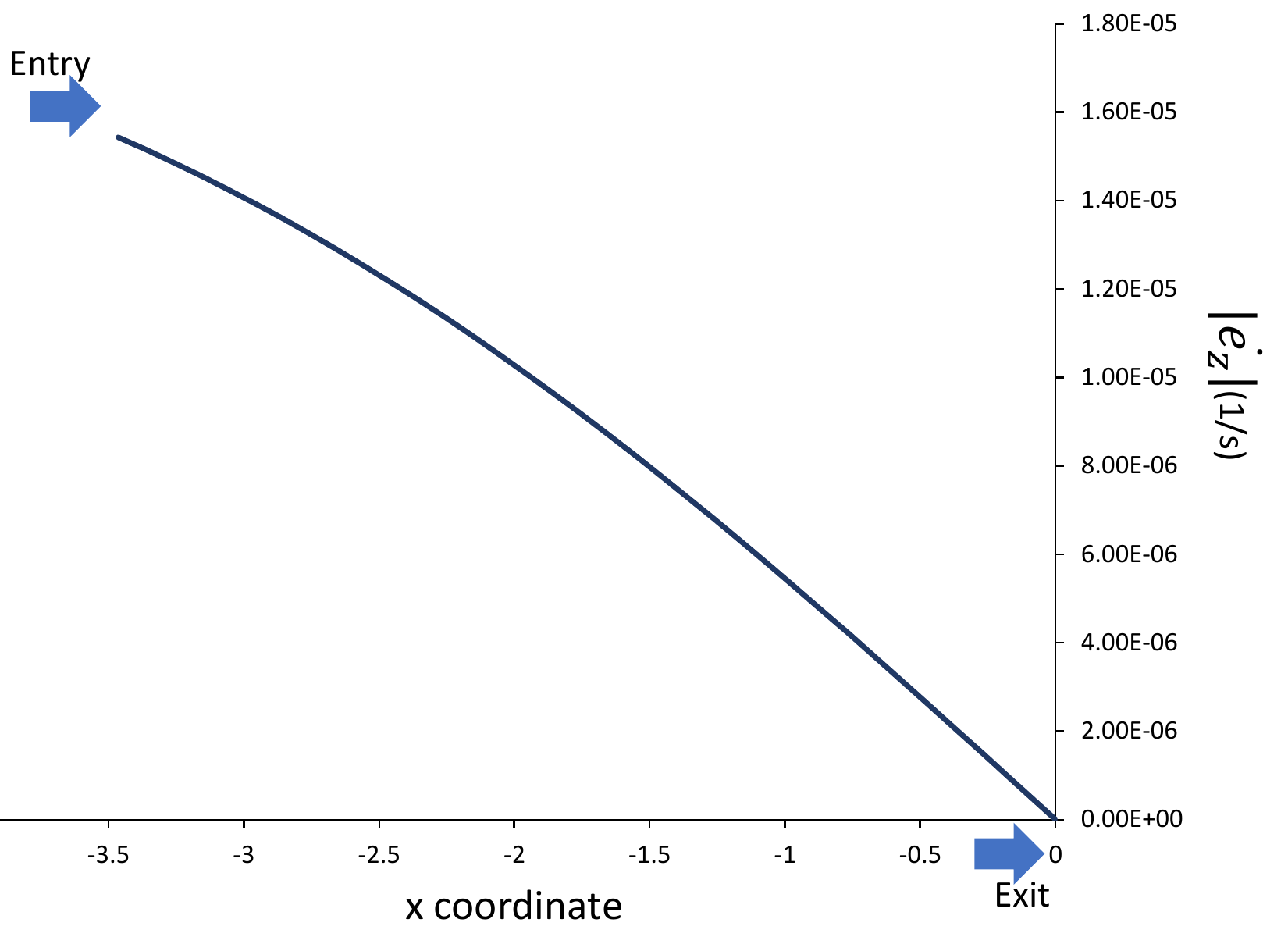}
\caption{Transverse strain rate plotted as a function of distance from the exit during deformation of film-stack in the rollers.
For this case, $h_1=0.6$ mm, $h_o=0.54$ mm (indicating 20\% nominal strain in thickness reduction),
and $v_o \approx 10.0$ mm/min.}
\label{fig:Rigid-plastic-analysis-rollers-1}
\end{figure}
\begin{figure}[tbp]
\centering
\includegraphics[scale=.45]{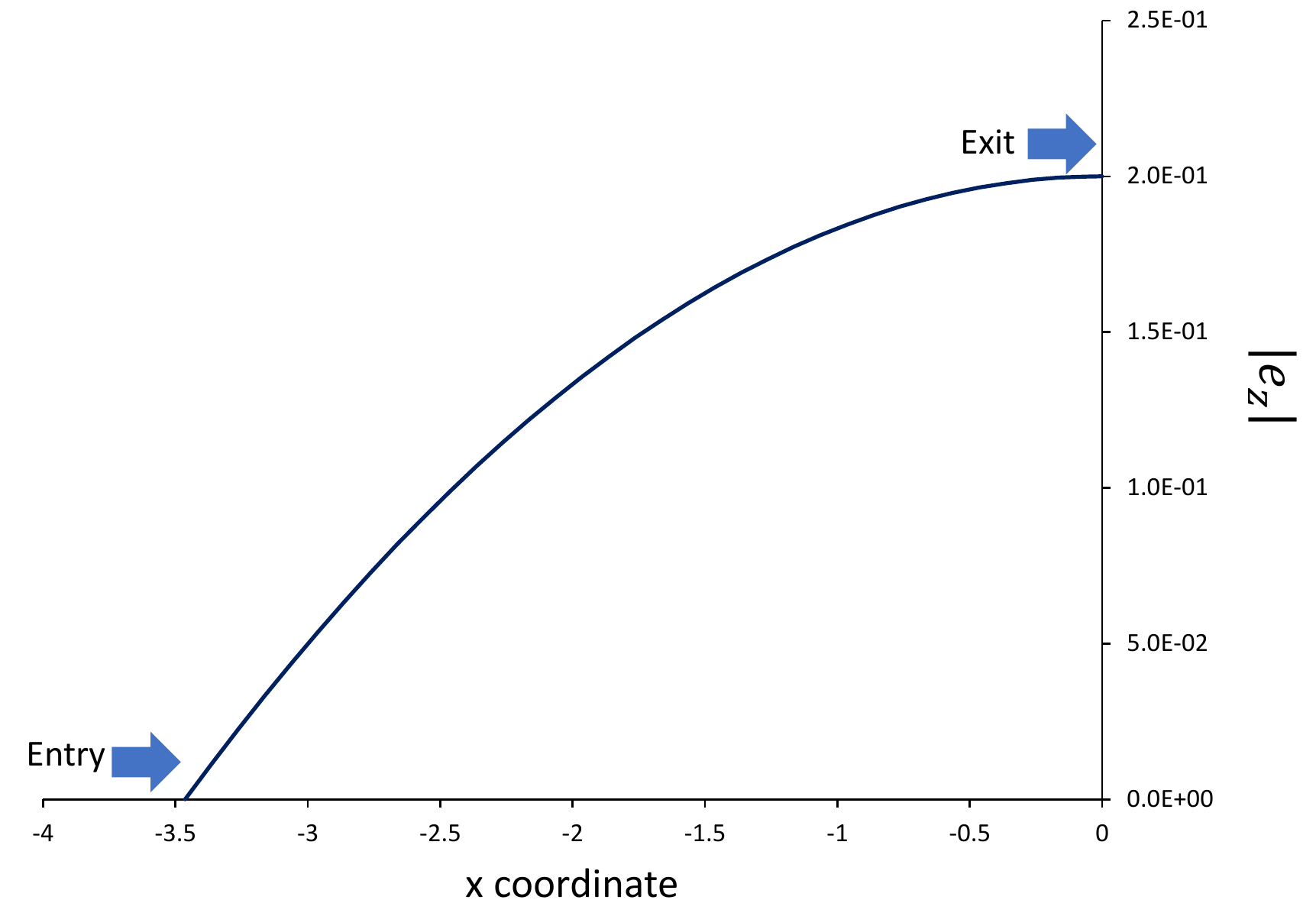}
\caption{Transverse strain plotted as a function of distance from the exit during deformation of film-stack in the rollers.
For this case, $h_1=0.6$ mm, $h_o=0.54$ mm (indicating 20\% nominal strain in thickness reduction),
and $v_o \approx 10.0$ mm/min.}
\label{fig:Ezz_vs_t}
\end{figure}

A rigid-plastic rolling scheme is shown in Figure ~\ref{fig:Rigid-rolling-plastic-free-body}, see 
\cite{KLJ} for complete details. In this rolling, the estimates of line loading $P_l$,
and torque $M$ per unit width, into the plane of the figure are given as
\begin{equation}
\label{eq:line-loading-2}
\frac{P_l}{ (\sigma_{y,film}/\sqrt{6}) a} = 2 + \frac{a}{ \bar{h}} \left( \frac{1}{2}-\frac{1}{3}\frac{a}{R}\right),
\end{equation}
and
\begin{equation}
\label{eq:moment-2}
\frac{M}{(\sigma_{y,film}/\sqrt{6})a^2} = 1 + \frac{a}{ 4\bar{h}} \left(  1 -\frac{a}{R}\right).
\end{equation}
In the above equations, $\sigma_{y,film}$ is the yield strength of the material, $a$ is the half contact width, $\bar{h}=\frac{h_1+h_o}{2}$ is the mean film-stack thickness, and $R$ is the radius of the rollers. 
From the geometry of deformation, if we assume that there is no slippage or
sliding between the rolled-stock and the rollers at the exit, then the
exit velocity $v_o$ of the rolled-stock is equal to tangential velocity
of the rollers $R\omega$, i.e., $v_o=R\omega$. 
The variation of an element height, in the roller bite, as a function of x-coordinate is given as
\begin{equation} 
h = h_o + 2R \left( 1- \sqrt{1-\frac{x^2}{R^2}} \right ).
\label{eq:height}
\end{equation}
The nominal compressive strain in the thickness (or transverse) direction is given as
\begin{equation}
|e_y| = (1-h/h_1),
\end{equation}
and accordingly 
the compressive strain rate is expressed as 
\begin{equation}
|\dot {e_z}| = \frac{1}{h_1} \frac{dh}{dt}.
\end{equation}
From the conversation of mass, during plane strain plastic flow in rolling, we can write
\begin{equation}
v_x  h = v_o h_o,
\end{equation}
where $v_x$ represents the velocity at any section, and thus using the above equations, 
we can re-write the expression for strain rate as
\begin{equation}
|\dot {e_z}| = \frac{2 x h_o v_o}{h_1 h \sqrt{R^2-x^2}}.
\end{equation}
In this rolling scheme, if  $R/h >> 1$, the time of compression $\tau$
under the rollers can be estimated as
\begin{equation}
\label{eq:contact-time}
\tau 	= \frac{\sqrt{R(h_i-h_o)}}{v_o}.
\end{equation}
According to the processing parameters provided in \cite{padhye-thesis-2015,padhye2014roll},
for $\sigma_{y,film}=6.0$ MPa, $h_1=0.6$ mm, $h_o=0.48$ mm, $2d=(h_1-h_o)=0.12$ mm 
(indicating 20\% plastic compression), $R=100$ mm,  and $v_o=10$ mm/min, we find 
$\tau=20.78$ seconds. For these settings the plots of strain rate and strain in the roller-bite are shown in Figures 
~\ref{fig:Rigid-plastic-analysis-rollers-1} and ~\ref{fig:Ezz_vs_t}, respectively.

Since we employed a rigid-plastic rolling model in the above calculations, the computed strain rates  are overestimates of the actual strain rates that would occur, as the elastic straining is neglected.
The strain rates from Figure ~\ref{fig:Rigid-plastic-analysis-rollers-1} lie in the range of $10^{-6}-10^{-5}$ $s^{-1}$, and are very small.  This is a key observation 
because solid-state polymers, when deformed at slow to moderate rates, show little rate-sensitivity in their 
mechanical response, and this simplifies the material modeling task by enabling us to use a rate-independent 
material model in the finite element simulations. As pointed out in \cite{ames2009thermo}, at strain-rates less 
than 0.01, hardly any rate sensitivity is noted. In fact the rate-dependent models at slow deformation 
rates reduce to their rate-independent counterparts. If the strain-rate effects are significant, such as in calendering of 
elastic-viscous materials \cite{paslay1955calendering}, then appropriate material behavior models must be 
considered. 
At a molecular level, well below the T$_g$, the resistance to plastic flow in polymers is mostly 
dominated by the resistance to molecular level re-orientations. The backstress due to entropic chain 
alignment is dormant well below T$_g$ when only moderately large deformation regimes 
are considered. Thus, 
effects due to network elasticity of the polymer matrix can be ignored all together at small to moderately large plastic strains. The plastic flow can be modeled using $J2$ flow criterion and normality flow rule. Pressure sensitivity can also be ignored well below T$_g$ for moderate stress states occurring in rolling. 
For comprehensive details on the mechanical behavior of polymers see
\cite{argon2013physics}. For popular polymer models see, e.g., \cite{boyce1988large,anand2003theory,arruda1995effects}.
Finally, we also take note of the fact that polymers can exhibit noticeable elastic deformations which cannot be 
ignored with respect to plastic strains, and our experimental results indeed reveal this behavior.


\section{Finite Strain Rate-Independent Elastic-Plastic Material Model for Polymeric Films}
\label{sec:material-modeling}
Based on the discussions presented in the previous section, we now work towards deriving 
a rate-independent deformation model to describe the polymer behavior at finite strains. We present two 
approaches to model this behavior, along with the necessary preliminaries, based on: (a) multiplicative decomposition of deformation gradient into elastic and plastic parts, and (b) hypoelastic formulation that employs an objective measure of stress-rate, and utilizes additive decomposition of spatial strain-rate into elastic
and plastic parts.


\subsubsection*{Kinematics}
%
A homogeneous body B is identified with the region of space that it occupies with respect to a fixed reference configuration. $\mathbf{X}$ denotes an arbitrary material point of B. The motion of 
B is then described as a smooth one-to-one mapping $\mathbf{x} = \bm{\chi} (\mathbf{X},t)$, and
the deformation gradient is defined as
\begin{equation}
\mathbf{F} = \triangledown \bm{\chi} =  \frac{\partial \bm{\chi}}{\partial \mathbf{X}}.
\label{deformation-gradient}
\end{equation}
The velocity gradient $\mathbf{L}$, i.e., gradient of the spatial velocity with respect to spatial coordinates, is related to the deformation gradient $\mathbf{F}$ through the identity
\begin{equation}
\mathbf{L} = \text{grad}\, \dot{\bm{\chi}} = \dot{\mathbf{F}} \mathbf{F}.\superscript{-1}
\label{velocity-gradient}
\end{equation}
The velocity gradient $\mathbf{L}$ can be decomposed into the symmetric part 
$\mathbf{D}$ and anti-symmetric part $\mathbf{W}$ as 
\begin{equation}
\mathbf{D} = \frac{1}{2} \big[ \mathbf{L} +  \mathbf{L}\superscript{\top}\big],
\label{total-stretching}
\end{equation}
and
\begin{equation}
\mathbf{W} = \frac{1}{2} \big[ \mathbf{L} -  \mathbf{L}\superscript{\top}\big].
\label{total-spin}
\end{equation}
$\mathbf{D}$ is the stretching rate tensor and $\mathbf{W}$ is the continuum spin. Within the framework of large deformations, the multiplicative decomposition \cite{lee1969elastic} of the deformation gradient $\mathbf{F}$ can be written as
\begin{equation}
\mathbf{F} = \mathbf{F}\superscript{e} \mathbf{F}\superscript{p},
\label{F-decomposition-into-Fe-Fp}
\end{equation}
where $\mathbf{F}\superscript{e}$ represents the local elastic deformation deformation of material in an infinitesimal neighborhood of $\mathbf{X}$ due to stretch and rotation. $\mathbf{F}\superscript{p}$ is the plastic distortion and represents the local deformation of material $\mathbf{X}$ in an infinitesimal neighborhood due to the irreversible or plastic deformation.
By substituting $\mathbf{F}$ from Equation \ref{F-decomposition-into-Fe-Fp} into Equation \ref{velocity-gradient}, we can obtain 
the following expression for 
$\mathbf{L}$ as
%
%

\begin{equation}
\mathbf{L} = \dot{\mathbf{F}}\superscript{e} \mathbf{F}\superscript{e-1} + \mathbf{F}\superscript{e} \dot{\mathbf{F}}\superscript{p} \mathbf{F}\superscript{p-1} \mathbf{F}\superscript{e-1}.
\label{L-in-terms-of-Fe-Fp}
\end{equation}
We define the elastic and plastic velocity gradients,  
$\mathbf{L}\superscript{e}$ and $\mathbf{L}\superscript{p}$, as follows
\begin{equation}
\mathbf{L}\superscript{e} = \dot{\mathbf{F}}\superscript{e} \mathbf{F}\superscript{e-1},
\label{elastic-part-of-velocity-gradient-decomp}
\end{equation}
and 
\begin{equation}
\mathbf{L}\superscript{p} = \dot{\mathbf{F}}\superscript{p} \mathbf{F}\superscript{p-1}.
\label{plastic-part-of-velocity-gradient-decomp}
\end{equation}
Using Equations ~\ref{elastic-part-of-velocity-gradient-decomp} and ~\ref{plastic-part-of-velocity-gradient-decomp}, we can rewrite Equation ~\ref{L-in-terms-of-Fe-Fp} as 
\begin{equation}
\mathbf{L} = \mathbf{L}\superscript{e} + \mathbf{F}\superscript{e} \mathbf{L}\superscript{p} \mathbf{F}\superscript{e-1}.
\label{velocity-gradient-decomposition}
\end{equation}
Now the elastic stretching $\mathbf{D}\superscript{e}$ and the elastic spin $\mathbf{W}\superscript{e}$ are defined as 
\begin{equation}
\mathbf{D}\superscript{e} = \frac{1}{2} \big[ \mathbf{L}\superscript{e} +  \mathbf{L}\superscript{e\top}\big],
\label{elastic-stretching}
\end{equation}
and 
\begin{equation}
\mathbf{W}\superscript{e} = \frac{1}{2} \big[ \mathbf{L}\superscript{e} -  \mathbf{L}\superscript{e\top}\big].
\label{elastic-spin}
\end{equation}
Similarly, the plastic stretching $\mathbf{D}\superscript{p}$ and the plastic spin $\mathbf{W}\superscript{p}$ are defined as
\begin{equation}
\mathbf{D}\superscript{p} = \frac{1}{2} \big[ \mathbf{L}\superscript{p} +  \mathbf{L}\superscript{p\top}\big],
\label{plastic-stretching}
\end{equation}
and
\begin{equation}
\mathbf{W}\superscript{p} = \frac{1}{2} \big[ \mathbf{L}\superscript{p} -  \mathbf{L}\superscript{p\top}\big].
\label{plastic-spin}
\end{equation}

We shall utilize two commonly employed kinematical assumptions concerning the plastic flow. First, the plastic flow is incompressible, i.e., plastic flow does not induce volume changes. 
This condition is achieved by assuming that $\mathbf{L}\superscript{p}$ and hence $\mathbf{D}\superscript{p}$ are deviatoric, i.e., tr($\mathbf{L}\superscript{p}$)=tr($\mathbf{D}\superscript{p}$)=0. This condition can be expressed as
\begin{equation}
J\superscript{p} = \text{det}\, \mathbf{F}\superscript{p} = 1 \indent \indent \text{and} \indent \indent \text{tr}\,\mathbf{L}\superscript{p} = 0.
\label{J=detFp}
\end{equation}
We can also decompose $J=$det$ \mathbf{F}$  into elastic and plastic components
as $J=$ det$\mathbf{F}\superscript{e}$det$\mathbf{F}\superscript{p}$.
Using $J\superscript{e}=$ det$\mathbf{F}\superscript{e}$ and  $J\superscript{p}=$
det$\mathbf{F}\superscript{p}=1$, we have $J = J\superscript{e}  J\superscript{p} = J\superscript{e}$.
Secondly, if we idealize the material to be isotropic, then plastic flow can be assumed to be irrotational \cite{gurtin2010mechanics}, i.e.,
\begin{equation}
\mathbf{W}\superscript{p} = \mathbf{0}.
\label{}
\end{equation}
Accordingly, $\mathbf{L}\superscript{p}$ is symmetric and 
\begin{equation}
\mathbf{D}\superscript{p} = \mathbf{L}\superscript{p}.
\label{Dp=Lp}
\end{equation}
From Equation \ref{plastic-part-of-velocity-gradient-decomp}, we can write
\begin{equation}
\mathbf{D}\superscript{p} = \mathbf{L}\superscript{p} = \dot{\mathbf{F}}\superscript{p} \mathbf{F}\superscript{p-1}.
\label{Dp-Lp-relation}
\end{equation}
Rearranging the terms in Equation \ref{Dp-Lp-relation}, we obtain
\begin{equation}
\dot{\mathbf{F}}\superscript{p} = \mathbf{D}\superscript{p} \mathbf{F}\superscript{p}.
\label{}
\end{equation}
The deformation gradient $\mathbf{F}$ admits a right and left polar decomposition given as
\begin{equation}
\mathbf{F} = \mathbf{R} \mathbf{U} = \mathbf{V} \mathbf{R}.
\label{F=RU=VR}
\end{equation}
The polar decomposition for the elastic and plastic part of deformation gradient can 
also be carried out as
\begin{equation}
\mathbf{F}\superscript{e} = \mathbf{R}\superscript{e} \mathbf{U}\superscript{e} = \mathbf{V}\superscript{e} \mathbf{R}\superscript{e},
\label{Fe=ReUe=VeRe}
\end{equation}
and
\begin{equation}
\mathbf{F}\superscript{p} = \mathbf{R}\superscript{p} \mathbf{U}\superscript{p} = \mathbf{V}\superscript{p} \mathbf{R}\superscript{p}.
\label{Fp=RpUp=VpRp}
\end{equation}

In what follows, we assume that the plastic part of the deformation gradient, i.e., $\mathbf{F}\superscript{p}$ (and therefore $\mathbf{L}\superscript{p}$) is invariant under frame transformation. This is a common approach \cite{anand2003theory,gurtin2010mechanics}, and simply implies that the fictitious intermediate configuration, which is obtained by elastic destressing, of a material point neighborhood, is 
invariant.  $\mathbf{F}\superscript{p}$ is a point wise map of each material point neighborhood from the reference configuration to the intermediate configuration (due to plastic distortion alone), and its invariance under the change of the observer can be assumed without the loss of generality; granted that the remaining kinematics of deformation is then described consistently under this assumption.



\subsection{ Rate-Independent Deformation Model Based on Multiplicative Decomposition of 
Deformation Gradient}
\label{sec:rate-independent-model}
Multiplicative decomposition of deformation gradient into elastic and plastic parts was first introduced 
in \cite{kroner1959allgemeine} and \cite{lee1969elastic}, and has gained wide popularity in modeling constitutive responses 
of variety of materials. In this section we summarize a deformation model based on the multiplicative decomposition, list the constitutive equations, and also provide the time integration algorithm.


Consider B to be the reference body. $P$ be an arbitrary part of the reference body which is represented by 
$P_t$ at any time $t$. $\mathbf{n}$ is the outward unit normal on the boundary of $P_t$. 
Let the traction field be represented by $\mathbf{t}$ (with the associated unit vector $\mathbf{n}$) and the body force by $\mathbf{b}\subscript{0}$. Accordingly, the body force including the effect of intertia 
can be written as

\begin{equation}
\mathbf{b} = \mathbf{b}\subscript{0} - \rho \ddot{\bm{\chi}}.
\label{body-force}
\end{equation}
Both $\mathbf{t}$ and $\mathbf{b}$ expend power 
over the velocity $\dot{\bm{\chi}}$, and the rate of external work can be written as
%
\begin{equation}
\mathscr{W} (P\subscript{t}) = \int_{\partial P\subscript{t}} \mathbf{t} (\mathbf{n}) . \dot{\bm{\chi}} da + \int_{P\subscript{t}} \mathbf{b} . \dot{\bm{\chi}} dv.
\label{W-external}
\end{equation}
Now we assume that the internal power inside the body is expended against two internal 
kinematical processes during deformation, elastic and plastic, and corresponding ``micro-stresses'' arising due to these processes are:

\begin{itemize}
\item An elastic stress $\mathbf{S}\superscript{e}$ (defined in the current configuration) that is power-conjugate to $\mathbf{L}\superscript{e}$. Thus, the power expended per unit volume of the current configuration is $\mathbf{S}\superscript{e}:\mathbf{L}\superscript{e}$.

\item A plastic stress $\mathbf{T}\superscript{p}$ (defined in the intermediate configuration) which is power-conjugate to $\mathbf{L}\superscript{p}$. Thus, the power expended per unit volume in the intermediate configuration is $\mathbf{T}\superscript{p}:\mathbf{L}\superscript{p}$.
Since $\mathbf{L}\superscript{p}$ is deviatoric, we assume that $\mathbf{T}\superscript{p}$ is also deviatoric. 
\end{itemize}
We note that the fictitious intermediate configuration and the reference configuration have same volume (due to plastic incompressibility). Thus, the total internal power expended in the body can be 
expressed as a volume integral over the current configuration as
\begin{equation}
\mathcal{I} (P\subscript{t}) = \int_{P\subscript{t}} (\mathbf{S}\superscript{e}:\mathbf{L}\superscript{e} + J\superscript{-1} \mathbf{T}\superscript{p}: \mathbf{L}\superscript{p}) dv.
\label{W-internal}
\end{equation}


\subsubsection{Principal of virtual power}

Let us consider that at any given fixed time, the deformation fields $\bm{\chi}$ and $\mathbf{F}\superscript{e}$ (hence $\mathbf{F}$ and $\mathbf{F}\superscript{p}$) are given. Now the virtual velocities corresponding to the fields $\mathbf{v}$, $\mathbf{L}\superscript{e}$ and $\mathbf{L}\superscript{p}$ be represented as the set $\mathscr{V} = (\tilde{\mathbf{v}}, \tilde{\mathbf{L}}\superscript{e}, \tilde{\mathbf{L}}\superscript{p})$. According to Equation ~\ref{velocity-gradient-decomposition}, the kinematically admissible virtual quantities must satisfy the following constraint
\begin{equation}
\text{grad} \tilde{\mathbf{v}} = \tilde{\mathbf{L}}\superscript{e} + \mathbf{F}\superscript{e} \tilde{\mathbf{L}}\superscript{p} \mathbf{F}\superscript{e-1}.
\label{chi-F-Fe-Dp-equation}
\end{equation}
Then, according to Equations ~\ref{W-external} and ~\ref{W-internal} we can write the 
external virtual power as
\begin{equation}
\mathscr{W} (P\subscript{t}, \mathscr{V}) = \int_{\partial P\subscript{t}} \mathbf{t} (\mathbf{n}) . \tilde{\mathbf{v}} da + \int_{P\subscript{t}} \mathbf{b} . \tilde{\mathbf{v}} dv,
\label{W-external-virtual}
\end{equation}
and the internal virtual work as
\begin{equation}
\mathscr{I} (P\subscript{t}, \mathscr{V}) = \int_{P\subscript{t}} (\mathbf{S}\superscript{e}:\tilde{\mathbf{L}}\superscript{e} + J\superscript{-1} \mathbf{T}\superscript{p}: \tilde{\mathbf{L}}\superscript{p}) dv.
\label{W-internal-virtual}
\end{equation}
According to the principle of virtual work, given any part P, $\mathscr{W} (P\subscript{t}, \mathscr{V}) = \mathscr{I} (P\subscript{t}, \mathscr{V})$ for all generalized and kinematically admissible 
virtual velocities $\mathscr{V}$, i.e.,
\begin{equation}
\int_{\partial P\subscript{t}} \mathbf{t} (\mathbf{n}) . \tilde{\mathbf{v}} da + \int_{P\subscript{t}} \mathbf{b} . \tilde{\mathbf{v}} dv = \int_{P\subscript{t}} (\mathbf{S}\superscript{e}:\tilde{\mathbf{L}}\superscript{e} + J\superscript{-1} \mathbf{T}\superscript{p}: \tilde{\mathbf{L}}\superscript{p}) dv.
\label{W-external-virtual = W-internal-virtual}
\end{equation}
In other words, the assumption (or choice) of work conjugate stress measures $\mathbf{S}\superscript{e}$ and $\mathbf{T}\superscript{p}$ is justified, if and only if, they satisfy Equation ~\ref{W-external-virtual = W-internal-virtual} for all kinematically admissible virtual velocities constrained by the  Equation 
~\ref{chi-F-Fe-Dp-equation}.  

\subsubsection{Consequences of frame indifference}

Based on the physical grounds, we require that the internal (virtual) work $\mathscr{I} (P\subscript{t}, \mathscr{V})$ (Equation ~\ref{W-internal-virtual}), which is a scalar quantity, must be invariant under a change of reference frame. 
Given a change in frame, if $P\subscript{t}\superscript{*}$ and $\mathscr{I}\superscript{*} (P\subscript{t}\superscript{*}, \mathscr{V}\superscript{*})$ represent the region and the internal work in the new frame, then the invariance of the internal work requires 

\begin{equation}
\mathscr{I} (P\subscript{t}, \mathscr{V}) = \mathscr{I}\superscript{*} (P\subscript{t}\superscript{*}, \mathscr{V}\superscript{*}),
\label{eq:frame-indiff-for-internal-work}
\end{equation}
where $\mathscr{V}\superscript{*}$ is the kinematically admissible virtual velocity in the new frame.
Also transformation of the virtual fields upon the change of reference follows identically their nonvirtual counterparts. Further from standard transformation laws under change of reference frame, we can write

\begin{equation}
\tilde{\mathbf{L}}\superscript{e*} = \mathbf{Q} \tilde{\mathbf{L}}\superscript{e} \mathbf{Q}\superscript{\top} + \bm{\Omega},
\label{le-star-frame-transformation}
\end{equation}
where $\tilde{\mathbf{L}}\superscript{e*}$ is the elastic distortion rate in the new frame, 
$\mathbf{Q}$ is the frame-rotation, and $\mathbf{\Omega}=\dot{\mathbf{Q}}\mathbf{Q}\superscript{T}$ is the frame-spin. Since, we have taken $\mathbf{F}\superscript{p}$ to be frame invariant under change of frame, thus, according to Equation ~\ref{plastic-part-of-velocity-gradient-decomp} $\mathbf{L}\superscript{p}$ is frame invariant, and so is $\tilde{\mathbf{L}}\superscript{p}$.
If we consider a virtual field with $\tilde{\mathbf{L}}\superscript{p}=\mathbf{0}$, then the invariance of internal 
virtual work Equations ~\ref{W-internal-virtual} and ~\ref{eq:frame-indiff-for-internal-work} reduces to
\begin{equation}
\int_{P\subscript{t}} (\mathbf{S}\superscript{e}:\tilde{\mathbf{L}}\superscript{e}) dv = \int_{P\subscript{t}\superscript{*}} (\mathbf{S}\superscript{e*}:\tilde{\mathbf{L}}\superscript{e*}), dv^*,
\label{eqn:internal-virtual-work-for-se}
\end{equation}
i.e., internal virtual work due to $\mathbf{S}\superscript{e}$ must be invariant. 
In Equation ~\ref{eqn:internal-virtual-work-for-se}, $dv=dv^*$ (since differential volume is unaffected upon change of reference), the integral on the right hand side can be transformed over $P\subscript{t}$, and since 
$P\subscript{t}$ is arbitrary the following should be satisfied pointwise
\begin{equation}
 \mathbf{S}\superscript{e}:\tilde{\mathbf{L}}\superscript{e} = \mathbf{S}\superscript{e*}:\tilde{\mathbf{L}}\superscript{e*}.
\label{point-wise-internal-elastic-work}
\end{equation}
Substituting $\tilde{\mathbf{L}}\superscript{e*}$ from Equation ~\ref{le-star-frame-transformation} into 
the right hand side of Equation ~\ref{point-wise-internal-elastic-work}, we obtain

\begin{equation}
 \mathbf{S}\superscript{e}:\tilde{\mathbf{L}}\superscript{e} = \mathbf{S}\superscript{e*}:\mathbf{Q} \tilde{\mathbf{L}}\superscript{e} \mathbf{Q}\superscript{\top} + \mathbf{S}\superscript{e*}:\bm{\Omega}.
\label{point-wise-internal-elastic-work-eq2}
\end{equation}
If we choose a frame whose spin $\mathbf{\Omega}$ is zero then above equation reduces to 
\begin{equation}
 \mathbf{S}\superscript{e}:\tilde{\mathbf{L}}\superscript{e} = \mathbf{S}\superscript{e*}:\mathbf{Q} \tilde{\mathbf{L}}\superscript{e} \mathbf{Q}\superscript{\top},
\label{point-wise-internal-elastic-work-eq3}
\end{equation}
or
\begin{equation}
 \mathbf{S}\superscript{e}:\tilde{\mathbf{L}}\superscript{e} =  \mathbf{Q}\superscript{\top}\mathbf{S}\superscript{e*}\mathbf{Q}:\tilde{\mathbf{L}}\superscript{e}.
\label{point-wise-internal-elastic-work-eq4}
\end{equation}
Since Equation ~\ref{point-wise-internal-elastic-work-eq4} must hold true for all kinematically admissible 
$\tilde{\mathbf{L}}\superscript{e}$, we conclude that

\begin{equation}
\mathbf{S}\superscript{e*}=\mathbf{Q} \mathbf{S}\superscript{e}\mathbf{Q}\superscript{\top}.
\label{point-wise-internal-elastic-work-eq5}
\end{equation}
Further, if we choose a rotating frame with $\mathbf{Q}=\mathbf{I}$, i.e., a frame that instantaneously coincides with the reference frame but with a non-zero spin then $\mathbf{S}\superscript{e*}= \mathbf{S}\superscript{e}$ at that instant, and Equation ~\ref{point-wise-internal-elastic-work-eq3} reduces to 
\begin{equation}
 \mathbf{S}\superscript{e}:\mathbf{\Omega} = 0.
\label{point-wise-internal-elastic-work-eq6}
\end{equation}
Since $\mathbf{\Omega}$ is arbitrary and skew, therefore Equation ~\ref{point-wise-internal-elastic-work-eq6} is satisfied for all $\mathbf{\Omega}$, if and only if, $\mathbf{S}\superscript{e}=\mathbf{S}\superscript{eT}$.  So far we utilized the fact that internal virtual work due to elastic stretching must be invariant by choosing
$\tilde{\mathbf{L}}\superscript{p}=0$, which is justified in an elastic-plastic constitutive model 
if one imagines deformation processes in the elastic regime only. The invariance of elastic part of
the internal virtual work implies that plastic part of the internal virtual work must also be invariant,
and since $\tilde{\mathbf{L}}\superscript{p}$ is invariant so should  $\tilde{\mathbf{T}}\superscript{p}$ be. 

\subsubsection{Macroscopic force balance}
We again consider a macroscopic virtual velocity $\mathscr{V}$ for which $\tilde{\mathbf{v}}$ is arbitrary and $\tilde{\mathbf{L}}\superscript{e} = \text{grad}\tilde{\mathbf{v}}$, 
with $\tilde{\mathbf{L}}\superscript{p} = 0$. By substituting $\tilde{\mathbf{L}}\superscript{p} = 0$ in Equation \ref{W-internal-virtual} and equating equations \ref{W-external-virtual} and \ref{W-internal-virtual}, we obtain

\begin{equation}
\int_{\partial P\subscript{t}} \mathbf{t} (\mathbf{n}) . \tilde{\mathbf{v}} da + \int_{P\subscript{t}} \mathbf{b} . \tilde{\mathbf{v}} dv = \int_{P\subscript{t}} (\mathbf{S}\superscript{e}:\tilde{\mathbf{L}}\superscript{e}) dv.
\label{macroforce-balance-equation-3}
\end{equation}
By using divergence theorem, we can rewrite the right hand term of the equation \ref{macroforce-balance-equation-3} as
\begin{equation}
\int_{P\subscript{t}} (\mathbf{S}\superscript{e}:\tilde{\mathbf{L}}\superscript{e}) dv = - \int_{P\subscript{t}} \text{div} \mathbf{S}\superscript{e} . \tilde{\mathbf{v}} dv + \int_{\partial P\subscript{t}} (\mathbf{S}\superscript{e} \mathbf{n}) . \tilde{\mathbf{v}} da.
\label{}
\end{equation}
Therefore, Equation \ref{macroforce-balance-equation-3} becomes

\begin{equation}
\int_{\partial P\subscript{t}} \mathbf{t} (\mathbf{n}) . \tilde{\mathbf{v}} da + \int_{P\subscript{t}} \mathbf{b} . \tilde{\mathbf{v}} dv = - \int_{P\subscript{t}} \text{div} \mathbf{S}\superscript{e} . \tilde{\mathbf{v}} dv + \int_{\partial P\subscript{t}} (\mathbf{S}\superscript{e} \mathbf{n}) . \tilde{\mathbf{v}} da,
\label{macroforce-balance-equation-4}
\end{equation}
upon re-writing
\begin{equation}
\int_{\partial P\subscript{t}} ( \mathbf{t} (\mathbf{n}) - \mathbf{S}\superscript{e} \mathbf{n}) . \tilde{\mathbf{v}} da + \int_{P\subscript{t}} (\text{div} \mathbf{S}\superscript{e} + \mathbf{b}) . \tilde{\mathbf{v}} dv = 0.
\label{macroforce-balance-equation-final}
\end{equation}
Since, Equation \ref{macroforce-balance-equation-final} must hold for all $P$ and all $\tilde{\mathbf{v}}$, the traction condition yields to
\begin{equation}
\mathbf{t} (\mathbf{n}) = \mathbf{S}\superscript{e} \mathbf{n},
\end{equation}
and the local force balance
\begin{equation}
\text{div} \mathbf{S}\superscript{e} + \mathbf{b} = \mathbf{0}.
\label{local-force-balance}
\end{equation}
This traction condition and the force balance and the symmetry, and frame-indifference of $\mathbf{S}\superscript{e}$ are classical conditions satisfied by the Cauchy stress $\mathbf{T}$, an observation that allow us to write
\begin{equation}
\mathbf{T} \stackrel{def}{=} \mathbf{S}\superscript{e},
\end{equation}
and to view 
\begin{equation}
\mathbf{T} = \mathbf{T}\superscript{\top}
\end{equation}
as the macroscopic stress and as the local macroscopic force balance. Granted that we are working in an inertial frame, so that Equation \ref{local-force-balance} reduces to the local balance law for linear momentum
\begin{equation}
\text{div} \mathbf{T} + \mathbf{b}\subscript{0} = \rho \dot{\mathbf{v}},
\label{local-balance-law}
\end{equation}
where $\mathbf{b}\subscript{0}$ is the non-inertial body force.


\subsubsection{Microscopic force balance}
Now we assume virtual fields $\mathscr{V}$ such that $\tilde{\mathbf{v}}=0$.
Then according to Equation ~\ref{chi-F-Fe-Dp-equation} we have 
\begin{equation}
\tilde{\mathbf{L}}\superscript{e} = - \mathbf{F}\superscript{e} \tilde{\mathbf{L}}\superscript{p} \mathbf{F}\superscript{e-1}.
\label{eq:virtual-field-constraint-1}
\end{equation}
Similarly, Equation ~\ref{W-external-virtual = W-internal-virtual} reduces to 
\begin{equation}
\int_{P\subscript{t}} (\mathbf{T}:\tilde{\mathbf{L}}\superscript{e} + J\superscript{-1} \mathbf{T}\superscript{p}:\tilde{\mathbf{L}}\superscript{p} ) dv = 0,
\label{micro-force-balance}
\end{equation}
for all $P_t$. This implies that quantity inside the integral is identically zero, i.e.,
\begin{equation}
J\superscript{-1} \mathbf{T}\superscript{p}:\tilde{\mathbf{L}}\superscript{p} = -\mathbf{T}:\tilde{\mathbf{L}}\superscript{e}.
\label{eq:virtual-field-constraint-2}
\end{equation}
Now substituting  $\tilde{\mathbf{L}}\superscript{e}$ from Equation ~\ref{eq:virtual-field-constraint-1}
into Equation ~\ref{eq:virtual-field-constraint-2} we obtain

\begin{equation}
J\superscript{-1} \mathbf{T}\superscript{p}:\tilde{\mathbf{L}}\superscript{p} = \mathbf{T}:(\mathbf{F}\superscript{e} \tilde{\mathbf{L}}\superscript{p} \mathbf{F}\superscript{e-1}).
\end{equation}
By performing some algebraic manipulation and utilizing the fact that 
$\tilde{\mathbf{L}}\superscript{p}$ is deviatoric and $\mathbf{T}$ is symmetric, 
we find that

%
%

\begin{equation}
J\superscript{-1} \mathbf{T}\superscript{p}:\tilde{\mathbf{L}}\superscript{p} = (\mathbf{F}\superscript{e\top} \mathbf{T}\subscript{0} \mathbf{F}\superscript{e-\top}) : \tilde{\mathbf{L}}\superscript{p}.
\end{equation}
Since $\tilde{\mathbf{L}}\superscript{p}$ is arbitrary, the microscopic force balance leads to
\begin{equation}
J \mathbf{F}\superscript{e\top} \mathbf{T}\subscript{0} \mathbf{F}\superscript{e-\top} = \mathbf{T}\superscript{p}.
\end{equation}
Next, we define the Mandel stress as 
\begin{equation}
\mathbf{M}\superscript{e}\stackrel{def}{=} J \mathbf{F}\superscript{e\top} \mathbf{T} \mathbf{F}\superscript{e-\top}
\label{define-mandel-stress}
\end{equation}
We note that 
%
%
%
%
\begin{equation}
\mathbf{T}\superscript{p}=\mathbf{M}\superscript{e}\subscript{0}.
\label{relation-between-tp-and-mandel}
\end{equation}

\subsubsection{Rate-independent elastic-plastic constitutive response}

We denote the free energy per unit volume of the intermediate configuration as $\varphi$, and make a  constitutive hypothesis that it only depends on the elastic part of the deformation 
$\mathbf{F}\superscript{e}$. Further, if we invoke the requirement of the frame invariance 
of free energy (which it must satisfy due to being a scalar quantity), then we can deduce 
%
%
%

\begin{equation}
\varphi = {{\hat{\varphi}}}(\mathbf{C}\superscript{e}).
\label{free-energy-function-1}
\end{equation}

Following the standard and well accepted principle of free-energy imbalance in solids \cite{coleman1963thermodynamics}, during the deformation, where 
$\mathscr{W}(P\subscript{t}) = \mathscr{I}(P\subscript{t})$, we require that 
\begin{equation}
{{\int_{P\subscript{t}} \dot{\varphi} J\superscript{-1} dv}} - \int_{P\subscript{t}} (\mathbf{T}:\mathbf{L}\superscript{e} + J\superscript{-1} \mathbf{T}\superscript{p}:\mathbf{L}\superscript{p}) dv \le 0.
\end{equation}
Since $P\subscript{t}$ is arbitrary the above equation implies
\begin{equation}
\dot{\varphi} - J \mathbf{T}:\mathbf{L}\superscript{e} - \mathbf{T}\superscript{p}:\mathbf{L}\superscript{p} = - \delta \le 0,
\label{free-energy-imbalace-dissipation-equation}
\end{equation}
where $\delta$ is the dissipation magnitude. We note that $\mathbf{T}$ is symmetric, thus 
\begin{equation}
\mathbf{T}:\mathbf{L}\superscript{e}=\mathbf{T}:\mathbf{D}\superscript{e}.
\label{elastic-work-conjugate-1}
\end{equation}
By defining
\begin{equation}
\mathbf{T}\superscript{e}\stackrel{def}{=} J \mathbf{F}\superscript{e-1} \mathbf{T} \mathbf{F}\superscript{e-\top},
\label{definition-te}
\end{equation}
we can show that 
\begin{equation}
J\mathbf{T}:\mathbf{D}\superscript{e}= \frac{1}{2}\mathbf{T}\superscript{e}: \dot{\mathbf{C}\superscript{e}}.
\label{elastic-work-conjugate-2}
\end{equation}
Also, using the definition of  $\mathbf{M}\superscript{e}$ 
from Equation ~\ref{define-mandel-stress} and that of
$\mathbf{T}\superscript{e}$ from Equation ~\ref{definition-te} we obtain 
\begin{equation}
\mathbf{M}\superscript{e}= \mathbf{C}\superscript{e}\mathbf{T}\superscript{e}.
\label{relation-between-te-me}
\end{equation}
Using Equation ~\ref{free-energy-function-1} we can write
\begin{equation}
\dot{\varphi} = \frac{\partial \hat{\varphi}(\mathbf{C}\superscript{e})}{\partial \mathbf{C}\superscript{e}} : \dot{\mathbf{C}\superscript{e}}
\label{free-energy-time-derivative-1}
\end{equation}
and, now using  Equations ~\ref{elastic-work-conjugate-1}, ~\ref{elastic-work-conjugate-2} and ~\ref{free-energy-time-derivative-1} in Equation ~\ref{free-energy-imbalace-dissipation-equation}, the free-energy imbalance can be written as
\begin{equation}
\Big[ \frac{1}{2}{\mathbf{T}}\superscript{e} - \frac{\partial \hat{\varphi}(\mathbf{C}\superscript{e})}{\partial \mathbf{C}\superscript{e}} \Big] : \dot{\mathbf{C}\superscript{e}} + {\mathbf{T}}\superscript{p} : \mathbf{L}\superscript{p} \ge 0.
\label{inequality-constitutive}
\end{equation}
If we consider kinematical processes in which $\mathbf{L}\superscript{p} = \mathbf{0}$ and 
\begin{equation}
\dot{\mathbf{C}\superscript{e}} = -\kappa \Big[\frac{1}{2}{\mathbf{T}}\superscript{e} - \frac{\partial \hat{\varphi}(\mathbf{C}\superscript{e})}{\partial \mathbf{C}\superscript{e}} \Big],
\end{equation}
where $\kappa$ is some positive constant then Equation ~\ref{inequality-constitutive} reduces to 
\begin{equation}
- \kappa \Big[ {\frac{1}{2}{\mathbf{T}\superscript{e}}} - \frac{\partial \hat{\varphi}(\mathbf{C}\superscript{e})}{\partial \mathbf{C}\superscript{e}} \Big] : \Big[ {\frac{1}{2}{\mathbf{T}}\superscript{e}} - \frac{\partial \hat{\varphi}(\mathbf{C}\superscript{e})}{\partial \mathbf{C}\superscript{e}} \Big]\ge 0.
\label{inequality-constitutive-2}
\end{equation}
The only possible way to satisfy equation ~\ref{inequality-constitutive-2}, for arbitrary pair of
nonzero $\mathbf{T}\superscript{e}$ and $\frac{\partial \hat{\varphi}(\mathbf{C}\superscript{e})}{\partial \mathbf{C}\superscript{e}}$, is to have 
\begin{equation}
 {\mathbf{T}}\superscript{e} = 2\frac{\partial \hat{\varphi}(\mathbf{C}\superscript{e})}{\partial \mathbf{C}\superscript{e}}.
\label{inequality-constitutive-3}
\end{equation}
An immediate consequence of Equation ~\ref{inequality-constitutive-3} in conjunction with Equation ~\ref{inequality-constitutive} is that 
\begin{equation}
\mathbf{T}\superscript{p}: \mathbf{L}\superscript{p} \ge 0.
\label{TpLpepge0}
\end{equation}
Equation ~\ref{TpLpepge0} is satisfied for all 
possible combinations of  $\mathbf{T}\superscript{p}$ and $\mathbf{L}\superscript{p}$, if and only if, the deviatoric plastic velocity gradient is colinear with   $\mathbf{T}\superscript{p}$. 
To proceed further, we need to specify the form of free energy function. If we assume that the material remains isotropic during deformation, then the dependence of the free energy $\varphi$ 
reduces to only the invariants of $\mathbf{C}\superscript{e}$, i.e.
\begin{equation}
\varphi =\tilde{\varphi} (\mathscr{I}\subscript{\mathbf{C}\superscript{e}}),
\label{free-energy-fun-on-invariant-set}
\end{equation}
where $\mathscr{I}\subscript{\mathbf{C}\superscript{e}}=(I_1(\mathbf{C}\superscript{e}),I_2(\mathbf{C}\superscript{e}),I_3(\mathbf{C}\superscript{e}))$ is set of invariants 
of $\mathbf{C}\superscript{e}$. The spectral decomposition of $\mathbf{C}\superscript{e}$ can be 
written as 
\begin{equation}
\mathbf{C}\superscript{e} = \sum_{i=1}^{3} \omega\subscript{i}^{e} \mathbf{r}\subscript{i}^{e} \otimes \mathbf{r}\subscript{i}^{e}, \indent \text{with} \indent \omega\subscript{i}^{e} = \lambda\subscript{i}^{e2},
\label{eq:spectral-decom-ce}
\end{equation}
where $(\mathbf{r}\subscript{1}^{e},\mathbf{r}\subscript{2}^{e},\mathbf{r}\subscript{3}^{e})$
are the orthonormal eigenvectors of $\mathbf{C}\superscript{e}$ and $\mathbf{U}\superscript{e}$,
and $(\lambda\subscript{1}^{e}, \lambda\subscript{2}^{e}, \lambda\subscript{3}^{e})$ are the positive eigen values of $\mathbf{U}\superscript{e}$. The free energy function 
~\ref{free-energy-fun-on-invariant-set} can then be written as a function of 
$\lambda\subscript{1}^{e}$, $\lambda\subscript{2}^{e}$ and $\lambda\subscript{3}^{e}$, i.e.,
\begin{equation}
\varphi = \hat{\psi}\superscript{} (\lambda\subscript{1}^{e}, \lambda\subscript{2}^{e}, \lambda\subscript{3}^{e}).
\label{free-energy-fun-on-invariant-set-2}
\end{equation}
Using Equation ~\ref{inequality-constitutive-3} and chain rule of differential we can write
\begin{equation}
\mathbf{T}\superscript{e} = 2 \frac{\partial \hat{\psi}\superscript{} (\lambda\subscript{1}^{e}, \lambda\subscript{2}^{e}, \lambda\subscript{3}^{e})}{\partial \mathbf{C}\superscript{e}} = 2 \sum_{i=1}^{3} \frac{\partial \hat{\psi}\superscript{} (\lambda\subscript{1}^{e}, \lambda\subscript{2}^{e}, \lambda\subscript{3}^{e})}{\partial \lambda\subscript{i}^{e}} \frac{\partial \lambda\subscript{i}^{e}}{\partial \mathbf{C}\superscript{e}} = \sum_{i=1}^{3} \frac{1}{\lambda\subscript{i}^{e}} \frac{\partial \hat{\psi}\superscript{} (\lambda\subscript{1}^{e}, \lambda\subscript{2}^{e}, \lambda\subscript{3}^{e})}{\partial \lambda\subscript{i}^{e}} \frac{\partial \omega\subscript{i}}{\partial \mathbf{C}\superscript{e}}.
\label{eq:expression-for-te}
\end{equation}
Using Equation ~\ref{eq:spectral-decom-ce} into Equation ~\ref{eq:expression-for-te},  
it can be proved that 
\begin{equation}
\frac{\partial \omega\subscript{i}^{e}}{\partial \mathbf{C}\superscript{e}} = \mathbf{r}\subscript{i}^{e} \otimes \mathbf{r}\subscript{i}^{e},
\label{eq:constitutive-relation}
\end{equation}
and now using Equations ~\ref{eq:constitutive-relation} and ~\ref{eq:expression-for-te} we obtain
\begin{equation}
\mathbf{T}\superscript{e} = \sum_{i=1}^{3} \frac{1}{\lambda\subscript{i}^{e}} \frac{\partial \hat{\psi}\superscript{} (\lambda\subscript{1}^{e}, \lambda\subscript{2}^{e}, \lambda\subscript{3}^{e})}{\partial \lambda\subscript{i}^{e}} \mathbf{r}\subscript{i}^{e} \otimes \mathbf{r}\subscript{i}^{e}.
\end{equation}
Using Equation ~\ref{definition-te}, we can now find the Cauchy stress  $\mathbf{T}$ as
\begin{equation}
\mathbf{T} = J\superscript{-1} \mathbf{F}\superscript{e} \mathbf{T}\superscript{e} \mathbf{F}\superscript{e\top} = J\superscript{-1} \mathbf{R}\superscript{e} \mathbf{U}\superscript{e} \mathbf{T}\superscript{e} \mathbf{U}\superscript{e} \mathbf{R}\superscript{e\top}  = J\superscript{-1} \mathbf{R}\superscript{e} \Big( \sum_{i=1}^{3} \lambda\subscript{i}^{e} \frac{\partial \hat{\psi}\superscript{} (\lambda\subscript{1}^{e}, \lambda\subscript{2}^{e}, \lambda\subscript{3}^{e})}{\partial \lambda\subscript{i}^{e}} \mathbf{r}\subscript{i}^{e} \otimes \mathbf{r}\subscript{i}^{e} \Big)\mathbf{R}\superscript{e\top}.
\end{equation}
Similarly, using Equation ~\ref{relation-between-te-me} we can obtain an expression for 
$\mathbf{M}\superscript{e}$ as
\begin{equation}
\mathbf{M}\superscript{e} = \sum_{i=1}^{3} \lambda\subscript{i}^{e} \frac{\partial \hat{\psi}\superscript{} (\lambda\subscript{1}^{e}, \lambda\subscript{2}^{e}, \lambda\subscript{3}^{e})}{\partial \lambda\subscript{i}^{e}} \mathbf{r}\subscript{i}^{e} \otimes \mathbf{r}\subscript{i}^{e}.
\end{equation}
Next, we define a measure of elastic strain $\mathbf{E}\superscript{e}$ as
\begin{equation}
\mathbf{E}\superscript{e} \stackrel{def}{=} \text{ln} \mathbf{U}\superscript{e} = \sum_{i=1}^{3}  E\subscript{i}\superscript{e} \mathbf{r}\subscript{i}^{e} \otimes \mathbf{r}\subscript{i}^{e}
\end{equation}
where, 
\begin{equation}
E\subscript{i}\superscript{e} \stackrel{def}{=} \text{ln}  \lambda\subscript{i}^{e}.
\end{equation}
The free energy function can now be re-written in terms of $E\subscript{1}^{e}$, $E\subscript{2}^{e}$ and $E\subscript{3}^{e}$ as 
\begin{equation}
\varphi=\hat{\psi}\superscript{} (\lambda\subscript{1}^{e}, \lambda\subscript{2}^{e}, \lambda\subscript{3}^{e}) = {\psi}\superscript{} (E\subscript{1}^{e}, E\subscript{2}^{e}, E\subscript{3}^{e}).
\end{equation}
\begin{equation}
\mathbf{M}\superscript{e} = \sum_{i=1}^{3} \frac{\partial {\psi}\superscript{} (E\subscript{1}^{e}, E\subscript{2}^{e}, E\subscript{3}^{e})}{\partial E\subscript{i}^{e}} \mathbf{r}\subscript{i}^{e} \otimes \mathbf{r}\subscript{i}^{e}.
\label{eq:mandel-in-terms-of-free-energy}
\end{equation}
A popular choice for free energy form that is valid for moderately large elastic deformation is 
as follows
\begin{equation}
{\psi}\superscript{} (\mathbf{E}\superscript{e}) = G |\mathbf{E}\superscript{e}\subscript{0}|\superscript{2} + \frac{1}{2} K (\text{tr} \mathbf{E}\superscript{e})\superscript{2},
\label{eq:free-energy-functional-form}
\end{equation}
where $G$ is the shear modulus and $K$ is the bulk modulus. If  Youngs modulus $E$ and Poisson's ratio $\nu$ are known, then bulk and shear modulus are obtained from following relations
\begin{equation}
G = \frac{E}{2 (1 + \nu)},
\end{equation}

\begin{equation}
K = \frac{2 G\nu}{1 - 2\nu}.
\end{equation}
The free energy is an isotropic function of 
$\mathbf{E}\superscript{e}$ (satisfying frame invariance and material isotropy). 
From Equations  ~\ref{eq:free-energy-functional-form} and ~\ref{eq:mandel-in-terms-of-free-energy}, we obtain an expression for Mandel stress in terms of $\mathbf{E}\superscript{e}$ as
\begin{equation}
\mathbf{M}\superscript{e} =  2 G \mathbf{E}\superscript{e}\subscript{0} + K (\text{tr} \mathbf{E}\superscript{e}) \mathbf{I}.
\label{mandel-stress}
\end{equation}
Using the relation in Equation ~\ref{definition-te}, now we can obtain the Cauchy stress in the 
current configuration as
\begin{equation}
\mathbf{T} = J\superscript{-1} \mathbf{R}\superscript{e} \mathbf{M}\superscript{e} \mathbf{R}\superscript{e\top}.
\label{cauchy-mandel-relation-1}
\end{equation}
%
%
%
%
%
%
%
%
%
%
%
We note that the free energy, Equation ~\ref{eq:free-energy-functional-form}, 
based on Hencky strain measure does not satisfy the property of polyconvexity, a sufficient condition for existence of minimizers for problems of nonlinear elasticity, see \cite{montella2016exponentiated,neff2015exponentiated,simo1998numerical}; however,
for small to moderately large strains in problems of nonlinear elasticity, and large strain plasticity of glassy polymers, its use has proven to be extremely effective.


\subsubsection{Plastic flow: yield and consistency condition}

Recall that earlier in Section ~\ref{sec:rate-independent-model}, we defined 
$\mathbf{T}\superscript{p}$ as the work conjugate to plastic flow which is 
characterized by $\mathbf{L}\superscript{p}$ (or $\mathbf{D}\superscript{p}$). 
Now we require a condition to determine the onset of yielding or active plastic flow. 
To this end, the yield surface is represented with a spherical surface of radius $Y(e\superscript{p}) > 0$ in the space of symmetric and deviatoric
tensors. Here $e\superscript{p}$ indicates the accumulated plastic strain, and radius of the
the yield surface depends on it.  The elastic range then identified as the closed ball with the radius $Y(e\superscript{p})$. Plastic flow can only happen when $\mathbf{T}\superscript{p}$ lies on the 
yield surface. From Equation ~\ref{relation-between-tp-and-mandel} we have
$|\mathbf{M}\subscript{0}\superscript{e}|=|\mathbf{T}\superscript{p}|$ \footnote{In our notation, the $|\mathbf{A}|$ of a tensor $\mathbf{A}$ is equal to $\sqrt{\frac{3}{2}\mathbf{A}:\mathbf{A}}$ }, and thus 
condition for yielding can be written as 

\begin{equation}
|\mathbf{M}\subscript{0}\superscript{e}| = Y(e\superscript{p}) \indent \text{for} \indent \mathbf{D}\superscript{p} \neq 0,
\label{yield-condition-1}
\end{equation}
and the condition
\begin{equation}
\mathbf{D}\superscript{p} = 0 \indent \text{for} \indent |\mathbf{M}\subscript{0}\superscript{e}| < Y(e\superscript{p}).
\label{yield-condition-2}
\end{equation}
We introduce a scalar yield function 
\begin{equation}
f = |\mathbf{M}\subscript{0}\superscript{e}| - Y(e\superscript{p}),
\label{yield-condition-3}
\end{equation}
and $f$ follows the following constraint
\begin{equation}
- Y(e\superscript{p}) \leq f \leq 0.
\end{equation}
The yield condition in Equation \ref{yield-condition-1} is then equivalent to the requirement that
\begin{equation}
f = 0 \indent \text{for} \indent \mathbf{D}\superscript{p} \neq 0,
\label{yield-condition-4}
\end{equation}
and Equation \ref{yield-condition-2} takes the form
\begin{equation}
\mathbf{D}\superscript{p} = 0 \indent \text{for} \indent f < 0.
\label{yield-condition-5}
\end{equation}
The yield function $f$ obeys the following additional restriction:
\begin{equation}
\text{if} \indent f = 0 \indent \text{then} \indent \dot{f} \leq 0.
\label{yield-condition-6}
\end{equation}
This leads to the no-flow condition
\begin{equation}
\mathbf{D}\superscript{p} = 0 \indent \text{if} \indent f < 0 \indent \text{or if} \indent f = 0 \indent \text{and} \indent \dot{f} < 0,
\label{yield-condition-7}
\end{equation}
and the consistency condition
\begin{equation}
\text{if} \indent \mathbf{D}\superscript{p} \neq 0 \indent \text{, then} \indent f = 0 \indent \text{and} \indent \dot{f} = 0.
\label{yield-condition-8}
\end{equation}
Next, let
\begin{equation}
Y\superscript{'} (e\superscript{p}) = \frac{d Y(e\superscript{p}) }{d e\superscript{p} },
\label{yield-condition-9}
\end{equation}
so that by Equation \ref{yield-condition-4}
\begin{equation}
\dot{\overline{Y(e\superscript{p})}} = Y\superscript{'} (e\superscript{p}) |\mathbf{D}\superscript{p}|;
\label{yield-condition-10}
\end{equation}
then, letting
\begin{equation}
H(e\superscript{p}) \stackrel{def}{=} Y\superscript{'} (e\superscript{p}),
\label{yield-condition-11}
\end{equation}
we arrive at 
\begin{equation}
\dot{\overline{Y(e\superscript{p})}} = H(e\superscript{p}) |\mathbf{D}\superscript{p}|.
\label{yield-condition-12}
\end{equation}
Thus, if we assume that $\mathbf{D}\superscript{p} \neq \mathbf{0}$, then by Equation \ref{yield-condition-1}, $|\mathbf{M}\subscript{0}\superscript{e}| - Y(e\superscript{p}) = 0$, hence Equation \ref{yield-condition-12} yields
\begin{equation}
\dot{\overline{|\mathbf{M}\subscript{0}\superscript{e}|}} = H(e\superscript{p}) |\mathbf{D}\superscript{p}|.
\label{yield-condition-13}
\end{equation}
First of all, a consequence of flow rule is that
\begin{equation}
\mathbf{N}\superscript{p} = \frac{\mathbf{M}\subscript{0}\superscript{e}}{|\mathbf{M}\subscript{0}\superscript{e}|},
\label{normality-flow-rule}
\end{equation}
i.e., the deviatoric Mandel tensor points in the direction of plastic flow. During plastic flow we have

\begin{equation}
\dot{f} = \dot{\overline{|\mathbf{M}\subscript{0}\superscript{e}|}} - \dot{\overline{Y(e\superscript{p})}},
\label{eq:yield-condition-derivative-1}
\end{equation}

\begin{equation}
\dot{f} =\frac{\mathbf{M}\subscript{0}\superscript{e}}{|\mathbf{M}\subscript{0}\superscript{e}|}:\dot{\mathbf{M}\subscript{0}\superscript{e}} - H(e\superscript{p}) |\mathbf{D}\superscript{p}|,
\end{equation}

\begin{equation}
\dot{f} = \mathbf{N}\superscript{p}:\dot{\mathbf{M}\subscript{0}\superscript{e}} - H(e\superscript{p}) |\mathbf{D}\superscript{p}|.
\end{equation}
Our next step is to compute the term $\mathbf{N}\superscript{p}:\dot{\mathbf{M}\subscript{0}\superscript{e}}$. Clearly,
\begin{equation}
\mathbf{N}\superscript{p}:\dot{\mathbf{M}\subscript{0}\superscript{e}} = \mathbf{N}\superscript{p}:\dot{\mathbf{M}\superscript{e}},
\end{equation}
because $\mathbf{N}\superscript{p}$ is deviatoric.

\subsubsection{Time Integration for Rate-Independent Multiplicative Plasticity}
Now we describe the time integration for the rate-independent and isotropic hardening deformation model with the multiplicative decomposition of the deformation gradient. The structure adopted here is that of a deformation driven problem, i.e., at each integration point of the finite element
an updated deformation gradient is provided, and we are required to compute all the desired quantities. 
Most quantities are updated explicitly, however, the yield condition (and  associated consistency parameter) are solved iteratively. \\

\noindent \textbf{Initialization: } At time step $n = 0$,

Compute the volume ratio $J\subscript{n+1}$ using

\begin{equation}
J\subscript{n+1} = det(\mathbf{F}\superscript{}\subscript{n+1}).
\label{J-n+1-t0}
\end{equation}

Initialize $\mathbf{N}\superscript{p} = 0$, $\Delta\overline{\varepsilon}\superscript{pl} = 0$, and 
\begin{equation}
\mathbf{F}\superscript{p}\subscript{n+1} = \mathbf{I}
\end{equation}
where $\mathbf{I}$ is the second order identity tensor. Then,

\begin{equation}
\mathbf{F}\superscript{e}\subscript{n+1} = \mathbf{F}\superscript{}\subscript{n+1} \mathbf{F}\superscript{p-1}\subscript{n+1} = \mathbf{F}\superscript{}\subscript{n+1}.
\label{F-e-n+1-t0}
\end{equation}

Decompose $\mathbf{F}\superscript{e}\subscript{n+1}$ from Equation \ref{F-e-n+1-t0} into $\mathbf{R}\superscript{e}\subscript{n+1}$ and $\mathbf{U}\superscript{e}\subscript{n+1}$ as

\begin{equation}
\mathbf{F}\superscript{e}\subscript{n+1} = \mathbf{R}\superscript{e}\subscript{n+1} \mathbf{U}\superscript{e}\subscript{n+1}.
\label{Fe=ReUet0}
\end{equation}

Compute the Hencky strain  $\mathbf{E}\superscript{e}\subscript{n+1}$ as

\begin{equation}
\mathbf{E}\superscript{e}\subscript{n+1} = ln (\mathbf{U}\superscript{e}\subscript{n+1}).
\label{E-e-n+1-t0}
\end{equation}

Compute the Mandel stress as

\begin{equation}
\mathbf{M}\superscript{e}\subscript{n+1} = 2\,G\,\mathbf{E}\superscript{e}\subscript{n+1} + K \left[ tr(\mathbf{E}\superscript{e}\subscript{n+1}) \right] \mathbf{I}.
\label{M-e-n+1}
\end{equation}

Now Cauchy stress $\mathbf{T}$ can be calculated by transforming the Mandel stress into the current configuration as

\begin{equation}
\mathbf{T}\subscript{n+1} = \frac{1}{J\subscript{n+1}} \mathbf{R}\subscript{n+1}\superscript{e}  \mathbf{M}\subscript{n+1}\superscript{e}  \mathbf{R}\subscript{n+1}\superscript{e\top}.
\end{equation}

 For any time step $n>0$, assume the following quantities are known from previous step \{$\mathbf{F}\superscript{}\subscript{n}$, $\mathbf{F}\subscript{n}\superscript{e}$, $\mathbf{F}\subscript{n}\superscript{p}$, $Y\subscript{n}$, $\mathbf{M}\subscript{n}\superscript{e}$, $\overline{\varepsilon}\subscript{n}\superscript{pl}$, $h\subscript{n}$\}. Given $\mathbf{F}\superscript{}\subscript{n+1}$, our goal is to find \{$\mathbf{F}\subscript{n+1}\superscript{e}$, $\mathbf{F}\subscript{n+1}\superscript{p}$, $Y\subscript{n+1}$, $\mathbf{M}\subscript{n+1}\superscript{e}$, $\overline{\varepsilon}\subscript{n+1}\superscript{pl}$, $h\subscript{n+1}$\}.
We carry out explicit integration for all the quantities except for the consistency parameter, which we solve through numerical iterations so as to accurately satisfy the 
consistency condition during yielding. Following steps elaborate the procedure (trial quantities are denoted by $^{**}$):\\


\noindent \textbf{Step 1:} Compute the volume ratio $J\subscript{n+1}$ using
\begin{equation}
J\subscript{n+1} = det(\mathbf{F}\superscript{}\subscript{n+1}).
\label{J-n+1}
\end{equation}

\noindent \textbf{Step 2:}  Compute 
\begin{equation}
\mathbf{F}\superscript{e**}\subscript{n+1} =\mathbf{F}\superscript{}\subscript{n+1} \mathbf{F}\superscript{p-1}\subscript{n} 
\label{F-e-n+1-trial}. 
\end{equation}

\textbf{Step 2.1}: 
\begin{equation}
\mathbf{F}\superscript{e**}\subscript{n+1} = \mathbf{R}\superscript{e**}\subscript{n+1} \mathbf{U}\superscript{e**}\subscript{n+1}.
\label{Fe=ReUe*}
\end{equation}

\textbf{Step 2.2}:  Compute the Hencky strain  $\mathbf{E}\superscript{e**}\subscript{n+1}$ as

\begin{equation}
\mathbf{E}\superscript{e**}\subscript{n+1} = ln (\mathbf{U}\superscript{e**}\subscript{n+1}).
\label{E-e-n+1-*}
\end{equation}

\textbf{Step 2.3}: Compute the Mandel stress as

\begin{equation}
\mathbf{M}\superscript{e**}\subscript{n+1} = 2\,G\,\mathbf{E}\superscript{e**}\subscript{n+1} + K \left[ tr(\mathbf{E}\superscript{e**}\subscript{n+1}) \right] \mathbf{I}.
\label{M-e-n+1*}
\end{equation}

\textbf{Step 2.4}: Compute the equivalent (von mises) stress in the intermediate configuration using the deviatoric part of the Mandel stress:

\begin{equation}
(\sigma\subscript{v}\superscript{**})\subscript{n+1} = \sqrt{\frac{3}{2} (\mathbf{M}\superscript{e**}\subscript{n+1})\superscript{}\subscript{0} : (\mathbf{M}\superscript{e**}\subscript{n+1})\superscript{}\subscript{0}}.
\label{von-mises-n*}
\end{equation}

In the above equation  $(\mathbf{M}\superscript{e**}\subscript{n+1})\superscript{}\subscript{0}$ denotes the 
deviatoric part of the Mandel stress $\mathbf{M}\superscript{e**}\subscript{n+1}$. 
$(\mathbf{M}\superscript{e**}\subscript{n+1})\superscript{}\subscript{0}$ is given by

\begin{equation}
(\mathbf{M}\superscript{e**}\subscript{n+1})\superscript{}\subscript{0} = \mathbf{M}\superscript{e**}\subscript{n+1} + \bar{p}\superscript{**}\subscript{n+1}\,\mathbf{I},
\label{Mandel-stress-deviatoric-n*}
\end{equation}

where 


\begin{equation}
\bar{p}\superscript{**}\subscript{n+1} = -\frac{1}{3} tr(\mathbf{M}\superscript{e**}\subscript{n+1}),
\end{equation}

and $\mathbf{I}$ is the second order identity tensor. Here we have not included any kinematic hardening (or back stress). Mandel stress is work conjugate to the plastic velocity gradient, and we have assumed plastic spin to be zero (on grounds of material isotropy), and plastic velocity gradient is essentially plastic stretching. \\

\noindent \textbf{Step 3:}
Now we check for yielding by comparing the equivalent stress 
$(\sigma\subscript{v}\superscript{**})\subscript{n+1} $, given by Equation \ref{von-mises-n*}, and current yield strength  $Y\subscript{n}$. If 
$(\sigma\subscript{v}\superscript{**})\subscript{n+1}  - Y\subscript{n} < 0$, then 
the current increment is only elastic, set equivalent plastic strain increment 
to be zero ($\Delta\overline{\varepsilon}\subscript{n}\superscript{pl} = 0$), 
and goto \textbf{Step 6}. \\


\noindent \textbf{Step 4:} If the yield condition is satisfied and plasticity has to occur, then  we calculate the direction of plastic flow ($\mathbf{N}\superscript{p}\subscript{n}$) 
 and take it to be co-linear with the deviatoric part of Mandel stress.

\begin{equation}
\mathbf{N}\superscript{p}\subscript{n} =  \frac{ (\mathbf{M}\superscript{e}\subscript{n})\superscript{}\subscript{0}}{ \sqrt{\frac{2}{3}  (\mathbf{M}\superscript{e}\subscript{n})\superscript{}\subscript{0} : (\mathbf{M}\superscript{e}\subscript{n})\superscript{}\subscript{0}} }.
\label{N-p-n}
\end{equation}


\noindent \textbf{Step 5:} The plastic stretching $\mathbf{D}\superscript{p}\subscript{n}$
is given as 

\begin{equation}
\mathbf{D}\superscript{p}\subscript{n} = \dot{\lambda} \mathbf{N}\superscript{p}\subscript{n},
\label{D-p-n}
\end{equation}

where $\dot{\lambda}$ is the equivalent plastic strain rate. If we assume that the current time step is small then $\dot{\lambda}dt = \Delta\overline{\varepsilon}\subscript{n}\superscript{pl} = d\lambda$. Now the goal is to find the incremental equivalent plastic strain such that consistency condition 
for the yield locus is satisfied.

%

\noindent \textbf{Step 5:} Here we demonstrate how to solve for the consistency parameter ($d\lambda$) through iterations. During this iterative search we use $(**)$ to denote trial quantities.
We start with some trial value for consistency parameter as 
$d\lambda\superscript{**}$.\\

\textbf{Step 5.1}: Compute trial plastic deformation gradient (recall that in a continuum setting we have $\dot{\mathbf{F}}\superscript{p}=\mathbf{D}\superscript{p}\mathbf{F}\superscript{p}$ )

\begin{equation}
\mathbf{F}\superscript{p**}\subscript{n+1} =\mathbf{F}\superscript{p}\subscript{n} + d\lambda\superscript{**} \mathbf{N}\superscript{p}\subscript{n} \mathbf{F}\superscript{p}\subscript{n} 
\label{F-p-n+1-*}. 
\end{equation}

\textbf{Step 5.2}: Compute trial elastic deformation gradient 

\begin{equation}
\mathbf{F}\superscript{e**}\subscript{n+1} = \mathbf{F}\superscript{}\subscript{n+1} \mathbf{F}\superscript{p**-1}\subscript{n+1}.
\label{F-e-n+1-*}
\end{equation}

\textbf{Step 5.3}: Decompose the obtained $\mathbf{F}\superscript{e**}\subscript{n+1}$ from Equation \ref{F-e-n+1-*} into $\mathbf{R}\superscript{e**}\subscript{n+1}$ and $\mathbf{U}\superscript{e**}\subscript{n+1}$ as

\begin{equation}
\mathbf{F}\superscript{e**}\subscript{n+1} = \mathbf{R}\superscript{e**}\subscript{n+1} \mathbf{U}\superscript{e**}\subscript{n+1}.
\label{Fe=ReUe*}
\end{equation}

\textbf{Step 5.4}: Compute the trial Henkcy strain $\mathbf{E}\superscript{e**}\subscript{n+1}$ as
\begin{equation}
\mathbf{E}\superscript{e**}\subscript{n+1} = ln (\mathbf{U}\superscript{e**}\subscript{n+1}).
\label{E-e-n+1*}
\end{equation}

\textbf{Step 5.5}: Compute the trial Mandel stress as
\begin{equation}
\mathbf{M}\superscript{e**}\subscript{n+1} = 2\,G\,\mathbf{E}\superscript{e**}\subscript{n+1} + K \left[ tr(\mathbf{E}\superscript{e**}\subscript{n+1}) \right] \mathbf{I},
\label{M-e-n+1}
\end{equation}

\textbf{Step 5.6}: Compute the trial equivalent (von Mises) stress in the intermediate configuration using the deviatoric part of the Mandel stress.

\begin{equation}
(\sigma\subscript{v})\subscript{n+1}\superscript{**} = \sqrt{\frac{3}{2} (\mathbf{M}\superscript{e**}\subscript{n+1})\superscript{}\subscript{0} : (\mathbf{M}\superscript{e**}\subscript{n+1})\superscript{}\subscript{0}},
\label{von-mises-n*}
\end{equation}

where

\begin{equation}
(\mathbf{M}\superscript{e**}\subscript{n+1})\superscript{}\subscript{0} = \mathbf{M}\superscript{e**}\subscript{n+1} + \bar{p}\subscript{n}\superscript{**}\,\mathbf{I}
\label{Mandel-stress-deviatoric-n*},
\end{equation}

and 

\begin{equation}
\bar{p}\subscript{n+1}\superscript{**} = -\frac{1}{3} tr(\mathbf{M}\superscript{e**}\subscript{n+1}).
\end{equation}

\textbf{Step 5.7}: Compute trial value of the new yield locus:

\begin{equation}
Y\subscript{n}\superscript{**} =Y\subscript{n}+ h\subscript{n+1}\superscript{**} d\lambda\superscript{**}.
\end{equation}

In the above $h\subscript{n}\superscript{**}$ is the hardening modulus (and depends upon the accumulated plastic strain).

\textbf{Step 5.8}: Check whether the consistency condition is satisfied:
\begin{equation}
(\sigma\subscript{v})\subscript{n+1}\superscript{**} - Y\subscript{n} - h\subscript{n+1}\superscript{**} d\lambda\superscript{**} = 0
\label{consistency-equation-*}
\end{equation}

If the above equation is not satisfied (according to some predefined numerical tolerance), then choose the different value of $d\lambda\superscript{**}$ and go to 
\textbf{Step 5.1}. We performed this iterative procedure using a Bisection method.

\textbf{Step 5.9}: The solution $d\lambda$ satisfying the Equation \ref{consistency-equation-*} is taken as the incremental equivalent plastic strain ($\Delta\overline{\varepsilon}\subscript{n}\superscript{pl}$) 
\begin{equation}
\Delta\overline{\varepsilon}\subscript{n}\superscript{pl} = d\lambda.
\label{incremental-peeq}
\end{equation}

Also, the value of the hardening modulus satisfying the consistency condition is updated as $h\subscript{n+1}$.
%
%
%

\noindent \textbf{Step 6:} Compute the plastic part of the deformation gradient at time $t\subscript{n+1}$ as 

\begin{equation}
\mathbf{F}\superscript{p}\subscript{n+1} =\mathbf{F}\superscript{p}\subscript{n} + \Delta\overline{\varepsilon}\subscript{n}\superscript{pl} \mathbf{N}\superscript{p}\subscript{n} \mathbf{F}\superscript{p}\subscript{n}. 
\label{F-p-n+1}
\end{equation}

\noindent \textbf{Step 7:} Compute the elastic part of the deformation gradient at time $t\subscript{n+1}$

\begin{equation}
\mathbf{F}\superscript{e}\subscript{n+1} = \mathbf{F}\superscript{}\subscript{n+1} \mathbf{F}\superscript{p-1}\subscript{n+1}.
\label{F-e-n+1}
\end{equation}

\noindent \textbf{Step 8:} Decompose $\mathbf{F}\superscript{e}\subscript{n+1}$ from Equation \ref{F-e-n+1} into $\mathbf{R}\superscript{e}\subscript{n+1}$ and $\mathbf{U}\superscript{e}\subscript{n+1}$ as

\begin{equation}
\mathbf{F}\superscript{e}\subscript{n+1} = \mathbf{R}\superscript{e}\subscript{n+1} \mathbf{U}\superscript{e}\subscript{n+1}.
\label{Fe=ReUe}
\end{equation}

\noindent \textbf{Step 9:} Compute the Hencky strain  $\mathbf{E}\superscript{e}\subscript{n+1}$ as

\begin{equation}
\mathbf{E}\superscript{e}\subscript{n+1} = ln (\mathbf{U}\superscript{e}\subscript{n+1}).
\label{E-e-n+1}
\end{equation}

\noindent \textbf{Step 10:} Compute the Mandel stress at time $t\subscript{n+1}$ as

\begin{equation}
\mathbf{M}\superscript{e}\subscript{n+1} = 2\,G\,\mathbf{E}\superscript{e}\subscript{n+1} + K \left[ tr(\mathbf{E}\superscript{e}\subscript{n+1}) \right] \mathbf{I}.
\label{M-e-n+1}
\end{equation}

\noindent \textbf{Step 10:}

Now Cauchy stress $\mathbf{T}\subscript{n+1}\superscript{}$ can be calculated
by transforming the Mandel stress into the current configuration as

\begin{equation}
\mathbf{T}\subscript{n+1}\superscript{} = \frac{1}{J\subscript{n+1}} \mathbf{R}\subscript{n+1}\superscript{e}  \mathbf{M}\subscript{n+1}\superscript{e}  \mathbf{R}\subscript{n+1}\superscript{e\top}.
\end{equation}

\noindent \textbf{Step 11:}

Updated yield strength 
\begin{equation}
Y\subscript{n+1} = Y\subscript{n} + h\subscript{n+1} \Delta\overline{\varepsilon}\subscript{n}\superscript{pl}.
\end{equation}

Update the total equivalent plastic strain \begin{equation}
\overline{\varepsilon}\subscript{n+1}\superscript{pl} = \overline{\varepsilon}\subscript{n}\superscript{pl} + \Delta\overline{\varepsilon}\subscript{n}\superscript{pl}.
\end{equation}

\subsection{Hypoelasticity: Additive Decomposition of Spatial Strain Rate}
\label{sec:hypoelasticity}
In a hypoelastic constitutive law, the time rate change in stress is expressed as some function of time derivatives of strains, and the integration procedure utilized should satisfy the criterion of ``incremental objectivity'', see \cite{hughes1980finite,rashid1993incremental}. The stresses are obtained by 
integration of these rate-type equations, see \cite{pinsky1983numerical,flanagan1987accurate,weber1990objective,nemat1992new,li1993explicit,
wang1994analysis}. Using Equation ~\ref{total-stretching}, we can write 
\begin{equation}
\mathbf{D} = \frac{1}{2} \big[ \mathbf{L} +  \mathbf{L}\superscript{\top}\big] = Sym (\mathbf{L}),
\label{expression-for-total-stretching}
\end{equation}
where $Sym$ yields the symmetric part of the operand. Substituting the expression for 
$\mathbf{L}$ from Equation ~\ref{velocity-gradient-decomposition} in Equation 
~\ref{expression-for-total-stretching}, we obtain 
\begin{equation}
\mathbf{D} =   Sym \big[  \mathbf{L}\superscript{e} + \mathbf{F}\superscript{e}(   \mathbf{D}\superscript{p}+ \mathbf{W}\superscript{p} ) 
\mathbf{F}\superscript{e-1} \big].
\label{expression-for-total-stretching-expansion-1}
\end{equation}
By setting plastic spin $\mathbf{W}\superscript{p}=\mathbf{0}$, substituting $\mathbf{F}\superscript{e}=\mathbf{R}\superscript{e} \mathbf{U}\superscript{e}$,
and utilizing $\mathbf{D}\superscript{e}=Sym\big[\mathbf{L}\superscript{e}]$ we can re-write Equation ~\ref{expression-for-total-stretching-expansion-1} as 
\begin{equation}
\mathbf{D} =  \mathbf{D}\superscript{e} +  Sym \big[ \mathbf{R}\superscript{e}\mathbf{U}\superscript{e}(   \mathbf{D}\superscript{p}) 
\mathbf{U}\superscript{e-1}\mathbf{R}\superscript{e-1} \big].
\label{expression-for-total-stretching-expansion-2}
\end{equation}
Further if elastic stretches are small, i.e. $\mathbf{U}\superscript{e} \approx 1$, then above equation can be reduced to 
\begin{equation}
\mathbf{D} =  \mathbf{D}\superscript{e} +  Sym \big[ \mathbf{R}\superscript{e}( \mathbf{D}\superscript{p})\mathbf{R}\superscript{-e} \big].
\label{expression-for-total-stretching-expansion-3}
\end{equation}
It can be shown that  $\mathbf{R}\superscript{e}( \mathbf{D}\superscript{p})\mathbf{R}\superscript{-e}$ is symmetric and hence 
we can re-write 
\begin{equation}
\mathbf{D} =  \mathbf{D}\superscript{e} +  {\bar{\mathbf{D}}\superscript{p}},
\label{expression-for-total-stretching-expansion-4}
\end{equation}
with 
\begin{equation}
\bar{\mathbf{D}}\superscript{p}=\mathbf{R}\superscript{e}\mathbf{D}\superscript{p}\mathbf{R}\superscript{-e} = \mathbf{R}\superscript{e}\mathbf{D}\superscript{p}\mathbf{R}\superscript{e\top}. 
\label{plastic-stretching-current-and-intermediate}
\end{equation}

Equation ~\ref{expression-for-total-stretching-expansion-4} is commonly known as the additive decomposition of the spatial strain rate. Material models based on above decomposition of 
require evolution rule for $\mathbf{D}\superscript{e}$ and $\mathbf{D}\superscript{p}$  (or $\bar{\mathbf{{D}}}\superscript{p}$). One of the commonly utilized measure of objective rate is known as the Jaumman  rate, and the Jaumman rate of cauchy stress (denoted by $\mathbf{T}\superscript{J}$) is given as

\begin{equation}
\label{eq:jaumman-rate}
\mathbf{T}\superscript{J} = \dot{\mathbf{T}} - \mathbf{W}\mathbf{T}+ \mathbf{T}\mathbf{W},
\end{equation}
 where $\dot{\mathbf{T}}$ is the material derivative of the Cauchy stress with respect to 
undeformed (fixed) basis. $\mathbf{T}\superscript{J}$ is also known co-rotational stress rate as it corresponds to the time rate of change with respect to the observer-frame attached and 
rigidly-rotating with the material point of interest. It should be noted that the hypoelastic relation given by
Equation ~\ref{eq:jaumman-rate} is not known to be thermodynamically consistent, in the sense
that constitutive equation for $\mathbf{T}$ is not derived from a free energy function, i.e., path independence 
aspect is missing from its definition. Issues with rate-type formulations have been well-known
\cite{simo1984remarks,leonov2000conditions}, however, given their simplicity, efficiency, and applicability for problems with small elastic strains they are still practiced \cite{maire2013nominally,fish1999computational,mourad2014incrementally}. 
The constitutive equation for \textit{small} strain elasticity is 
\begin{equation}
\mathbf{T}=\mathds{C}:\mathbf{E\superscript{e}},
\label{sigma-ij}
\end{equation}
where $\mathds{C}$ is the fourth order elasticity tensor
\begin{equation}
\mathds{C} = \left(K- \frac{2}{3}G\right) \mathbf{I} \otimes \mathbf{I} +2 G \mathds{I}.
\end{equation}
Equation \ref{sigma-ij} can also be written in a Jaumann (corotational) \textit{rate form} as
\begin{equation}
{\mathbf{T}\superscript{J}} = \mathds{C}:{\mathbf{D}\superscript{e}}.
\label{sigma-ij-jaumann}
\end{equation}
This finally reduces to
\begin{equation}
\mathbf{T}\superscript{J} = (K-\frac{2}{3}G) (\mathbf{I}:\mathbf{D}\superscript{e})\mathbf{I}+ 2 G \mathbf{D}\superscript{e}.
\label{sigma-ij-jaumann-2}
\end{equation}
It should be noted that the above constitutive law describing the elastic deformation satisfies the requirement of ``frame-indifference,'' or objectivity. Now we need to specify the evolution law for the plastic flow $\bar{{\mathbf{D}}}\superscript{p}$. As shown in the previous section, Equation ~\ref{normality-flow-rule} (the normality flow rule) can be used to write 
\begin{equation}
\mathbf{D}\superscript{p} = \dot{\lambda} \mathbf{N}\superscript{p} = 
\dot{\lambda} \frac{\mathbf{M}\subscript{0}\superscript{e}}{|\mathbf{M}\subscript{0}\superscript{e}|},
\label{flow-rule-hypoelasticity}
\end{equation}
where $\dot{\lambda}$ is the consistency parameter,  and we have used 
Equation ~\ref{normality-flow-rule} for the direction of plastic flow. From Equation 
~\ref{yield-condition-1}, we know that during yielding $|\mathbf{M}\subscript{0}\superscript{e}| =Y(e\superscript{p})$. In hypoelasticity the relation between Cauchy stress and Mandel stress (Equation 
~\ref{cauchy-mandel-relation-1}) can be 
reduced to (by choosing $\mathbf{U}\superscript{e}\approx 1$ and $J \approx 1$) as
\begin{equation}
\mathbf{T} = \mathbf{R}\superscript{e} \mathbf{M}\superscript{e}\mathbf{R}\superscript{e\top}.
\label{cauchy-mandel-relation-2}
\end{equation}
Accordingly, it can be shown that $\mathbf{T}\subscript{0} = \mathbf{R}\superscript{e} \mathbf{M}\superscript{e}\subscript{0}\mathbf{R}$, and substituting Mandel stress in terms of Cauchy stress, and $|\mathbf{M}\subscript{0}\superscript{e}| =Y(e\superscript{p})$ in Equation ~\ref{flow-rule-hypoelasticity}, we can obtain
\begin{equation}
\mathbf{D}\superscript{p} = \dot{\lambda} \left ( \frac{ \mathbf{R}\superscript{-e} \mathbf{T}\subscript{0} \mathbf{R}\superscript{-e\top}}{Y(e\superscript{p})} \right).
\label{flow-rule-hypoelasticity-3}
\end{equation}
Using Equation ~\ref{plastic-stretching-current-and-intermediate},  Equation ~\ref{flow-rule-hypoelasticity-3} can be re-written as
\begin{equation}
\mathbf{\bar{D}}\superscript{p} = \dot{\lambda} \left( \frac{ \mathbf{T}\subscript{0} }{Y(e\superscript{p})} \right).
\label{flow-rule-hypoelasticity-4}
\end{equation}
Thus, we have obtained the flow rule in terms of Cauchy stress. Also, 
$\dot{\lambda} = {\dot{e}\superscript{p}} = |\mathbf{\bar{D}}|$. 
In case of hypoelasticity, it is easy to show that 
$|\mathbf{M}\subscript{0}\superscript{e}| = |\mathbf{T}\subscript{0}|$, thus the 
yield function given in Equation ~\ref{yield-condition-3} can be re-written as
\begin{equation}
f_h = |\mathbf{T}\subscript{0}| - Y(e\superscript{p}),
\label{yield-condition-hypo}
\end{equation}
where $f_h$ indicates specialization of the yield function in case of hypoelasticity. Now during plastic yielding ($f_h=0$ and $\dot{f_h}=0$), i.e.,
\begin{equation}
\dot{f_h} = \dot{\overline{|\mathbf{T}\subscript{0}|}} - \dot{\overline{Y(e\superscript{p})}},
\label{eq:yield-condition-derivative-1-hypoelasticity}
\end{equation}
\begin{equation}
\dot{f_h} =\frac{\mathbf{T}\subscript{0}}{|\mathbf{T}\subscript{0}|}:\dot{\mathbf{T}\subscript{0}} - H(e\superscript{p}) |\mathbf{\bar{D}}\superscript{p}|.
\label{eq:yield-condition-derivative-2-hypoelasticity}
\end{equation}
Now we define the direction of plastic flow in current configuration as 
\begin{equation}
\mathbf{\bar{N}}\superscript{p}
=\frac{\mathbf{T}\subscript{0}}{|\mathbf{T}\subscript{0}|},
\label{plastic-flow-hypoelasticity}
\end{equation}
and accordingly re-write the Equation ~\ref{eq:yield-condition-derivative-2-hypoelasticity} as
\begin{equation}
\dot{f_h} = \mathbf{\bar{N}}\superscript{p}:\dot{\mathbf{T}\subscript{0}} - H(e\superscript{p}) |\mathbf{\bar{D}}\superscript{p}|.
\label{eq:yield-condition-derivative-2-hypoelasticity}
\end{equation}
Since $\mathbf{\bar{N}}\superscript{p}$ is deviatoric and $\dot{\mathbf{T}}$ is symmetric, therefore 
$\mathbf{\bar{N}}\superscript{p}:\dot{\mathbf{T}\subscript{0}}=\mathbf{\bar{N}}\superscript{p}:\dot{\mathbf{T}}$. Thus during 
active yielding, we have $\mathbf{\bar{D}}\superscript{p} \ne \mathbf{0}$, and 
$\dot{f_h}=0$, therefore Equation ~\ref{eq:yield-condition-derivative-2-hypoelasticity} becomes
\begin{equation}
\mathbf{\bar{N}}\superscript{p}:\dot{\mathbf{T}} - H(e\superscript{p}) |\mathbf{\bar{D}}\superscript{p}| = 0.
\label{eq:yield-condition-derivative-3-hypoelasticity}
\end{equation}
From Equation ~\ref{eq:jaumman-rate}, we can substitute $\dot{\mathbf{T}}$ in terms of $\mathbf{T}\superscript{J}$, $\mathbf{T}$ and $\mathbf{W}$, in Equation ~\ref{eq:yield-condition-derivative-3-hypoelasticity}, to finally obtain
\begin{equation}
\mathbf{\bar{N}}\superscript{p}:{\mathbf{T}\superscript{J}} - H(e\superscript{p}) |\mathbf{\bar{D}}\superscript{p}| = 0.
\label{eq:yield-condition-derivative-4-hypoelasticity}
\end{equation}
Equation ~\ref{sigma-ij-jaumann} can be re-written as
\begin{equation}
{\mathbf{T}\superscript{J}} = \mathds{C}:({\mathbf{D}-\mathbf{\bar{D}}\superscript{p}}),
\label{sigma-ij-jaumann-3}
\end{equation}
and substituting $\mathbf{\bar{D}}= \dot{\lambda} \mathbf{\bar{N}}\superscript{p}$, we have
\begin{equation}
{\mathbf{T}\superscript{J}} = \mathds{C}:{\mathbf{D}-
\dot{\lambda} \mathds{C}:\mathbf{\bar{N}}\superscript{p}}.
\label{sigma-ij-jaumann-4}
\end{equation}
Subsituting ${\mathbf{T}\superscript{J}}$ from Equation 
~\ref{sigma-ij-jaumann-4} into Equation ~\ref{eq:yield-condition-derivative-4-hypoelasticity}, and writing $|\mathbf{\bar{D}}\superscript{p}|=\dot{\lambda}$, 
we can solve for $\dot{\lambda}$ as
\begin{equation}
\dot{\lambda} =\frac{ \mathbf{\bar{N}}\superscript{p}: \mathds{C}:\mathbf{D}}{	H(e\superscript{p})+\mathbf{\bar{N}}\superscript{p}:\mathds{C}:\mathbf{\bar{N}}\superscript{p}}.
\label{consistency-parameter-equation-1}
\end{equation}
Finally back substituting $\dot{\lambda}$ from Equation ~\ref{consistency-parameter-equation-1} in Equation ~\ref{sigma-ij-jaumann-4}, we obtain 
\begin{equation}
{\mathbf{T}\superscript{J}} = \mathds{C}:{\mathbf{D}-
\frac{ \mathbf{\bar{N}}\superscript{p}: \mathds{C}:\mathbf{D}}{	H(e\superscript{p})+\mathbf{\bar{N}}\superscript{p}:\mathds{C}:\mathbf{\bar{N}}\superscript{p}}		 \mathds{C}:\mathbf{\bar{N}}\superscript{p}},
\label{sigma-ij-jaumann-5}
\end{equation}
and upon rearranging we can rewrite above equation as
\begin{equation}
{\mathbf{T}\superscript{J}} = 
\left ( \mathds{C} - \frac{
\mathds{C}:\mathbf{\bar{N}}\superscript{p} \otimes  \mathbf{\bar{N}}\superscript{p}: \mathds{C}}{H(e\superscript{p})+\mathbf{\bar{N}}\superscript{p}:\mathds{C}:\mathbf{\bar{N}}\superscript{p}}
\right):\mathbf{D}.
\label{sigma-ij-jaumann-6}
\end{equation}
Accordingly, the elasto-plastic tangent modulus is given as
\begin{equation}
\mathds{C}\superscript{ep} =\left ( \mathds{C} - \frac{
\mathds{C}:\mathbf{\bar{N}}\superscript{p} \otimes  \mathbf{\bar{N}}\superscript{p}: \mathds{C}}{H(e\superscript{p})+\mathbf{\bar{N}}\superscript{p}:\mathds{C}:\mathbf{\bar{N}}\superscript{p}}
\right).
\label{elasto-plastic-tangent}
\end{equation}
Equation ~\ref{elasto-plastic-tangent} provides Jaumann stress rate in terms of spatial stretching,
and now we need to integrate these equations with respect to time to obtain stresses and other deformation fields. 

\subsubsection{Time Integration Rate-Independent Hypoelasticity}
We again follow a deformation driven problem structure
and the computation of strain increment for the co-rotational updates is computed using mid-point rule as proposed by Hughes and Winget \cite{hughes1980finite} \footnote{We assume that we have the solution to the problem up to time $t_n$ (when $\mathbf{x}_n$ is the spatial location), and at time 
$t_{n+1}$ the prescribed deformation takes a material element to the spatial point $\mathbf{x}_{n+1}$. This is equivalent way of saying that solution up to $\mathbf{F}_{n}$ is known, and 
we need to find the solution corresponding to $\mathbf{F}_{n+1}$. We denote 
 the incremental displacement as $\mathbf{u}=\mathbf{x}_{n+1}-\mathbf{x}_{n}$, and 
a mid-point configuration $\mathbf{x}_{n+\frac{1}{2}}=\frac{1}{2}(\mathbf{x}_{n+1}+\mathbf{x}_{n+1})$.
Now the increment in the strain, from $t_n$ to $t_{n+1}$, are computed with 
respect to this mid-point configuration as:\\

$\triangle \mathbf{E}\subscript{n+1}= \frac{1}{2} \left(  \frac{ \mathbf{\partial u}}{ \mathbf{\partial {x}_{n+\frac{1}{2}}}} + (\frac{ \mathbf{\partial u}}{ \mathbf{\partial {x}_{n+\frac{1}{2}}}} )\superscript{T}\right )$. This maintains second order accuracy as long as time steps are \textit{small}.  
}\\


\noindent \textbf{Initialization: } For  $n = 0$,\\

Initialize $\mathbf{N}\superscript{p}_{n+1} = \mathbf{0}$, $\Delta\overline{\varepsilon}\superscript{pl}_{n+1} = \mathbf{0}$,
$\mathbf{T}_{n+1}=\mathbf{0}$, $\mathbf{T}\superscript{J}_{n+1}=\mathbf{0}$.\\

\noindent \textbf{At any time ${n+1} > 0$} \\

We assume that quantities at time the beginning of the time increment $t_{n+1}$ (i.e. at the end of $t_n$) are known: Cauchy stress $(\mathbf{T}\subscript{n})$, co-rotational stress $(\mathbf{T}\superscript{J}\subscript{n})$ , yield strength $Y\subscript{n}$,
$\overline{\varepsilon}\subscript{n}\superscript{pl}$, $h\subscript{n}$. Now given 
$\mathbf{F}\superscript{}\subscript{n+1}$, $\triangle t_{n+1}$ ($= t_{n+1}-t_{n}$), the goal is to find the updated quantities as 
$\mathbf{T}\subscript{n+1}$, co-rotational stress $\mathbf{T}\superscript{J}\subscript{n+1}$, yield strength $Y\subscript{n+1}$, $\overline{\varepsilon}\subscript{n+1}\superscript{pl}$, $h\subscript{n+1}$. A radial-return algorithm is used to satisfy the yield condition. In what follows, the trial quantities are denoted $**$. For a given strain increment 
$\triangle \mathbf{E}\subscript{n+1}$, the goal is to find the additive split of 
the strain increment into elastic and plastic parts as

\begin{equation}
\triangle \mathbf{E}\superscript{}\subscript{n+1} = \triangle \mathbf{E}\superscript{e}\subscript{n+1}+
\triangle \mathbf{E}\superscript{p}\subscript{n+1}
\label{eqn:additvesplit}
\end{equation}

\noindent \textbf{Step 1:} Compute the trial corrotational stress using the strain increment as 

\begin{equation}
\mathbf{T}\superscript{J**}\subscript{n+1}=\mathbf{T}\superscript{J}\subscript{n}	
+	(K-\frac{2}{3}G) tr(\triangle \mathbf{E}\superscript{}\subscript{n+1}) + 2 G \triangle (\mathbf{E}\superscript{}\subscript{n+1})_0.
\label{eqn:trial-elastic-stress}
\end{equation}

\noindent \textbf{Step 2:} Compute the trial equivalent von-mises stress

\begin{equation}
(\sigma\subscript{v})\subscript{n+1}\superscript{**} = \sqrt{\frac{3}{2} (\mathbf{T}\superscript{J**}\subscript{n+1})\superscript{}\subscript{0} : (\mathbf{T}\superscript{J**}\subscript{n+1})\superscript{}\subscript{0}}. 
\label{von-mises-n-hypoelasticity*}
\end{equation}

\noindent \textbf{Step 3:} If $(\sigma\subscript{v})\subscript{n+1}\superscript{**}$ computed in Equation ~\ref{von-mises-n-hypoelasticity*} is less than $Y_n$ then it is an elastic increment,
set $\triangle \mathbf{E}\superscript{p}\subscript{n+1}=0$, and go to Step 4.5, else continue with following.

\noindent \textbf{Step 4:} Radial return algorithm is employed to compute the incremental plastic strain. Following steps describe how this is achieved.

\textbf{Step 4.1:} The direction of plastic flow is computed as

\begin{equation}
\mathbf{N}\superscript{p}\subscript{n+1} =  \frac{ (\mathbf{T}\superscript{J**}\subscript{n+1})\superscript{}\subscript{0}}{ \sqrt{\frac{2}{3}  (\mathbf{T}\superscript{J**}\subscript{n+1})\superscript{}\subscript{0} : (\mathbf{T}\superscript{J**}\subscript{n+1})\superscript{}\subscript{0}} }.
\label{N-p-n-corot}
\end{equation}

\textbf{Step 4.2:} The magnitude of the incremental equivalent plastic strain ($\Delta \bar{\varepsilon}\superscript{pl}$) is obtained by consistently linearlizing and satisfying the yield condition, which yields 
 
\begin{equation}
(\Delta \bar{\varepsilon}\superscript{pl})\subscript{n+1} = \frac{(\sigma\subscript{v})\subscript{n+1}\superscript{**} - Y\subscript{n}}{3\mu + h\subscript{n}}. 
\label{delta-eqpl-n+1}
\end{equation}

\textbf{Step 4.3:} The incremental plastic strain is obtained as 

\begin{equation}
\triangle \mathbf{E}\superscript{p}\subscript{n+1} = \Delta \bar{\varepsilon}\superscript{pl}\subscript{n+1}\mathbf{N}\superscript{p}\subscript{n+1}.
\label{delatep-corot}
\end{equation}

\textbf{Step 4.4:} By using Equation \ref{delta-eqpl-n+1}, the yield strength at $t_{n+1}$ is updated as 

\begin{equation}
Y\subscript{n+1} =Y\subscript{n} + h\subscript{n} \times (\Delta \bar{\varepsilon}\superscript{pl})\subscript{n+1}
\label{yield-strength-n+1}
\end{equation}

The hardening modulus is also updated in accordance with the experimental data. 

\textbf{Step 4.5:} Compute the elastic strain by subtracting the plastic strain increment given by Equation ~\ref{delatep-corot} from the total strain increment as
\begin{equation}
\triangle \mathbf{E}\superscript{e}\subscript{n+1}=\triangle \mathbf{E}\superscript{}\subscript{n+1}- \triangle \mathbf{E}\superscript{p}\subscript{n+1}.
\label{eqn:incremental-el-strain}
\end{equation}
\textbf{Step 5:} Compute the updated corrotational stress using the incremental elastic strains from Equation ~\ref{eqn:incremental-el-strain}. 
\begin{equation}
\mathbf{T}\superscript{J}\subscript{n+1}=\mathbf{T}\superscript{J}\subscript{n}	
+	\left(K-\frac{2}{3}G\right) tr(\triangle \mathbf{E}\superscript{e}\subscript{n+1}) + 2 G (\triangle \mathbf{E}\superscript{e}\subscript{n+1})_0
\label{eqn:updated-elastic-stress}
\end{equation}
\textbf{Step 6:} Obtain the Cauchy stress by transforming the corrotational stress into the current 
frame as
\begin{equation}
\mathbf{T}\superscript{}\subscript{n+1}= {\mathbf{R}\subscript{n+1}}\mathbf{T}\superscript{}\subscript{n+1}	\mathbf{R}\superscript{T}\subscript{n+1},
\label{eqn:cauchy-stress}
\end{equation}
where  $ {\mathbf{R}\subscript{n+1}}$ is obtained from polar decomposition of 
$\mathbf{F}\subscript{n+1}$ as
\begin{equation}
\mathbf{F}\subscript{n+1} = \mathbf{R}\subscript{n+1} \mathbf{U}\subscript{n+1}.
\label{Fe=ReUe*}
\end{equation}

\subsection{Solution to Global Equilibrium Equations Using Abaqus Explicit Dynamics}

The Explicit dynamicS solver used by Abaqus is based on the central difference formulas for the velocity and the acceleration, and the use of diagonal or lumped element mass matrices is adopted. The quantities such as displacement, velocities and accelerations are specified at nodal points. Let 
$\mathbf{\mathscr{F}}\subscript{n}$ and $\mathbf{\mathscr{I}}\subscript{n}$ denote
the external and internal (due to resultant of internal stresses) nodal point force vectors \footnote{
Nodal values are interpolated at the Gauss integration points of the element via shape functions of the used element. 
Obtained interpolated quantities at integration points are passed either to VUMAT or Abaqus material library to obtain 
stresses at the integration points which are further integrated for the whole element. This integrated stresses for the 
element is the internal force $\mathbf{\mathscr{I}}$ which is assembled for all the elements.}.  Then 
nodal point acceleration vector ($\ddot{\mathbf{\mathscr{U}}}\subscript{n}$) is given as
\begin{equation}
\ddot{\mathbf{\mathscr{U}}}\subscript{n} = \mathbf{\mathscr{M}}\superscript{-1} (\mathbf{\mathscr{F}}\subscript{n} - \mathbf{\mathscr{I}}\subscript{n}),
\label{u-double-dot-equation}
\end{equation}
where $\mathbf{\mathscr{M}}$ is the diagonal lumped mass matrix. $\mathbf{\mathscr{I}}$ is obtained from the VUMAT implementation of the material model. 
The nodal velocity vector, at mid-point time, is then computed as
\begin{equation}
\dot{\mathbf{\mathscr{U}}}\subscript{n+\frac{1}{2}} = \dot{\mathbf{\mathscr{U}}}\subscript{n-\frac{1}{2}} + \frac{\triangle t\subscript{n+\frac{1}{2}} + \triangle t\subscript{n}}{2} \ddot{\mathbf{\mathscr{U}}}\subscript{n},
\label{u-i+half-equation}
\end{equation}
and the updated nodal point displacement vector is then computed as
\begin{equation}
\mathbf{\mathscr{U}}\subscript{n+1} = \mathbf{\mathscr{U}}\subscript{n} + \triangle t\subscript{n+1} \dot{\mathbf{\mathscr{U}}}\subscript{n+\frac{1}{2}}.
\label{u-n+1-equation}
\end{equation}

where $\triangle t$ is defined as

\begin{equation}
\triangle t\subscript{n+1} = t\subscript{n+\frac{1}{2}} - t\subscript{n-\frac{1}{2}}.
\label{delta-t-equation}
\end{equation}


As stated earlier, the explicit procedure does not require any tangent stiffness, and for quasi-static problems (where acceleration terms are negligible) the $\mathscr{M}$ matrix can be scaled. However, caution must be 
taken that mass scaling does not lead to excessively high and unrealistic kinetic energies.

\subsection{Contact Interaction and Friction}


We employed the default kinematic constraint algorithm for the contact interaction. In particular, the tangential 
behavior was modeled through a penalty formulation (with a very large penalty factor of $10^{10}$),
and in absence of slip the relative tangential displacement is computed by dividing the frictional force by the chosen 
penalty factor. Such an approach is helpful when friction can cause excessive local distortion. 
The contact algorithm in a pure 
master-slave surface formulation, detects contact when the slave node penetrates into the master segment, therefore a 
sufficiently fine mesh for both bodies are chosen so that no spurious interpenetrations of the rigid-roller into the 
ductile film substrate occur. Additionally, large roller radii, slow angular speeds, and sufficiently fine mesh enable robust performance of the contact algorithm. Although not necessarily required, we chose to work with a finite sliding formulation, i.e., detect a change in master segment for every slave node (through a global search) after a set number of time increments. For the sake of completeness, a summary of the contact algorithm for rigid (master) and soft surface (slave) is given in Algorithm ~\ref{contact-pseudo-code}.
\begin{algorithm}
\caption{Pseudo code for Abaqus/Explicit Contact for rigid (master) segment and soft (slave) node interaction. 
Assumed that master surface follows a displacement specified motion.}
\label{contact-pseudo-code}
\begin{algorithmic}[1]
\Procedure{Contact}{}
\State Time $\rightarrow$ $t_{n}$ \Comment All quantities known, need to find explicit updates from $t_n$ to $t_{n+1}$
\State $\bm{S}^{t_{n+1}}$ \Comment Perform trial kinematic update of the slave node
\State $\bm{M}^{t_{n+1}}$ \Comment Perform the kinematic update of the current master segment
\If {(no penetration)} \Comment Check penetration of slave node into the current master segment
\State  return false; \Comment No contact and proceed to the next slave node
\Else \Comment There is penetration and corrector phase starts
\State $\bm{N}^{t_n}$ \Comment Calculate normal to the current master segment
\State $\bm{\tau}^{{t_n}}$ \Comment Calculate tangent to the current master segment
\State $\bm{RP}^{t_n}$ \Comment Calculate nearest point on the master segment w.r.t. slave node at $t_n$
\State $\bm{RP}^{ t_{n+1} }$ \Comment Calculate position of $\bm{RP}^{t_n}$ at time $t_{n+1}$ on the master segment
\State $_{+}\bm{S}^{t_{n+1}}$ \Comment Project slave node from position $\bm{S}^{t_{n+1}}$ along $\bm{N}^{t_n}$
\State $\bm{F}_N^{t_n}$ \Comment Solve equation of motions to find normal force needed to prevent interpenetration
\If {(no friction)} \Comment Check if friction is specified for the problem
\State  return false; \Comment No tangential force and proceed for the next slave node
\Else \Comment Friction is specified
\State $\bm{\delta U}$ \Comment Compute relative displacement in tangential direction
\If {($| \bm{\delta U} | > 0$)}  \Comment Slip will occur
\State $\bm{F}_{fr}^{t_n}$ \Comment Compute tangential force $\bm{F}_{fr}^{t_n}$ assuming no slip
\If {($\bm{F}_{fr}^{t_n} > \mu \bm{F}_N^{t_n}$)}  \Comment Slip will occur
\State $_{++}\bm{S}^{{t_{n+1}}} \leftarrow _{+}\bm{S}^{{t_{n+1}}}$ \Comment Update the projected location of the slave node
\State $\bm{F}_{fr}^{t_n} \leftarrow \mu \bm{F}_N^{t_n}$ \Comment Maximum value of $\bm{F}_{fr}^{t_n}$ can be $\mu \bm{F}_N^{t_n}$
\Else
\State return false; \Comment No slip (if penalty formulation compute small tangential motion)
\EndIf
\Else \Comment No relative displacement
\State return false; \Comment No tangential force and proceed for the next slave node
\EndIf
\EndIf
\EndIf
\State ${t_n}$ $\leftarrow$ ${t_{n+1}}$ \Comment Update time
\State If specified, update master segment for the slave node \Comment At the end of time increment
\EndProcedure
\end{algorithmic}
\end{algorithm}
\section{Experimental and Simulation Results}
\label{sec:performance-results}

\begin{figure}[tbp]
\centering
\includegraphics[scale=.5]{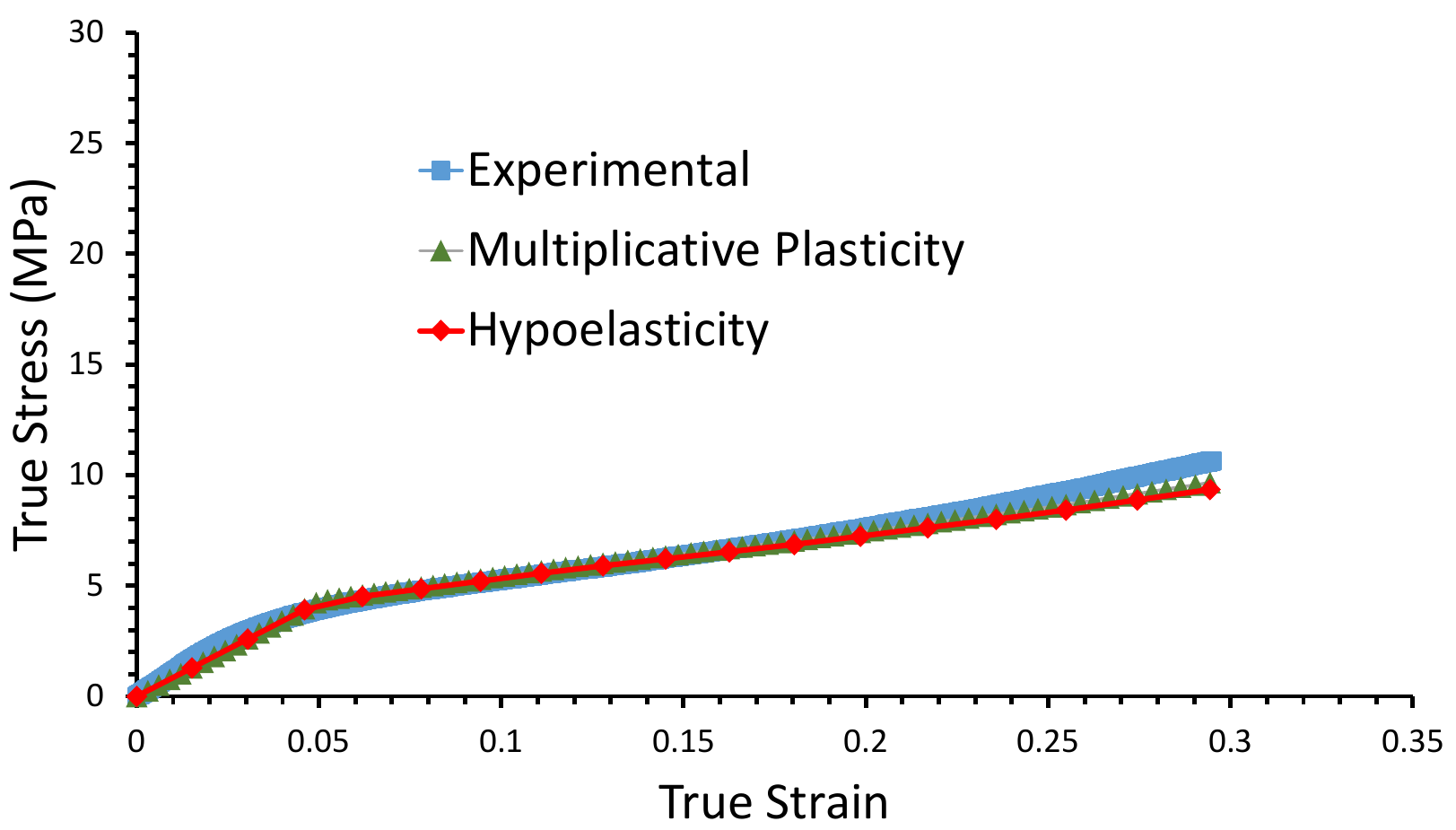}
\caption{Comparison of true stress-true strain curves based on experimental data, hypoelastic formulation, and 
$\mathbf{F^e}\mathbf{F^p}$ based decomposition. For both hypoelastic and $\mathbf{F^e}\mathbf{F^p}$ based computation, a rate-independent and isotropic hardening deformation model was employed. }
\label{fig:tensile-experiment-abaqus-comparison}
\end{figure}

\noindent \textbf{Material calibration:} Uniaxial tensile tests were carried out on polymer films at a strain rate of 
$0.0025$ sec$^{-1}$. The yield strength and elastic modulus of the films were found to be 
$\sigma_{y,film}=4$ MPa and $E_{film}=78$ MPa, respectively. Accordingly, the elastic limit,
$\epsilon_e = \sigma_{y,film}/E_{film}$, was estimated as
 5\% strain, which is a major contrast with respect to rigid-plastic model.   
The elastic modulus, yield strength, and hardening data were used to calibrate hypoelastic and multiplicative plasticity models. The performance of the multiplicative plasticity based model and hypoelastic formulation, calibrated against tensile test, is shown in Figure ~\ref{fig:tensile-experiment-abaqus-comparison}. It is clear polymer films have noticeable elastic range, and harden with increasing plastic deformation. As will become clear, such type of material behavior (which is quite distinct from metals) is not accounted in classical rolling theories, and therefore an appropriate finite element model is required for polymer rolling. The calibrated material models were used in rolling simulations. \\\\

\begin{figure}[tbp]
\centering
\includegraphics[scale=.5]{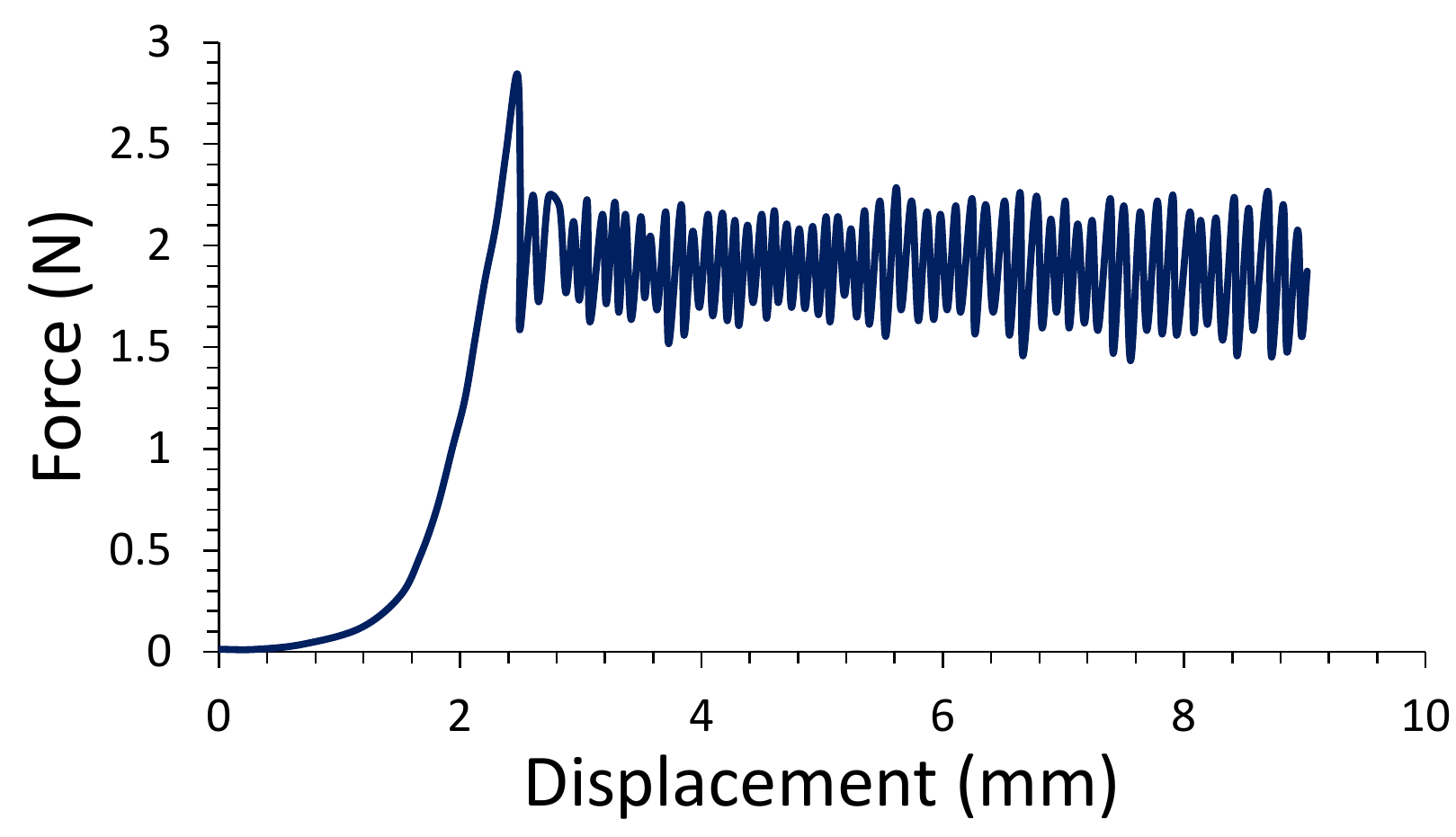}
\caption{\label{fig:Coulombic-friction-cropped} Force versus displacement curve during measurement of coefficient of friction on Instron mechanical tester. 
The coefficient of friction is estimated from the average force (F$_{friction}$), in the steady state sliding,  as 
$\mu=F_{friction}/m'g$, where $m'g$ is
the weight of the block. For $m'=0.5$ kg and F$_{friction}=2$N (based on this graph), $\mu_k$ is estimated to be $0.4$.}
\end{figure}

\begin{figure}[htbp]
\centering
\includegraphics[scale=.5]{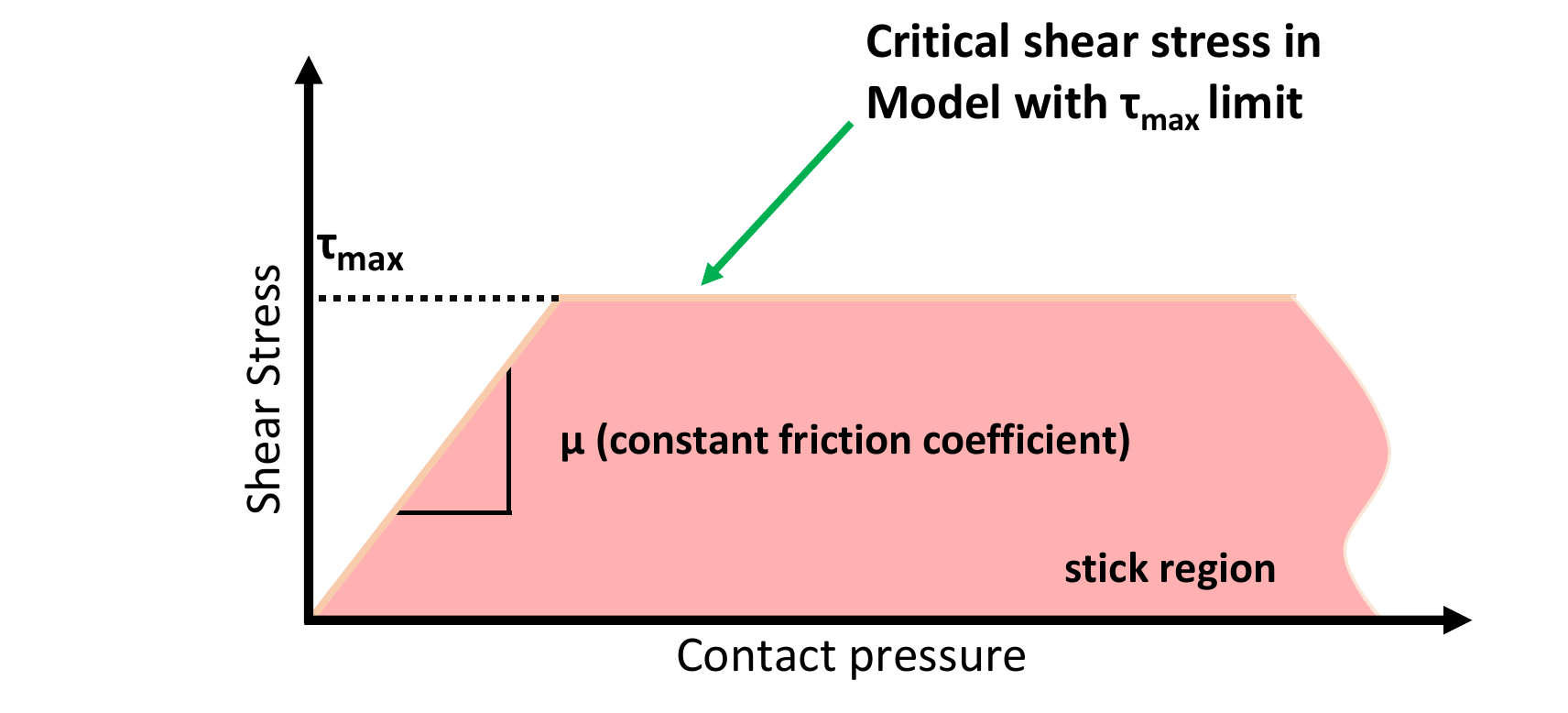}
\caption{Rate-independent Coulombic friction model).}
\label{fig:Coulombic-friction-cropped.pdf}
\end{figure}

\noindent \textbf{Frictional interaction:} We will treat the incoming stack of films as a single sheet under the assumption that there is no relative slippage between them, and will verify this assumption. To model the interaction between the stainless steel rollers and the film stack with Coulombic friction, we need to estimate the coefficient of friction. For this purpose,  the coefficient of friction between a stainless steel block 
(made of same material and surface properties as the rollers), and
polymer films was estimated using a friction-fixture on Instron mechanical tester \cite{instron-friction}.
Stainless steel block of approximately $500$ g was used, and the load vs. displacement during sliding motion of the 
block on the film was recorded, shown in Figure ~\ref{fig:Coulombic-friction-cropped}. The coefficient of 
friction $\mu$ was estimated to be $0.4$. In the finite element simulations we used the 
isotropic Coulomb friction model, with coefficient of friction of $0.4$, and 
limited the maximum value of the shear-traction equal to the shear yield strength of the material. The Coulombic 
friction model is schematically shown in Figure ~\ref{fig:Coulombic-friction-cropped.pdf}, where tangential traction is dependent on the normal traction but limited by the material's yield strength. \\\\

\noindent \textbf{Rolling experiments:} A snapshot of the roll-bonding is shown in 
Figure ~\ref{fig:roll-bonding-of-films}, where multiple layers are fed through the rollers. 
Several specimens comprising of a film-stack were rolled at different levels of compression loads,
leading to different levels of plastic strain. The initial widths of specimens (which were approximately $15$ mm) before rolling, were found to be measurably unchanged after rolling, and therefore rolling passes were 
consistent with the plane-strain scenario. The initial thickness of film-stacks were approximately $0.6$ mm, and the average final thickness of the rolled stock depended on the level of plastic-strain imposed. For a detailed correlation between imposed plastic strain and degree of  bonding see 
\cite{padhye-main-bonding,padhye-thesis-2015}. \\\\

\begin{figure}[]
  \centering
  \begin{turn}{0}
      \begin{minipage}{\textwidth}
   \centering
\includegraphics[scale=.6]{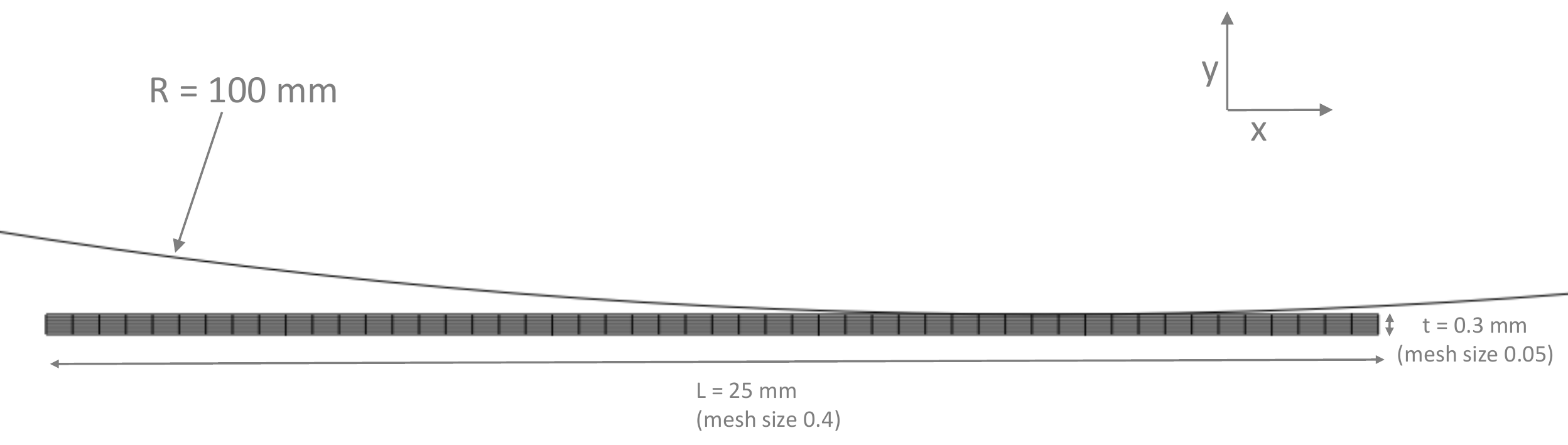}
\caption{The undeformed mesh in elastic-plastic rolling, only one-half of the model is considered while utilizing symmetry about the XZ symmetry plane.}
\label{fig:undeformed-mesh}
    \end{minipage}
  \end{turn}
\end{figure}

\begin{figure}[tbp]
\centering
\includegraphics[scale=.7]{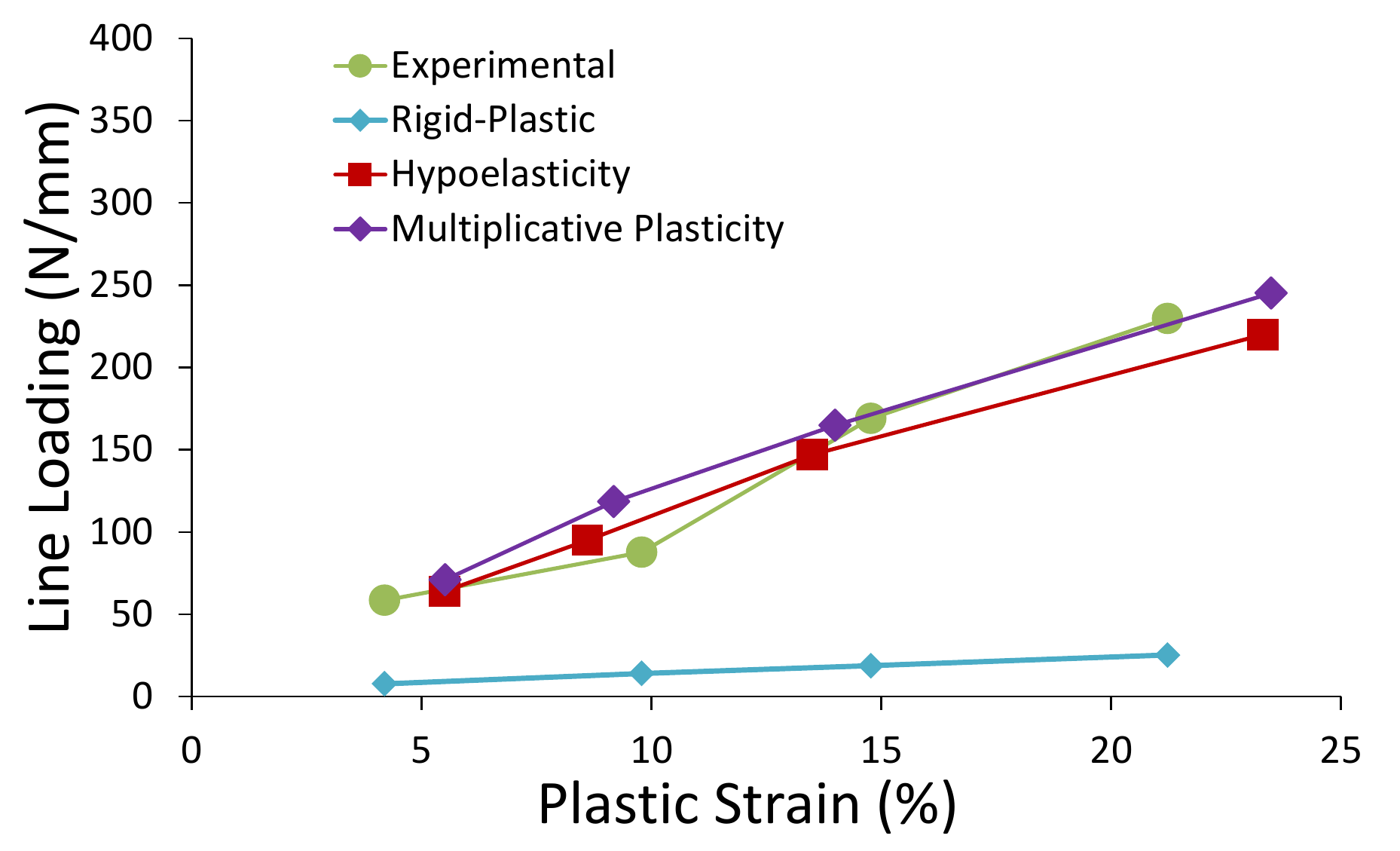}
\caption{Line loading (N/mm) with respect to \% plastic strain curves based on experimental data, and predictions from rigid-perfectly plastic model, and finite strain elastic plastic FEA analysis (both hypoelasticity and multiplicative plasticity).}
\label{fig:line-loading-cropped.pdf}
\end{figure}

\noindent \textbf{Rolling simulations:} Plane strain rolling simulations were carried out in Abaqus 
Explicit to study the deformation of polymer films in the roller-bite. A film-stack of initial thickness $0.6$ 
mm was modeled with four-noded plane strain elements with reduced integration (CPE4R), and rollers 
were treated as rigid surfaces.  We utilized the symmetry in rolling, and therefore constructed the 
simulation model for only one-half of the rolling process, as shown in Figure ~\ref{fig:undeformed-mesh}. Mechanical properties of the film, and coefficient of friction between the stack and rollers were 
used as stated earlier. The incoming films were modeled as a single stack, because 
the incoming stack had a very small thickness compared to the radius of the rollers (due to which 
it was expected that there will be negligible shearing through the thickness of the stack, and no slipping or sliding between 
the film layers).  The turning rollers were brought closer, with films in-between them, and the desired level of plastic strains were imposed.  Each rolling simulation was continued until rolling loads reached a steady-state value. 

\begin{figure}[]
  \centering
  \begin{turn}{0}
      \begin{minipage}{\textwidth}
   \centering
\includegraphics[scale=.8]{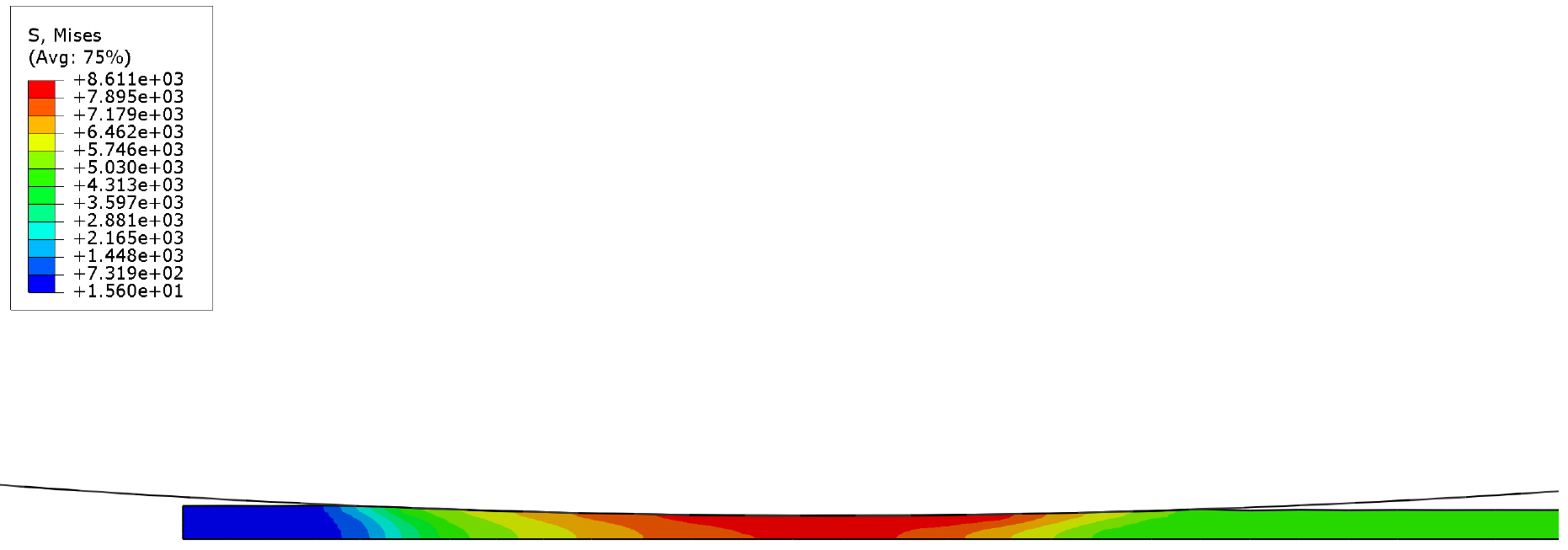}
\caption{Von Mises stresses during elastic-plastic rolling at steady state for approximately $13.5$\% nominal strain.
The peak S, Mises is $8.611$ MPa. }
\label{fig:S-4-30-cropped-2.pdf}
    \end{minipage}
  \end{turn}
\end{figure}

\begin{figure}[]
  \centering
\begin{turn}{0}
      \begin{minipage}{\textwidth}
   \centering
\includegraphics[scale=.8]{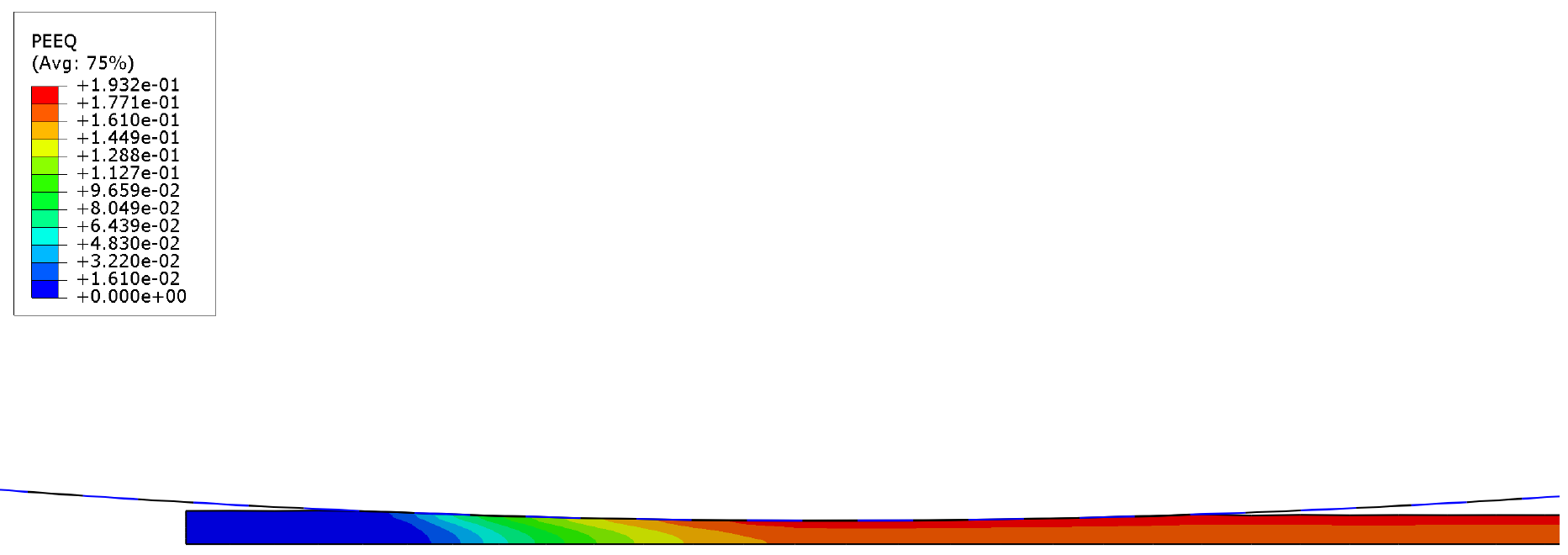}
\caption{Equivalent plastic strain (PEEQ) during elastic-plastic rolling at steady state for approximately $13.5$\% nominal strain.}
\label{fig:PEEQ-rolling-4-30-cropped-2.pdf}
    \end{minipage}
\end{turn}
\end{figure}

\begin{figure}[]
  \centering
\begin{turn}{0}
      \begin{minipage}{\textwidth}
   \centering
\includegraphics[scale=.8]{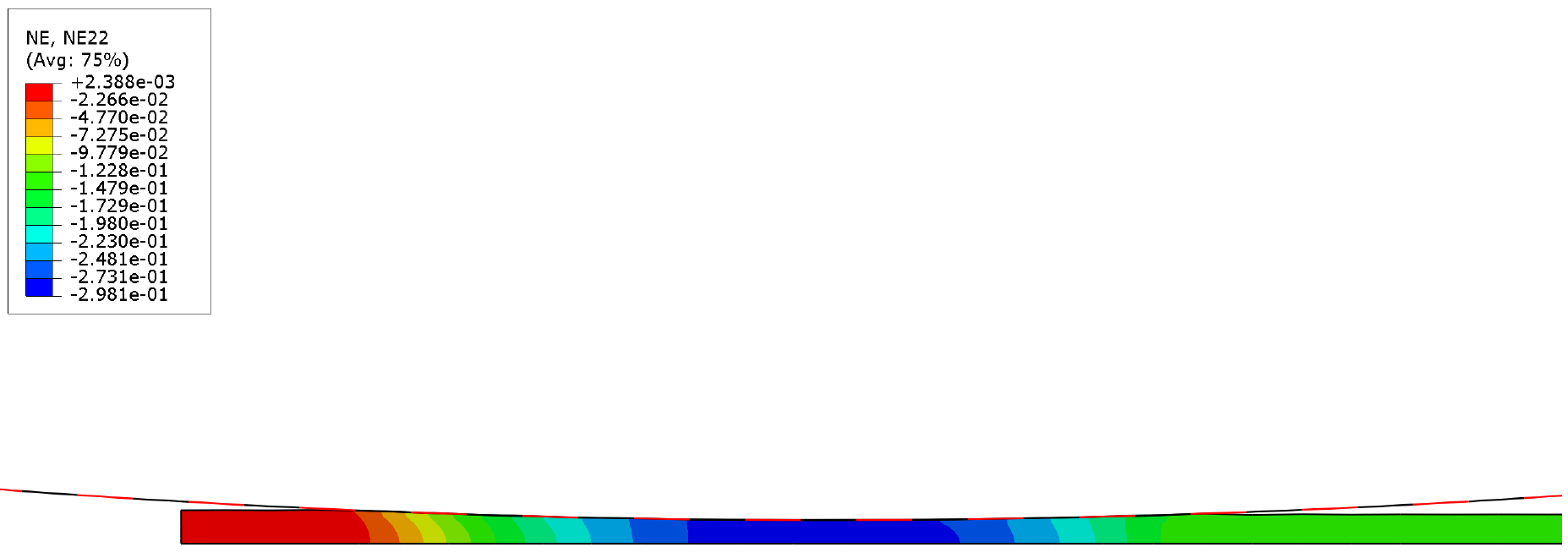}
\caption{Nominal strain (NE22) in the direction of thickness reduction during elastic-plastic rolling at steady state for approximately $13.5$\% nominal strain.}
\label{fig:NE22-rolling-4-30-cropped-2.pdf}
    \end{minipage}
  \end{turn}
\end{figure}

\begin{figure}[]
  \centering
\begin{turn}{0}
      \begin{minipage}{\textwidth}
   \centering
\includegraphics[scale=.8]{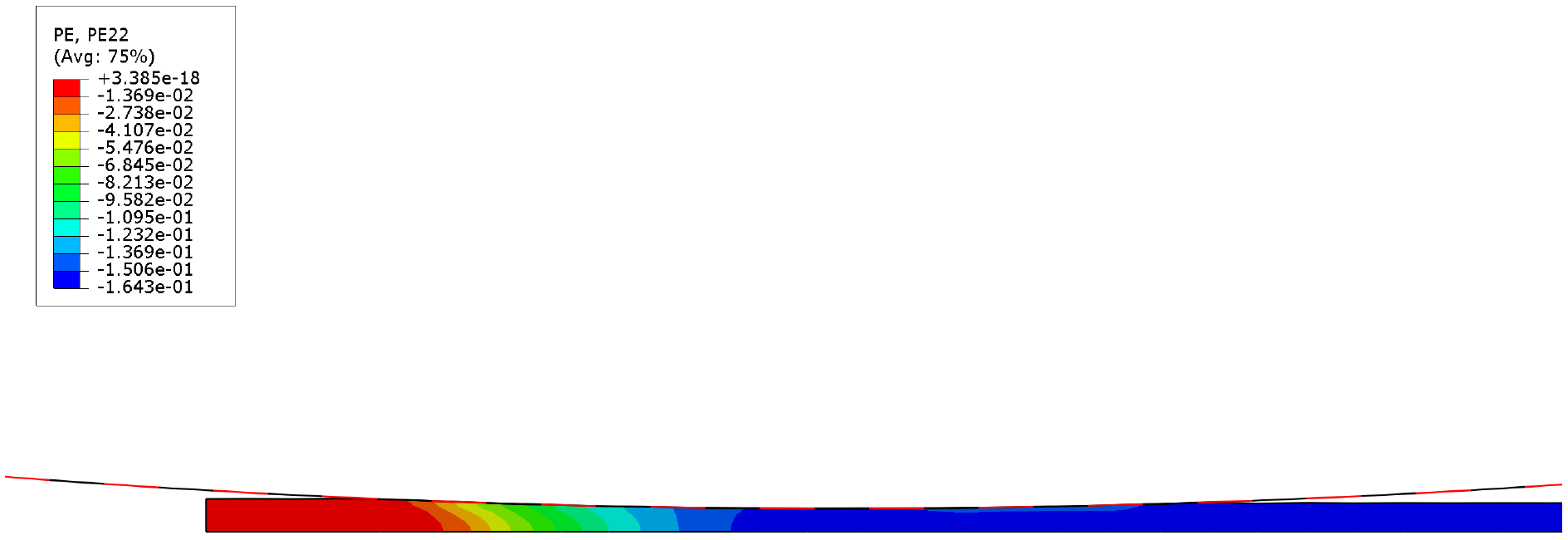}
\caption{Plastic strain (PE22) in direction of thickness reduction during elastic-plastic rolling at steady state for approximately $13.5$\% nominal strain.}
\label{fig:PE-22-rolling-4-30-cropped-2.pdf}
    \end{minipage}
\end{turn}
\end{figure}

Rolling loads based on experiments, rigid-plastic analysis, and elato-plastic finite element analysis (both, hypoelastic and multiplicative decomposition formulation) are show in Figure  ~\ref{fig:line-loading-cropped.pdf}.
The assumption of rigid-perfectly plastic model, assuming a constant yield stress, is found to greatly underestimate the experimental rolling loads for a desired  level of plastic strain,
whereas elasto-plastic finite element simulations exhibit an excellent match.
Figures ~\ref{fig:S-4-30-cropped-2.pdf} and ~\ref{fig:PEEQ-rolling-4-30-cropped-2.pdf}, 
show the plots of von Mises stresses and equivalent plastic strain, for a scenario
in which approximately 13.5\% nominal thickness reduction was achieved.  
Clearly, the plots indicate through-thickness and almost homogeneous plastic deformation. The corresponding plots of nominal strain and plastic strain in thickness reduction direction are also shown in Figures ~\ref{fig:NE22-rolling-4-30-cropped-2.pdf} and ~\ref{fig:PE-22-rolling-4-30-cropped-2.pdf}, respectively. 

\begin{figure}[]
  \centering
      \begin{minipage}{\textwidth}
   \centering
\includegraphics[scale=.9,angle=0]{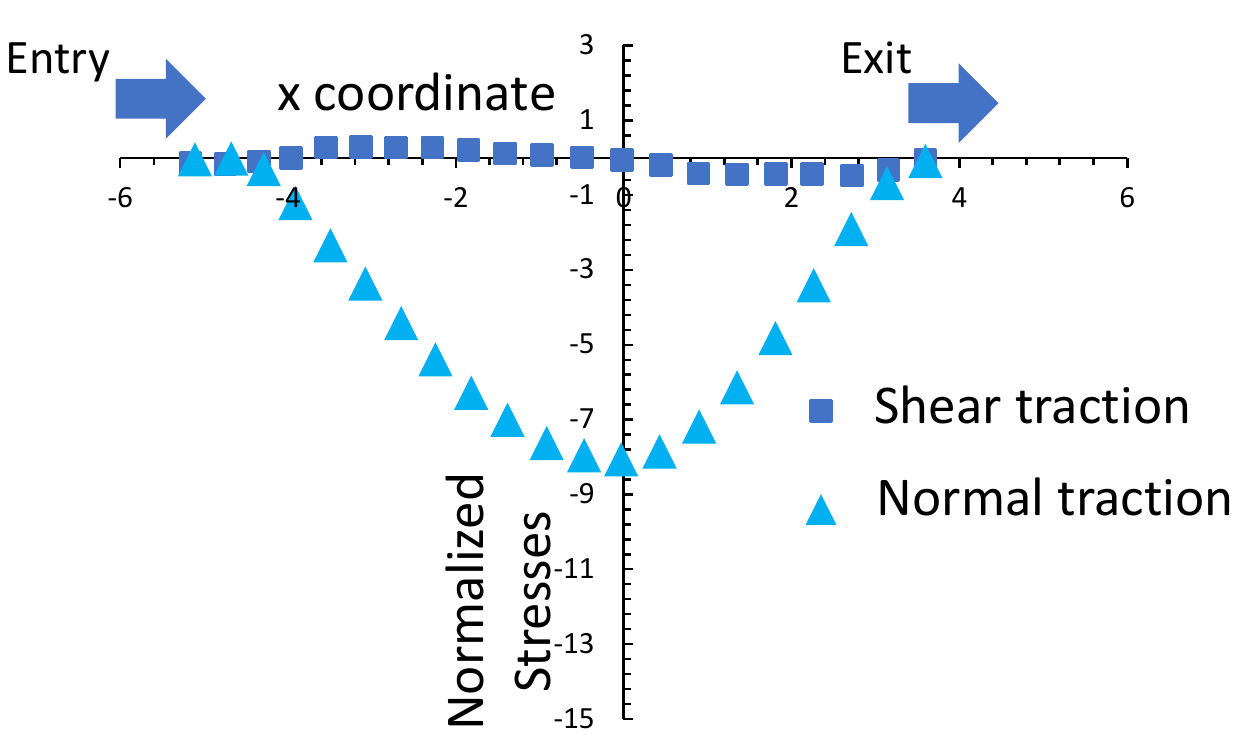}
\caption{Variation of normalized shear and normal stresses with respect to rolling direction, at steady state for approximately $13.5$\% nominal strain,
between the roller and the film surface. The stresses have been normalized with respect to the yield strength of the films ($\sigma_{y,film}$). The shear stresses are
maximum between the roller and the film interface, and zero on the symmetry plane in the thickness direction. The normal stresses were found to be almost uniform 
across the stock thickness.}
\label{fig:abaqus-analysis-cropped.pdf}
    \end{minipage}
\end{figure}

Since we treated the incoming stack of films as a single strip, to verify our no-slip assumption between the film layers, we plotted the variation of normal stresses and 
shear stresses along the rolling direction, at the roller-film interface, where the shear stresses attain 
maximum values (and shear stresses go to zero as the symmetry plane is approached). 
The variation of shear stresses and normal tractions between roller and film interface is shown in Figure 
~\ref{fig:abaqus-analysis-cropped.pdf}. The shear stresses are found to be significantly lower 
than the normal stresses, and much lower than the shear yield strength of the material. This quantitative data validates the assumption of treating incoming stack of film layers as a single strip, and that an incoming material element deforms homogeneously under normal compression without shearing with respect to the axes of the reference configuration. Furthermore, during rolling and active plastic deformation, as bonding occurs in the rolling bite, the resistance against slipping or sliding will only increase. 

\begin{figure*}[t!]
    \centering
     \begin{subfigure}[b]{0.4\textwidth}
        \centering
        \includegraphics[scale=0.5]{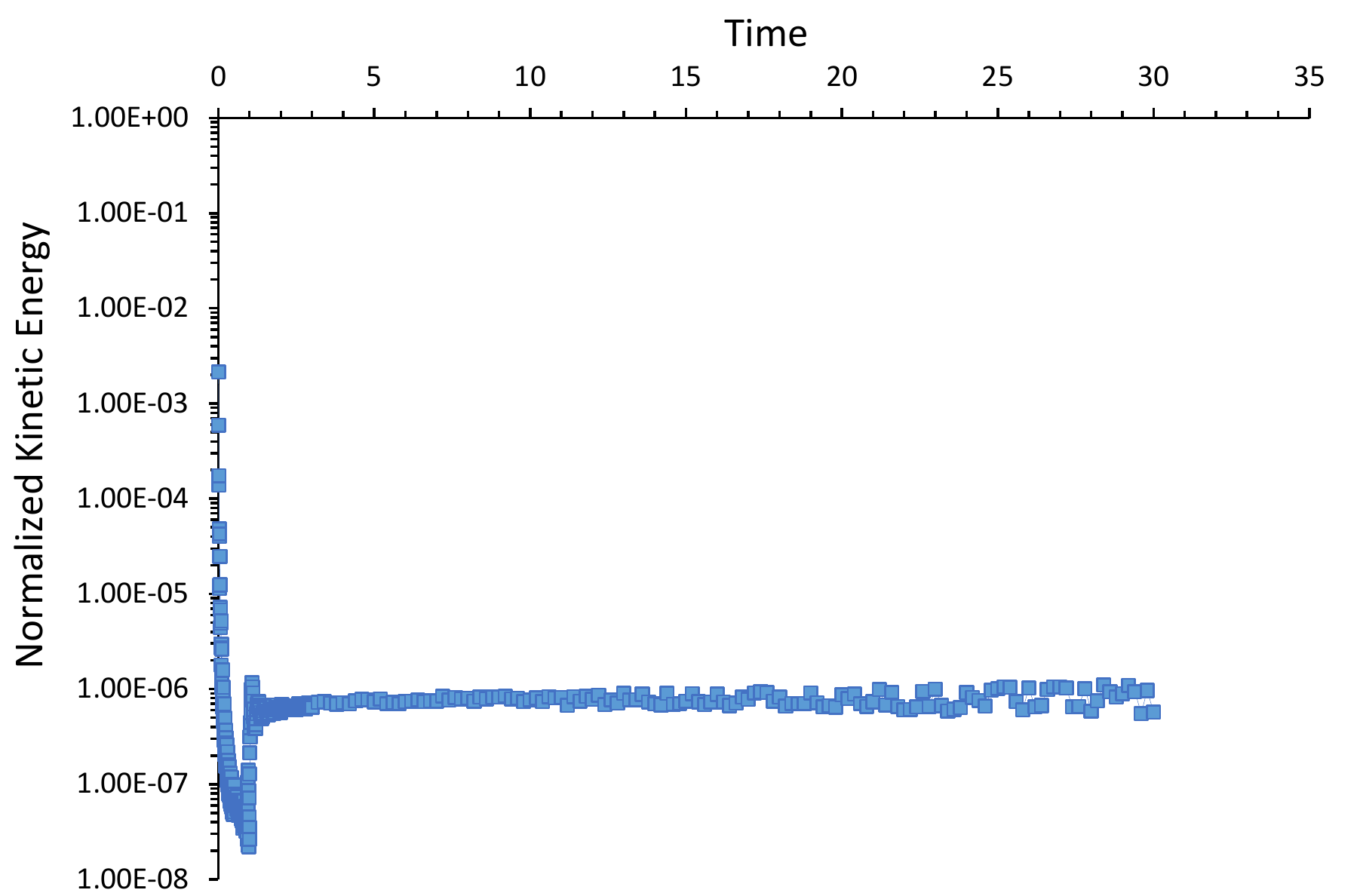}
        \caption{Normalized kinetic energy of finite element model. }
    \end{subfigure}\\%
     \begin{subfigure}[b]{0.4\textwidth}
        \centering
        \includegraphics[scale=0.5]{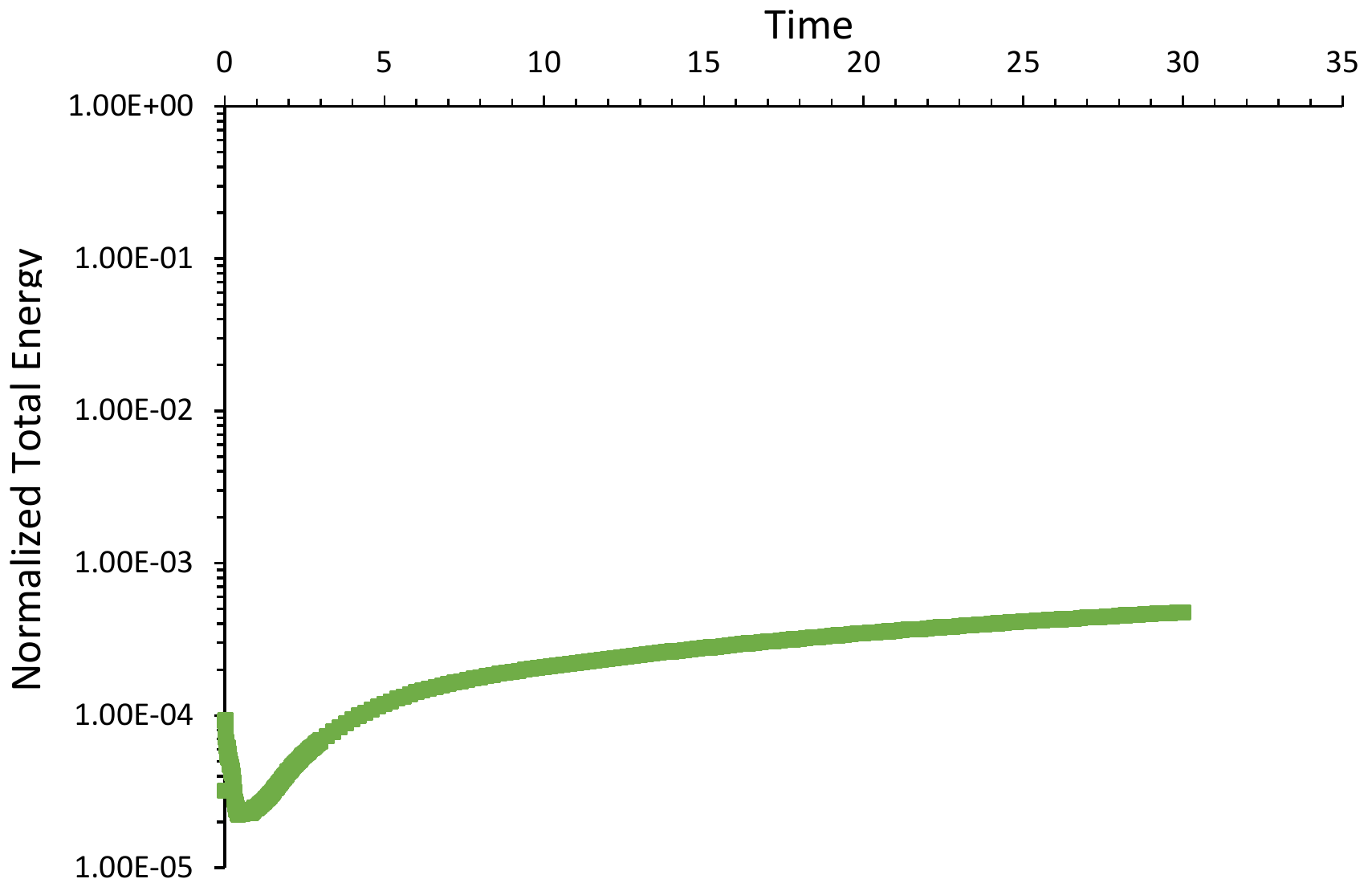}
        \caption{Normalized total energy of finite element model.}
    \end{subfigure}
    \caption{\label{fig-normalized-energy} Normalization of energies has been done with respect to the strain energy of the model at any given time. Both kinetic energy and total energy were initially zero and after \textit{long} time when steady-state has been reached they are still orders of magnitude smaller compared
to the dominant strain energy in the model. }
\end{figure*}

To compare the overall accuracy of our explicit dynamics simulations for approximating the quasi-static 
rolling process, we also compared the plots of total energy of the system, and kinetic energy for the 
whole model, as function of the simulation time. Total energy should ideally remain constant. Initially the 
kinetic energy of the model, and total energy, are both zero, and their magnitudes 
after \textit{long} times, when steady states have been reached, should also be close to zero. 
The plots of the total energy and kinetic energy, normalized with respect to strain energy, are shown in Figure ~\ref{fig-normalized-energy}, and are small. These are not exactly zero owing to intrinsic numerical errors during the explicit time integration.
The overall excellent energy behavior suggests that explicit integration scheme based on 
central difference (using automatic time increments), and kinematic constraint algorithm are working 
accurately over long times. Loss of energy during contact/impact in dynamic problems can be significant 
\cite{cirak2005decomposition,ryckman2012explicit}, but for our rolling cases the material velocities are 
negligible and therefore concerns related to dissipation during contact, or explicit integration over long 
times do raise any concern.

The classical model of plastic rolling of a strip is valid when the incoming strip is passed through rollers to produce appreciable reduction in thickness, and when elastic deformations (if any) are negligible compared to  plastic strains. This assumption of totally ignoring elastic strains, is often well-suited for metals, since elastic strains typically amount to only 0.5\% or so.
However, solid state polymers can exhibit elastic strains up to (or greater than) 5\% and, therefore such an idealization is expected to yield unsatisfactory results for polymer rolling.  In the rigid-plastic model, the maximum nominal strain in thickness is observed at the location where the roller-gap is minimum,
and the compression zone spans the region between the the point where the 
stack  first encounters the roller bite and the minimum roller-gap location. In contrast, if the material has sufficient elasticity then for the same level of nominal plastic thickness reduction, the contact zone is much larger on the entry side, and also extends  beyond the minimum roller-gap location on the exit side. For the rolling case discussed, where a nominal plastic strain of 13.5\% is achieved, we plotted the normalized contact pressure based on the finite element simulation and rigid-plastic model, shown in Figure ~\ref{fig:abaqus-analysis-cropped.pdf}. The minimum rolling gap location is at $x=0$. Clearly the contact zone and normalized contact pressure for elasto-plastic analysis are significantly larger then the rigid-plastic scheme. This is the primary reason for large polymer rolling loads encountered, which cannot be captured by the classical rolling theories. Such aspects of rolling mechanics are  necessarily relevant if polymer films, as those exhibiting hyper-elasticity and plasticity, are to be roll-bonded in a similar fashion.

\begin{figure}[]
  \centering
      \begin{minipage}{\textwidth}
   \centering
\includegraphics[scale=1.0]{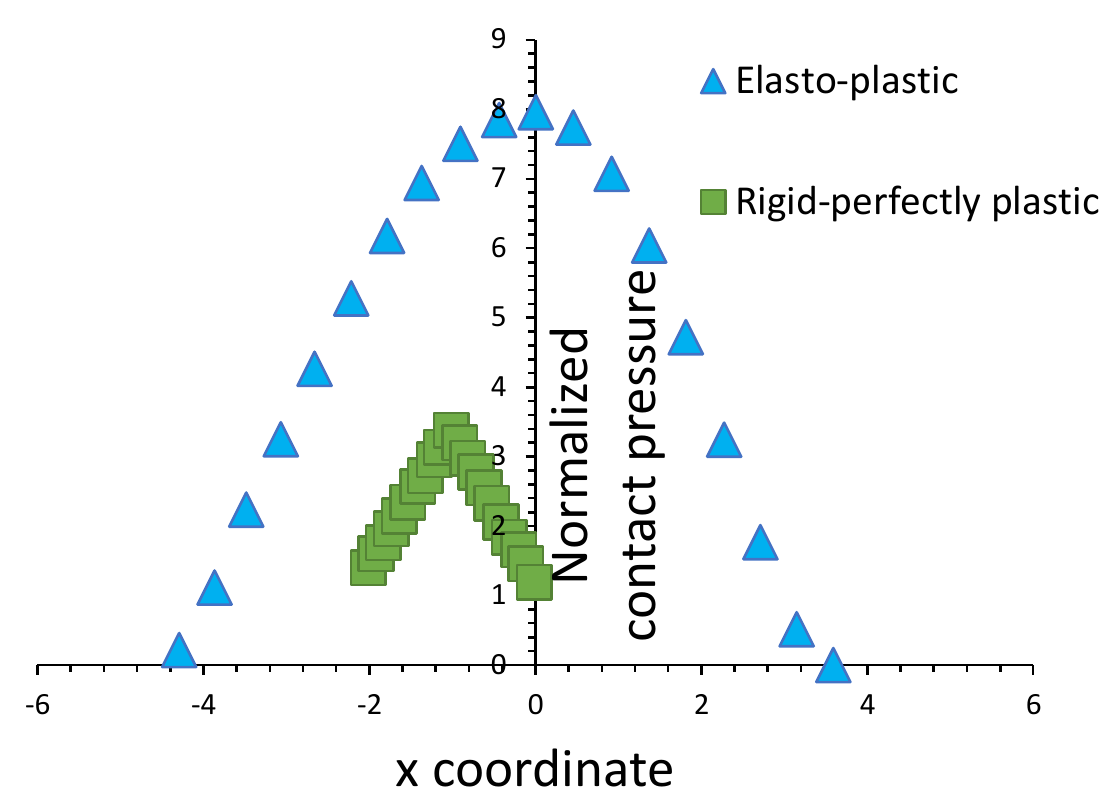}
\caption{Variation of contact pressure in the roller bite for $13.5$\% nominal strain based on elasto-plastic finite deformation in ABAQUS and rigid-perfect-plastic model.
The contact pressure has been normalized by yield strength of the material. }
\label{fig:abaqus-analysis-shear-and-normal-stress-cropped}
    \end{minipage}
\end{figure}

%
%

\section{Equivalency between hypoelasticity and multiplicative decomposition of deformation gradient formulations}

In Section ~\ref{sec:hypoelasticity}, we showed that hypoelastic formulation and associated additive decomposition of elastic and plastic spatial strain rates can we derived from $\mathbf{F}\superscript{e} \mathbf{F}\superscript{p}$ formulation under the assumptions of small elastic stretches, i.e, 
$\mathbf{U}\superscript{e} \approx \mathbf{I}$. However, for moderately large rolling strains, the 
elastic stretches are non-negligible, and the hypoelastic formulation is still found to yield satisfactory 
performance. This behavior can be understood by noting that even though elastic stretches are 
moderately large, the rotation tensors for material elements during rolling are close to identity when 
stack thicknesses are very small compared to the roller radii. 
Consider Equation ~\ref{expression-for-total-stretching-expansion-2}, where the spatial strain rate for 
$\mathbf{F}\superscript{e} \mathbf{F}\superscript{p}$ formulation is given as
\begin{equation}
\mathbf{D} =  \mathbf{D}\superscript{e} +  Sym \big[ \mathbf{R}\superscript{e}\mathbf{U}\superscript{e}(   \mathbf{D}\superscript{p}) 
\mathbf{U}\superscript{e-1}\mathbf{R}\superscript{e-1} \big], \nonumber
\end{equation}
now if we approximate, both, $\mathbf{R}\superscript{e}$ and $\mathbf{R}\superscript{p}$ equal to  
$\mathbf{I}$, and assume that material undergoes homogeneous plane strain compression such that
$\mathbf{U}\superscript{e}$, $\mathbf{U}\superscript{p}$, $\mathbf{D}\superscript{e}$
and $\mathbf{D}\superscript{p}$ are diagonal, then
\begin{equation}
\mathbf{D} =  \mathbf{D}\superscript{e} +  Sym \big[ \mathbf{U}\superscript{e}(   \mathbf{D}\superscript{p}) \mathbf{U}\superscript{e-1}\big].
\label{equivalence-2}
\end{equation}
Accordingly $\mathbf{U}\superscript{e}(\mathbf{D}\superscript{p})\mathbf{U}\superscript{e-1}$ is now symmetric, then Equation ~\ref{equivalence-2} reduces to 
\begin{equation}
\mathbf{D} =  \mathbf{D}\superscript{e} +  \mathbf{U}\superscript{e}(   \mathbf{D}\superscript{p}) \mathbf{U}\superscript{e-1},
\label{equivalence-3}
\end{equation}
or
\begin{equation}
\mathbf{D} =  \mathbf{D}\superscript{e} +   \mathbf{D}\superscript{p},
\label{equivalence-4}
\end{equation}
where we have used the fact that 
$\mathbf{U}\superscript{e}(\mathbf{D}\superscript{p}) \mathbf{U}\superscript{e-1}=\mathbf{D}\superscript{p}$, when $\mathbf{U}\superscript{e}$ and  $\mathbf{D}\superscript{p}$ are diagonal. 
Equation ~\ref{equivalence-4} is the additive decomposition of spatial strain-rate, and same as
Equation ~\ref{expression-for-total-stretching-expansion-4} when elastic rotations are identity. We take 
note of the fact that although we have assumed plastic spin to be zero, in general that does not imply 
that plastic part of rotation tensor $\mathbf{R}\superscript{p}$ is identity, or even constant. In the present rolling scenarios, the 
geometry and mechanics of deformation, yield all rotation tensors close to identity. Secondly, in the 
hypoelastic formulation of Section ~\ref{sec:hypoelasticity}, we constitutively 
related the objective stress-rate with 
elastic stretching Equation ~\ref{sigma-ij-jaumann}, and in $\mathbf{F}\superscript{e} \mathbf{F}\superscript{p}$ formulation we constitutively related Mandel stress to Hencky strain, and then Cauchy stress with Mandel stress, Equations ~\ref{mandel-stress} and ~\ref{cauchy-mandel-relation-1}, 
respectively. Clearly, in Equation ~\ref{cauchy-mandel-relation-1} the Mandel stress is equal to Cauchy 
stress under identity elastic rotation tensors but for a factor of $J$ (capturing volume change between elastically 
and plastically deformed states), whereas in hypoelasticity  the Cauchy stress relates to Hencky strain 
according to Equation ~\ref{sigma-ij} without accounting for such volume changes between incremental elastic and plastic deformations.  These differences also explain the small differences between loads predicted by hypoelastic and $\mathbf{F}\superscript{e} \mathbf{F}\superscript{p}$ formulation.

\section{Conclusions}
\label{sec:conclusion}
Cold roll-bonding of polymeric films is a new process, which requires through-thickness and 
homogeneous plastic deformation of the incoming material strip. In this paper, we have proposed 
a computationally efficient material modeling approach to carry out cold-rolling finite element simulations. Although analyses of rolling processes have been extensively carried out over 
last several decades, majority of the classical efforts have focused on applications related to metal 
deformation, where rigid-perfectly plastic type models are applicable. The distinction between the 
polymer rolling vs metal rolling is that the polymers can exhibit large elastic stretches, which are 
completely ignored in the classical rolling theories, and therefore the classical rolling theories cannot be 
used for making accurate predictions of rolling loads for polymers in general. We noted that in case of polymer rolling, the 
contact-width and contact-pressure in the roller bite, comprising both elastic and elastic-plastic regions, 
are substantially larger than those predicted by the rigid-plastic rolling scheme. 
Finite element simulations, taking elastic deformations into account,  up to 20\%
thickness reductions, give an accurate prediction of the 
actual rolling loads. While a sophisticated  polymer model, with hyperelastic and visco-plastic effects, 
can be chosen in polymer rolling, we exploited the fact that at low strain-rates and temperatures well-below T$_g$ 
the polymers do not exhibit rate-sensitivity, thereby allowing us to work with a rate-independent material model which is computationally much faster than simulating a rolling process with 
rate-dependent effects.  Both, rate-independent hypoelastic and multiplicative decomposition, yielded 
similar predictions and compared well against the experimentally observed loads. In principle, advanced polymer models can also be calibrated and applied, however, the computational cost involved in various steps during the time-integration of these sophisticated material-models can be significant, without any foreseeable benefits in predictive capability, especially when we are interested in long-time steady-state rolling. An analysis was 
presented which revealed that hypoelastic material model (which is usually valid for small elastic 
stretches) is applicable to moderately large elastic stretches, as encountered in rolling of polymers in this 
study, when principle axes of deformation show negligible rotation (due to large 
roller radii compared to film thickness). Under such circumstances using a hypoelastic model can also be 
considered as a more efficient approach compared to $\mathbf{F}\superscript{e} \mathbf{F}\superscript{p}$ 
formulation.  We also found that modeling frictional interaction through a rate-dependent  Coulombic law, consistent with the rate-independent material model, gave satisfactory results in the rolling simulations.

\section{Acknowledgments}
Nikhil Padhye gratefully thanks David Parks for pointing out that rate-independent elasto-plastic analysis could suffice for modeling the plane-strain rolling at low temperatures and strain-rates.

\bibliographystyle{acm}
\bibliography{main}

%

\end{document}